\documentclass[final,3p,12pt]{elsarticle}

\usepackage{graphicx,multirow}
\usepackage{amssymb}
\usepackage{amsthm}

\usepackage{nomencl}
\usepackage{amsmath}
\usepackage{xcolor}
\usepackage{hyperref}
\usepackage{import}
\usepackage{natbib}
\usepackage{xfrac}
\usepackage{pgfplots}
\usepackage{pifont}
\usepackage{tikz}
\usetikzlibrary{arrows.meta}
\usepackage{siunitx}
\usetikzlibrary{shapes.multipart}

\usepackage{caption} 
\newcommand\sbullet[1][.5]{\mathbin{\vcenter{\hbox{\scalebox{#1}{$\bullet$}}}}}
\newcommand{\ccirc}{\vcenter{\hbox{$\scriptstyle{\circ}$}}}

\newcommand{\Fref}[1]{Figure~\ref{#1}}
\newcommand{\Sref}[1]{Section~\ref{#1}}
\newcommand{\Tref}[1]{Table~\ref{#1}}

\journal{arXiv.org}

\pgfplotsset{compat=1.16}

\begin{document}

\begin{frontmatter}

\title{Experimental study on the gradual fracture of layers\\ in multi-layer laminated glass plates under low-velocity impact}

\author{Alena Zemanová, Petr Konrád, Petr Hála, Radoslav Sovják, \\Radim Hlůžek, Jan Zeman, Michal Šejnoha
}

\address{Czech Technical University in Prague, Czech Republic}

\begin{abstract}
Through pendulum impact testing on suspended samples, we demonstrate the effect of the multi-layer layout on the low-velocity impact response of laminated glass plates consisting of three or four glass layers and PVB interlayers. Non-destructive tests proved the repeatability and the consistency of performed experiments. Destructive tests revealed significant differences in impact energies and breakage forces leading to glass fracture for individual specimens under impacts of increasing height. Nevertheless, the fracture consistently initiated in outer glass layers around the impact point and vibrations of partially fractured samples exhibited similar first natural frequencies. Partially fractured samples also withstood in many cases higher contact forces than those leading to the fracture of the previous glass layer, and PVB interlayers provided a stiff shear connection between glass layers for all impact heights. Altogether, our experimental results provide comprehensive information on the pre-fracture, fracture, and post-fracture response of laminated glass, suitable particularly for validating computational models for multi-layer glass samples under low-velocity impact.

\end{abstract}

\begin{keyword}
Laminated glass \sep Multi-layer plate \sep Experimental testing \sep Low-velocity impact \sep Hard-body impact \sep Gradual fracture of layers 

\end{keyword}

\end{frontmatter}

\section{Introduction}
\label{S:Intro}

Laminated glass is a multi-layer composite that consists in its basic configuration of two glass layers connected with an interlayer, typically made of polymers. The softer interlayer provides damping of vibrations and connects glass shards after fracture, thus its presence improves the post-fracture resistance of brittle glass elements and thereby maintains the structural integrity of laminated glass components after glass breakage. 
The basic three-layer configuration was further extended to multi-layer laminated glass that has found its application for fail-safe, blast-resistant, or impact-resistant transparent structures~\citep{mori2016design, wever2016swimming, dotan2016zhangjiajie}. 

Despite the improvement in global fracture and post-critical structural response of laminated components, glass layers remain fragile and
vulnerable to damage. Accidental or intentional impact events represent one of reasons for the fracture of glass layers or destruction of the whole structural elements~\citep{overend2007diagnostic}. Based on the range of the initial velocity of the impactor, 
low-velocity impacts characterised by velocities (less than 25~m$\cdot$s$^{-1}$) corresponding to freely falling objects, impacts by a drop tower facility or a pendulum device, and forced-entry attacks by punching, kicking, and hitting with hard objects \citep{backman1978mechanics, mohagheghian2017quasi}. 
High-velocity impacts exhibit velocities (greater than 25~m$\cdot$s$^{-1}$) that are reached by projectiles fired by guns or other laboratory apparatus.
According to this classification, the laboratory tests performed on laminated glass specimens and presented 
in this study represent the category of low-velocity impact tests.

A significant part of research studies focused on laminated glass under dynamic loading is devoted to blast-loaded glass elements or to bullet impacts belonging to high-velocity impacts~\citep{larcher2012experimental, zhang2013parametric, pelfrene2016critical, liu2016energy, chen2021experimental, osnes2019dynamic, osnes2019fracture, osnes2021perforation}.
For low-velocity impacts, two categories referred to as soft or hard impacts can be found in the literature differing in the deformation of the impactor. During the contact, the deformation of the impactor is negligible compared to that of the target for hard body impacts, while its extensive deformation is observed for soft body impacts~\citep{mohagheghian2017quasi}. The largest group of low-velocity impact studies on laminated glass is motivated by the needs of car industry such as head or other impacts on a windshield~\citep{zhao2006analysis, untaroiu2007design, timmel2007finite, xu2010characteristics, pyttel2011failure, chen2017numerical, alter2017enhanced}, where thin three-layer laminated glass samples were tested and their fracture response and pattern were simulated numerically. For aircraft windshields, chemically strengthened glass is employed for glass layers and 
the performance of these components under quasi-static bending and low velocity impact is discussed in~\citep{mohagheghian2017quasi}.
Findings on windshields are often transferred to architectural glass, but it must be taken into account that the thicknesses of specimens presented in all mentioned studies are relatively small, and therefore, they may represent glazing infill in building rather than  building structural components with a load-bearing function. 

From the area of architecture, laminated glass window vulnerability to debris impact was studied in~\citep{zhang2013laboratory}, where the effect of interlayer thickness on the window capacity to prevent debris penetration was observed. 
Finite element models providing stresses in laminated glass units
under wind-borne debris impacts were analysed in~\citep{behr1999dynamic, flocker1997stresses}. 
The hard impact damage mechanisms of laminated glass were evaluated in~\citep{zhang2019temperature, zhang2020impact}.
Their observations proved the temperature-, time-, and strain rate-dependent properties of interlayers significantly influencing the response of laminated glass samples.
For three-layer laminated glass, the effect of thicknesses of individual glass layers on the impact resistance was examined in~\citep{grant1998damage}.
Their impact tests, using chippings accelerated by a catapult system, indicated that the thickness of the impacted glass layer was the key parameter determining the critical velocity for fracture initiation. The authors also identified different types of cracking associated with the impact velocity, radius of curvature of the impactor, impact angle and panel thickness, e.g., star cracking was observed in samples with thin impacted layers whereas cone cracking in those with thick impacted layers. They also reported significant influence of stone geometry impacting the plate because a sharp corner produced more severe damage than a flat face.
Finally, pendulum impact tests, employed in design practice to verify the impact resistance and strength of glass balustrades, have been discussed in the literature, and analytical or numerical models were introduced to replace the experimental testing~\citep{froling2014reduced, pelfrene2016numerical, kozlowski2019experimental, bez2021calibrated, viviani2021engineered}. Nevertheless, the numerical simulation of laminated glass is burdened by the fact that the tensile strength of glass is stochastic, driven by the initial surface flaw pattern, and stress-rate-dependent due to the sub-critical growth of initial cracks~\citep{overend2012computer, alter2017enhanced, osnes2019fracture}.

Regarding the post-fracture performance of three-layer laminated glass plates under impact loading, so called sacrificial-glass-ply design is often emphasised in the literature~\citep{kaiser2000impact, saxe2002effects, foraboschi2013hybrid,  wang2020post, wang2021optimal}. The idea of this concept is that one glass layer (usually the impacted one) is allowed to fracture to dissipate impact energy, whereas the other glass layer is preserved to further sustain other loads. Thus, the sacrificial ply is not taken into account for the load-bearing capacity during the design of a laminated glass element~\citep{foraboschi2013hybrid}.

Limited amount of works can also be found for the impact response of multi-layer laminated glass in the last decade, identifying the critical impact load for laminated glass panels with a few broken glass layers.
\citeauthor{wang2018experimental} pointed out, in their experimental study on low-velocity impact resistance of multi-layer glass panels \citep{wang2018experimental}, that the peak impact force at the second breakage was in some cases larger than that at the first breakage.
Similar conclusion was reached by~\citeauthor{overend2007diagnostic} in~\citep{overend2007diagnostic} for laminated glass beams under quasi-static bending, where the post-fracture bearing capacity of tested multi-layer laminated glass units was significant and also exceeded in some cases the strength at first fracture.
In~\citep{wang2018experimental}, the authors also reported an improvement in post-breakage stiffness of samples with heat strengthened glass layers compared to fully tempered glass layers due to their beneficial fragmentation pattern.
Further, experimental results in~\citep{zhao2019experimental} confirmed that the annealed glass plies provided the highest resistance in the post-breakage phase.
The post-fracture impact performance of laminated glass has been recently analysed in~\citep{wang2020post}, where the results indicated 
that the intact middle glass layer in a three-glass-ply laminate unit was crucial to sustain impact and dynamic stiffness at a post‐fracture stage.

These conclusions indicate that the cracked multi-layered panels might still provide a significant load bearing capacity for consecutive loading. Therefore, the timely replacement of laminated glass plates with a few cracked glass layers might not always be necessary because the damaged multi-layered plates are able to transfer a certain level of loading and  the building envelope integrity is preserved.
Nevertheless, let us emphasise that the multiple layers in laminated glass do not automatically provide an improved performance of a laminated glass sample. As observed in numerical analyses~\citep{flocker1998low},  three-layer samples may offer better impact resistance than five-layer or seven-layer laminated glass of the same overall thickness under certain conditions.

The literature review indicates that the experimental studies focusing on the response of multi-layer laminated glass panels are still rare even though these elements have been employed in buildings and other applications. 
Recently, the authors in~\citep{wang2020post} have emphasised the lack of data on the dynamic performance of fractured laminated glass panels at different breakage stages.
Therefore, this paper tries to narrow this gap and summarises the results of laboratory low-velocity impact tests on two types of \textit{multi-layer laminated glass} samples. 
In contrast with~\citep{wang2020post}, also laminated glass samples with four glass layers were represented in our experimental study and different boundary conditions were employed.
We aim to discuss the effect of \textit{different cross-section layout} of glass and polymer layers and  the effect of  \textit{gradual fracture} of independent glass layers on the response of the multi-layer composite to an impact event.
Except for the different cross-section geometry, the response of laminated glass is pronouncedly modified by the polymer material employed for interlayer foils. 
For this study, the industry-standard PVB-based laminated glass plates made of annealed float glass were selected, representing the most common material combination of structural laminated glass. 

The results presented here provide comprehensive information about the pre-fracture, fracture, and post-fracture response of laminated glass samples and thus may serve as a basis for validation of numerical models of laminated glass plates under the low-velocity impact loading 
for other researchers from the glass community.
To this purpose, all recorded and processed data are available in the GitLab repository~\citep{git_data}. 

\section{Experiments}
\label{S:EXP}

\subsection{Laminated glass samples}
\label{S:Samples}

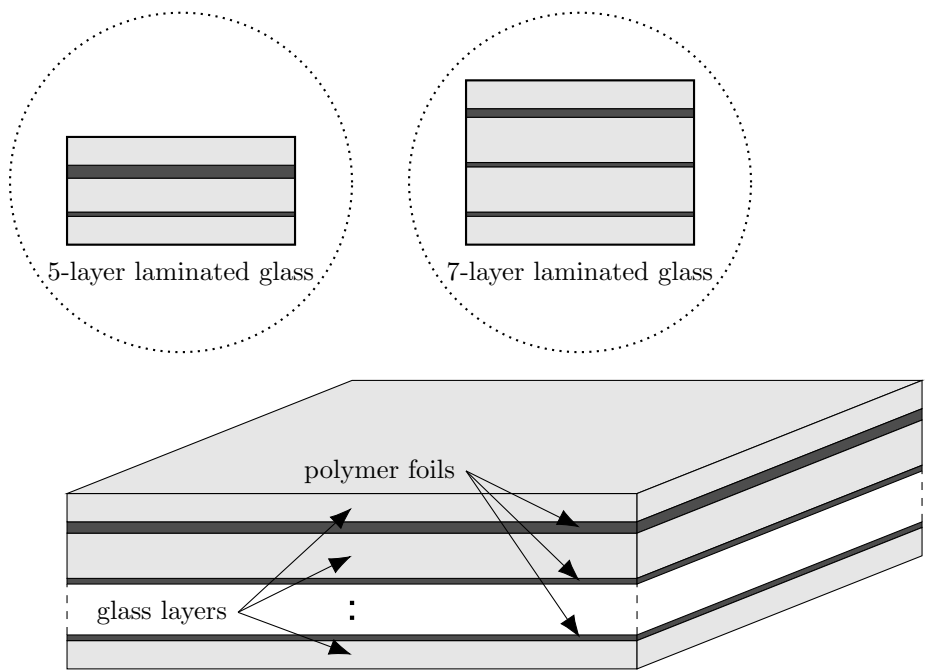
\begin{figure}[ht]
    \centering
{\begin{tikzpicture}[scale=0.75]
            \filldraw[fill=gray!20] (0mm,0mm) rectangle (100mm,5mm);
            \filldraw[fill=black!70] (0mm,5mm) rectangle (100mm,6mm);
            \draw[dashed] (0mm,6mm) -- (0mm,15mm);
            \draw[dashed] (100mm,6mm) -- (100mm,15mm);
            \draw[dashed] (150mm,26mm) -- (150mm,35mm);
            \filldraw[fill=black!70] (0mm,15mm) rectangle (100mm,16mm);
            \filldraw[fill=gray!20] (0mm,16mm) rectangle (100mm,24mm);
            \filldraw[fill=black!70] (0mm,24mm) rectangle (100mm,26mm);
            \filldraw[fill=gray!20] (0mm,26mm) rectangle (100mm,31mm);
            \draw[ultra thick, loosely dotted,shorten >=2mm, shorten <=2mm] (50mm,6mm) -- + (0, 9mm);
            \draw[{Latex[length=3mm]}-] (50mm, 2.5mm) -- (30mm, 10mm)  node[left]{\footnotesize glass layers};
            \draw[{Latex[length=3mm]}-] (50mm, 20mm) -- (30mm, 10mm);
            \draw[{Latex[length=3mm]}-] (50mm, 28.5mm) -- (30mm, 10mm);
            \draw[fill=gray!20] (0mm,31mm) -- (100mm,31mm) -- (150mm,51mm) -- (50mm,51mm) -- cycle;
            \draw[fill=gray!20] (100mm,31mm) -- (150mm,51mm) -- (150mm,46mm) -- (100mm,26mm) -- cycle;
            \draw[fill=black!70] (100mm,26mm) -- (150mm,46mm) -- (150mm,44mm) -- (100mm,24mm) -- cycle;
            \draw[fill=gray!20] (100mm,24mm) -- (150mm,44mm) -- (150mm,36mm) -- (100mm,16mm) -- cycle;
            \draw[fill=black!70] (100mm,16mm) -- (150mm,36mm) -- (150mm,35mm) -- (100mm,15mm) -- cycle;
            \draw[fill=black!70] (100mm,6mm) -- (150mm,26mm) -- (150mm,25mm) -- (100mm,5mm) -- cycle;
            \draw[fill=gray!20] (100mm,5mm) -- (150mm,25mm) -- (150mm,20mm) -- (100mm,0mm) -- cycle;
            \draw[{Latex[length=3mm]}-] (90mm, 5.5mm) -- (70mm, 35mm) node[left]{\footnotesize polymer foils};
            \draw[{Latex[length=3mm]}-] (90mm, 15.5mm) -- (70mm, 35mm);
            \draw[{Latex[length=3mm]}-] (90mm, 25mm) -- (70mm, 35mm);
            
            \draw[thick, dotted] (20mm,86mm) circle (30mm);
            \node at (20mm,70mm) {\footnotesize 5-layer laminated glass};
            \filldraw[fill=gray!20] (0mm,75mm) rectangle (40mm,80mm);
            \filldraw[fill=black!70] (0mm,80mm) rectangle (40mm,80.76mm);
            \filldraw[fill=gray!20] (0mm,80.76mm) rectangle (40mm,86.76mm);
            \filldraw[fill=black!70] (0mm,86.76mm) rectangle (40mm,89.04mm);
            \filldraw[fill=gray!20] (0mm,89.04mm) rectangle (40mm,94.04mm);
            \draw[thick] (0mm, 75mm) rectangle (40mm,94.04mm);
            \draw[thick, dotted] (90mm,86mm) circle (30mm);
            \node at (90mm,70mm) {\footnotesize 7-layer laminated glass};
            \fill[fill=gray!20] (70mm,75mm) rectangle (110mm,80mm);
            \filldraw[fill=black!70] (70mm,80mm) rectangle (110mm,80.76mm);
            \filldraw[fill=gray!20] (70mm,80.76mm) rectangle (110mm,88.76mm);
            \filldraw[fill=black!70] (70mm,88.76mm) rectangle (110mm,89.52mm);
            \filldraw[fill=gray!20] (70mm,89.52mm) rectangle (110mm,97.52mm);
            \filldraw[fill=black!70] (70mm,97.52mm) rectangle (110mm,99.04mm);
            \filldraw[fill=gray!20] (70mm,99.04mm) rectangle (110mm,104.04mm);
            \draw[thick] (70mm, 75mm) rectangle (110mm,104.04mm);
\end{tikzpicture} }
    \caption{Scheme of multi-layer laminated glass plates.}
    \label{fig:LGscheme}
\end{figure}

Non-destructive and destructive impact tests were carried out on laminated glass units, where two different geometries (\Fref{fig:LGscheme}) were selected for the experimental campaign, i.e., 5-layer laminated glass (5LG) containing three glass plies and two interlayers and 7-layer laminated glass (7LG) made from four glass plies connected with three interlayers. 

Motivated by the sacrificial-glass-ply concept~\citep{kaiser2000impact} and the key role of the middle glass layer~\citep{wang2020post}, we designed the thicknesses of both external glass layers to be thinner, whereas the inner
selected
glass layer/s in tested samples were slightly thicker. 
This setup might be also convenient for the post-fracture response of laminated samples because the thicker glass layers are protected by a polymer foil and an additional glass layer.
Therefore, fewer micro-defects due to the transport and manipulation before the impact tests should occur on the protected (inner) glass surfaces implying that the tensile strength of inner layers should exceed the outer ones~\citep{alter2017enhanced}.
Finally, a thin outer glass ply reduces the possibility of undesirable extensive interlayer debonding~\citep{flocker1998low}.

\Tref{tab:samples} summarises the dimensions of samples. Glass layers were arranged symmetrically in the laminated cross-section because both external surfaces could be the impacted ones, whereas a thickness of an interlayer foil was increased.
Three sets of values are presented for the thicknesses: nominal corresponding to the thicknesses given by the manufacturer, reduced assuming the largest manufacturing tolerances, and a range of overall thicknesses measured on four tested samples (i.e., an average from eight values measured on all edges of a sample). Apparently, the measured thicknesses consistently approach the minimum values allowed by manufacturing tolerances.

\begin{table}[ht]
\caption{Configuration and geometry of 5-layer and 7-layer laminated glass samples, denoted (5LG) and (7LG) respectively. The two set of values for thicknesses correspond to the nominal values given by manufacturer or to their reduced counterparts considering the manufacturing tolerances for glass and polymer layers.}
    \centering
    \footnotesize
    \centering
    \begin{tabular}{lll}
         \hline
         & 5LG & 7LG \\
         \hline
         number of layers & 5 & 7 \\
         glass type & float annealed & float annealed\\
         interlayer material & PVB (TROSIFOL BG R20) & PVB (TROSIFOL BG R20)\\
         number of tested samples & 4 & 4\\
         width $b$ (mm) & 500 & 500 \\
         length $l$ (mm) & 500 & 500 \\
         nominal thicknesses of layers (mm) & 5/2.28/6/0.76/5 & 5/1.52/8/0.76/8/0.76/5 \\
         overall nominal thickness (mm) & 19.04 & 29.04 \\
         reduced thicknesses of layers (mm) & 4.8/2.08/5.8/0.76/4.8 & 4.8/1.52/7.7/0.76/7.7/0.76/4.8 \\
         reduced overall thickness $h$ (mm) & 18.24 & 28.04 \\
         measured overall thicknesses (mm) & 18.2--18.5 & 28.0--28.2 \\
         \hline
    \end{tabular}
    \label{tab:samples}
\end{table}

\subsection{Pendulum impactor device}
\label{S:Setup}

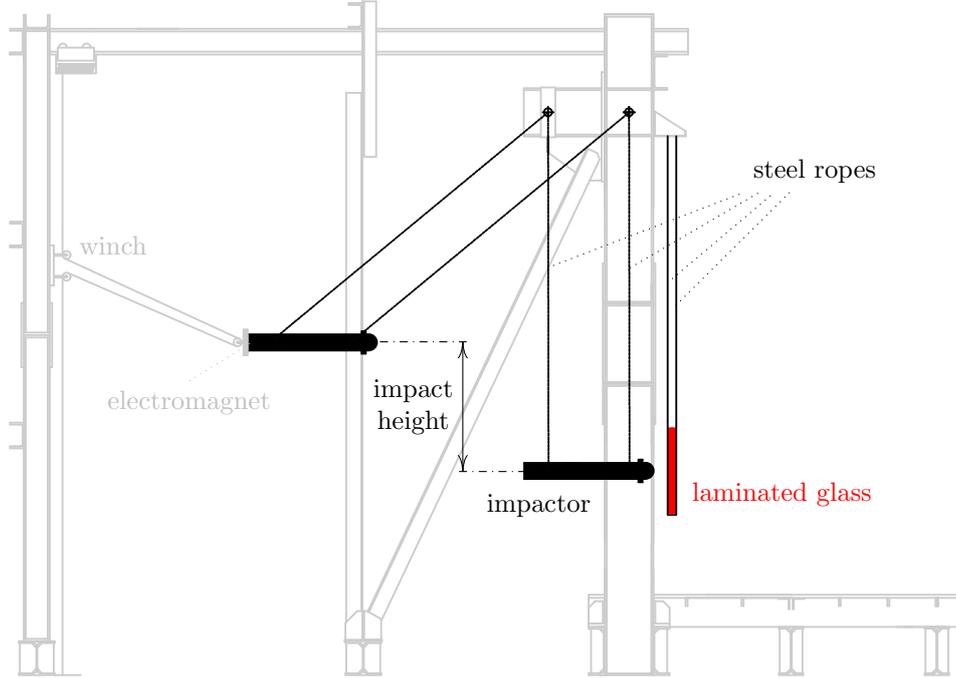
\begin{figure}[ht]
    \centering
    \begin{tikzpicture}[every text node part/.style={align=center}]
    \begin{footnotesize}
\node[]{\includegraphics[width=0.8\textwidth]{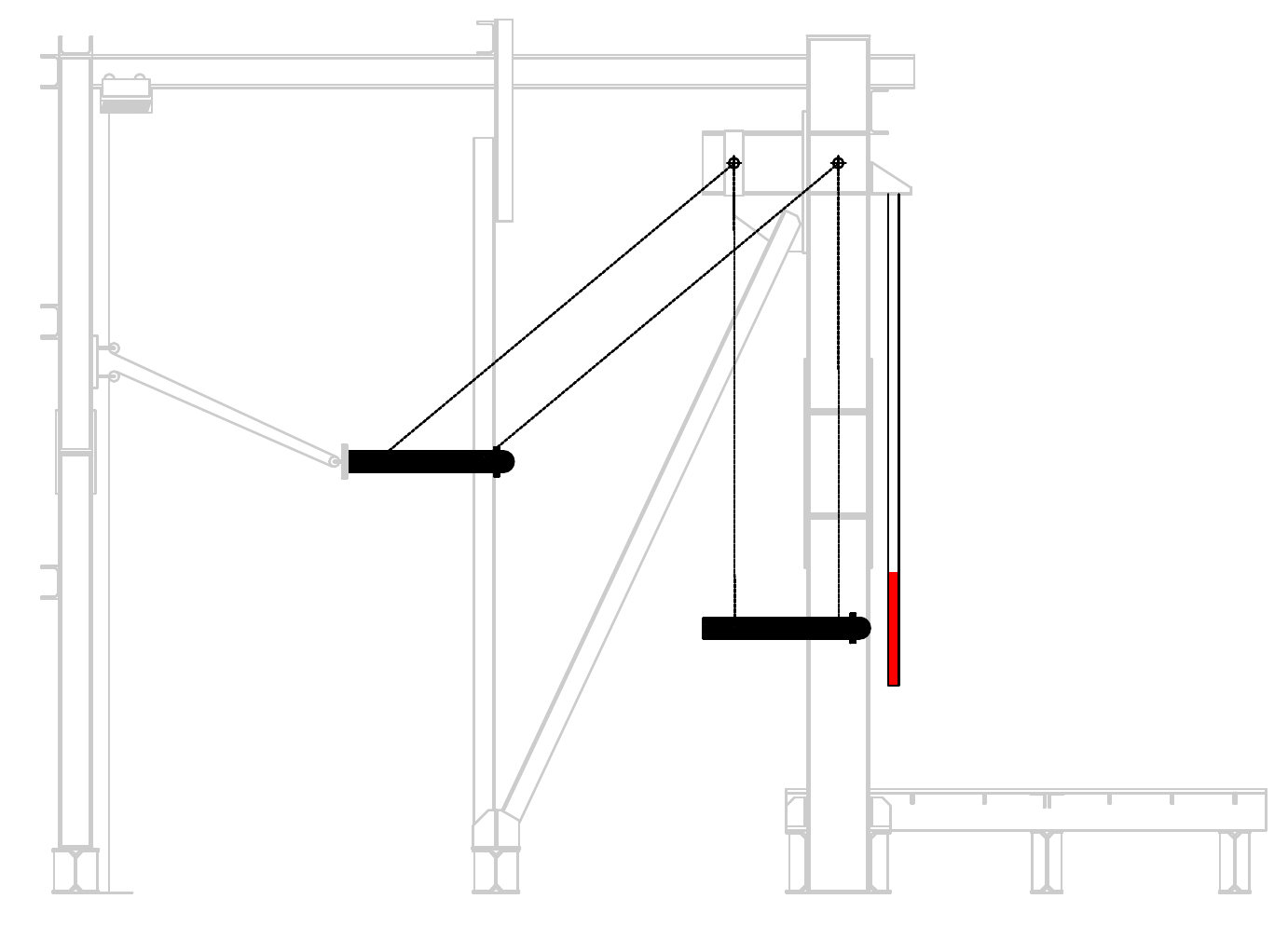}};
    \node[red] at (40mm,-20mm) {laminated glass};
    \node[] at (8mm,-21.5mm) {impactor};
    \node[lightgray] at (-48mm,13mm) {winch};
    \draw[-, lightgray, dotted] (-30mm, 0mm) -- (-38mm, -5mm) node[below, lightgray]{electromagnet};
    \draw[-, dotted] (9.1mm, 10mm) -- (35mm, 20mm) node[above right]{steel ropes};
    \draw[-, dotted] (20mm, 10mm) -- (37mm, 20mm);
    \draw[-, dotted] (24.9mm, 8mm) -- (39mm, 20mm);
    \draw[-, dotted] (26mm, 5mm) -- (41mm, 20mm);
    \draw[{<[scale=2.5,length=2.5, width=1.5]}-{>[scale=2.5,length=2.5, width=1.5]}] (-2mm,-17mm) -- (-2mm, 0.2mm) node [midway, left] {impact \\ height};
    \draw[-, dash dot] (-2.5mm, -17mm) -- (8mm, -17mm);
    \draw[-, dash dot] (-12.9mm, 0.2mm) -- (-1.5mm, 0.2mm);
	\end{footnotesize}
    \end{tikzpicture}
    \caption{Side view of the pendulum impactor device.}
    \label{fig:SCH}
\end{figure}

The low-velocity impact was induced by a pendulum device, which works on the principle of a free-falling weight following a circular path (\Fref{fig:SCH}).
The sample of the laminated glass was suspended on a pair of steel ropes 
placed along the vertical sides of the glass sample with both sides approximately 50~mm from the vertical edges of the sample.
In this way, steel ropes were wrapped around the bottom edge of the sample (\Fref{fig:exp_photo}). 
Such type of supports is utilised in experimental modal or other dynamic analysis of structures to approximate the free boundary conditions~\citep{pyttel2011failure}.

The impactor had the shape of a cylinder with a hemispherical nose with a radius of 50~mm 
and 800~mm in length.
The impact was carried out by lifting the 48.2~kg impactor to the prescribed height with a winch and an electromagnet and then releasing it onto a glass sample. 
After the impact, the sample started to move in the direction of the impulse along the circular path
where it was captured 
to prevent multiple impacts.

The vertical distance between the impactor's axis of symmetry and the centre of a sample (impact height) was initially 5~cm. After each test, this distance was increased by 5~cm. Most of the samples were tested for the range of the impact heights 5--50~cm, except the sample 5LG--4 that exhibited higher impact resistance and was tested up to the impact height of 80~cm. 
However, performing the experiments for impact heights over 50~cm turned out to be dangerous for the staff and was not repeated for the other 7LG-samples with unfractured glass layers.
The largest impact heights $h_\text{imp}$ of 50~cm or 80~cm resulted in initial velocities of the impactor $v_\text{init}$ up to 3 m$\cdot$s$^{-1}$ and 4 m$\cdot$s$^{-1}$, respectively, (according to $v_\text{init}=\sqrt{2 g h_\text{imp}}$ with the gravitational acceleration $g=$~9.81~m$\cdot$s$^{-2}$)
and impact energies $E_\text{imp}$ up to 236~J and 378~J, respectively, (defined as the initial kinetic energy of the impactor $E_\text{imp} = 0.5 m_\text{imp} v_\text{init}^2$ with the impactor's mass $m_\text{imp} = 48.2$~kg). 
The critical impact energy and breakage force for fracture initiation might have been affected by the repeated impact tests due to the fatigue growth of pre-existing micro-cracks. However, the loading scheme, i.e., the initial impact height, its gradual increase, and parameters of the impactor, were fixed for all tested samples, and therefore, the possible effect of multiple impacts on the critical breakage  energy and force should be consistent for all samples at the same impact height. 
All impact tests were performed at room temperature (25\textpm 1$\,^\circ$C).

\subsection{Equipment used to record the response}

\begin{figure}[ht]
    \centering
    \begin{tabular}{cc}
{\begin{tikzpicture}[scale=1.2]
\filldraw[fill=gray!20] (5mm,7mm) rectangle (55mm,57mm);
\draw[very thick] (10mm,7mm) -- (10mm,69mm);
\draw[very thick] (50mm,7mm) -- (50mm,69mm);
\filldraw[red] (6mm,8mm) circle (0.7mm) node [red,below=1mm] {$C$};
\filldraw[red] (6mm,32mm) circle (0.7mm) node [red, below left] {$M'$};
\filldraw[red] (17.5mm,8mm) circle (0.7mm) node [red,below=1mm] {$Q$};
\filldraw[red] (30mm,8mm) circle (0.7mm) node [red,below=1mm] {$M$};
\draw[{<[scale=2.5,length=2.5, width=1.5]}-{>[scale=2.5,length=2.5, width=1.5]}] (5mm,-2mm) -- (55mm,-2mm) node [midway, above] {\footnotesize 500~mm};
\draw[{<[scale=2.5,length=2.5, width=1.5]}-{>[scale=2.5,length=2.5, width=1.5]}] (62mm,7mm) -- (62mm,57mm) node [midway, above, sloped] {\footnotesize 500~mm};
\draw[{<[scale=2.5,length=2.5, width=1.5]}-{>[scale=2.5,length=2.5, width=1.5]}] (10mm,60mm) -- (50mm,60mm) node [midway, above] {\footnotesize 400~mm};
\draw[{<[scale=2.5,length=2.5, width=1.5]}-{>[scale=2.5,length=2.5, width=1.5]}] (5mm,12mm) -- (17.5mm,12mm) node [midway, above] {\footnotesize \hspace{3mm} 125};
\draw[{<[scale=2.5,length=2.5, width=1.5]}-{>[scale=2.5,length=2.5, width=1.5]}] (17.5mm,12mm) -- (30mm,12mm) node [midway, above] {\footnotesize 125};
\draw[{<[scale=2.5,length=2.5, width=1.5]}-{>[scale=2.5,length=2.5, width=1.5]}] (0mm,7mm) -- (0mm,32mm) node [midway, above, sloped] {\footnotesize 250~mm};
\draw[dotted] (5mm,-2mm) -- (5mm,7mm);
\draw[dotted] (55mm,-2mm) -- (55mm,7mm);
\draw[dotted] (55mm,7mm) -- (62mm,7mm);
\draw[dotted] (55mm,57mm) -- (62mm,57mm);
\draw[dotted] (0mm,7mm) -- (5mm,7mm);
\draw[dotted] (0mm,32mm) -- (5mm,32mm);
\draw[dotted] (17.5mm,7mm) -- (17.5mm,12mm);
\draw[dotted] (30mm,7mm) -- (30mm,12mm);
\draw[-{Latex[length=3mm]}, red, thick] (30mm,32mm) -- (15mm,32mm);
\draw[red] (15mm,35mm) node {\text{$x$}};
\draw[-{Latex[length=3mm]}, red, thick] (30mm,32mm) -- (30mm,17mm);
\draw[red] (33mm,17mm) node {\text{$y$}};
\draw[-, dotted] (10mm, 40mm) -- (20mm, 50mm) node[above]{\footnotesize steel ropes};
\draw[-, dotted] (50mm, 40mm) -- (20mm, 50mm);
\draw[dashed] (30mm,32mm) circle (5mm);
\draw[red] (30mm,38.2mm) circle (0.7mm) node [red,above left] {$I$};
\draw[dashed] (28.8mm,39.4mm) -- (31mm,39.4mm);
\draw[dashed] (28.8mm,39.4mm) -- (28.8mm,36.9mm);
\draw[dashed] (31.2mm,39.4mm) -- (31.2mm,36.9mm);
\node[] at  (41mm,28mm) {\footnotesize impactor};
\node[] at  (45mm,4.5mm) {\footnotesize glass sample};
\end{tikzpicture} }
         &  
    {\begin{tikzpicture}[scale=1]
\node[]{\includegraphics[width=0.4\textwidth]{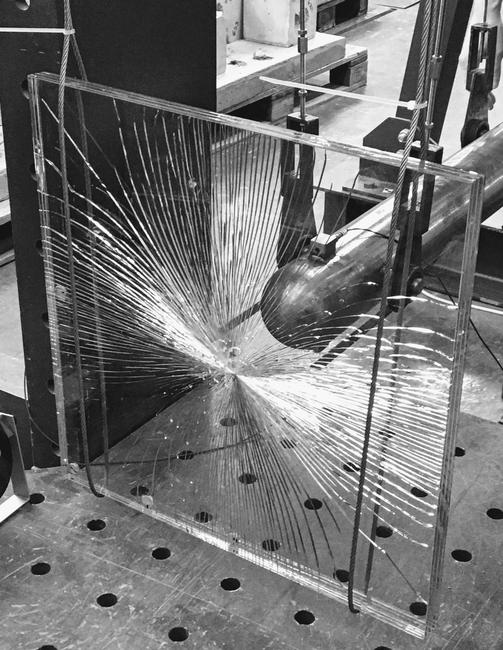}};
\draw[red, thick] (9.5mm,10.5mm) circle (3mm) node [red, above left=1mm] {\text{$I$}};
\draw[red, thick] (-2.5mm,-28mm) circle (2mm) node [red, below=1.5mm] {\text{$M$}};
\draw[red, thick] (-13.5mm,-23mm) circle (2mm) node [red, below=1.5mm] {\text{$Q$}};
\draw[red, thick] (-23.5mm,-18.5mm) circle (2mm) node [red, below=1.5mm] {\text{$C$}};
\draw[red, thick] (-25.5mm,4mm) circle (2mm) node [red, above left] {\text{$M'$}};
\end{tikzpicture} }
    \\
    (a) & (b)
    \end{tabular}
    \caption{Positions of sensors: (a) scheme of the positions of four sensors $C$, $Q$, $M$, and $M'$ on the impacted surface of a sample together with one sensor $I$ on the impactor behind, (b) photo of an experimental setup with all sensors highlighted.}
    \label{fig:exp_photo}
\end{figure}
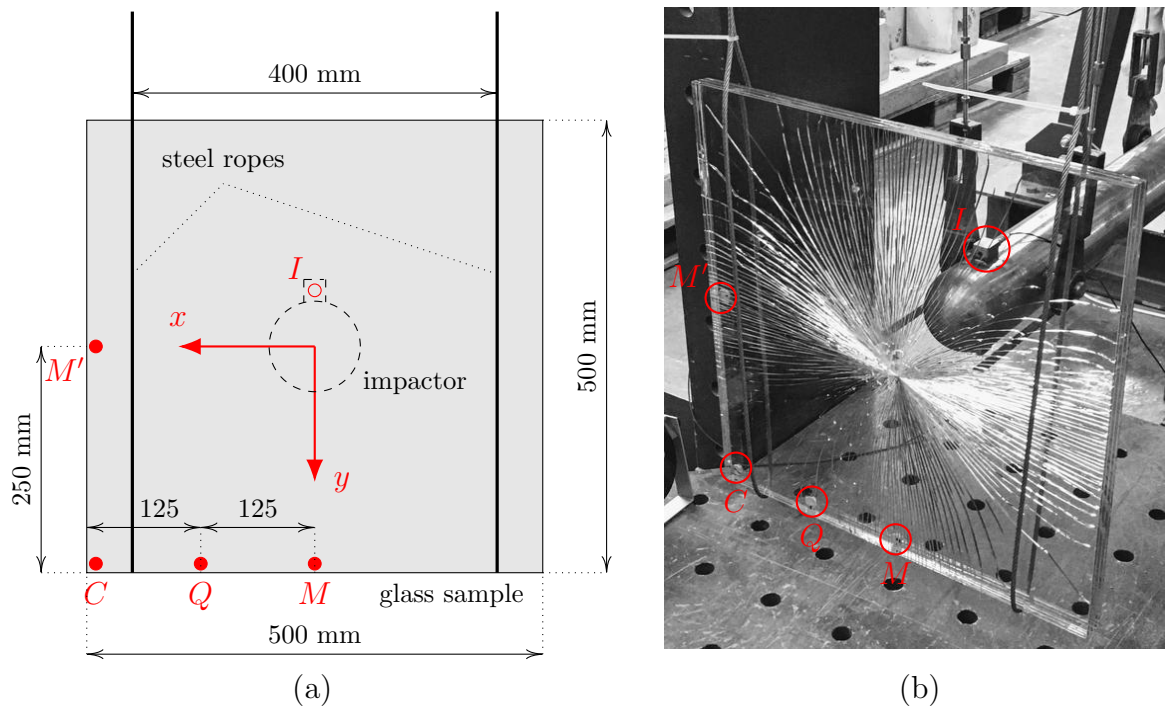

Five accelerometers with integral electronics (Model 350B04 Shear ICP® Shock Sensor, PCB Piezotronics) recorded the response of a laminated glass sample. Four of them were glued on the back face (the non-impacted outer surface) of each laminated sample (\Fref{fig:exp_photo}), and one additional sensor was bolted to the impactor (\Fref{fig:exp_photo}~(b)) to monitor the contact force resulting from impact. A data acquisition unit collected the data with a sampling frequency of 500~kHz. 

For four samples (5LG--2, 5LG--3, 7LG--3 and 7LG--4), a high-speed camera (i-SPEED 726, iX Cameras) with a pixel resolution of 2,048$\times$1,536 recorded the propagation of cracks in glass plies (frame rate 5,000 frames per second). 
After each impact event, we assessed visually the state of glass layers, identified the cracked ones by a naked-eye observations, 
and took pictures of fracture patterns.

\section{Response of laminated glass samples}
\label{S:ERes}

\subsection{Overview of impact tests}

\Tref{tab:EXP_fracture} summarises the response of all samples after individual impacts. The testing samples were initially intact, and the failure sequence of glass layers is schematically visualised for specific samples by circular marks. The impacted layer is located on the left side of each cross-section scheme. The empty circle $\ccirc$ corresponds to the initiation of fracture during the test, whereas the filled circle $\sbullet[.72]$ marks a glass layer fractured during previous impacts.
We complemented the table by providing the measured overall thicknesses of laminated glass samples, the room temperature that corresponds to the averaged value during the whole test, and the impact energies (impactor's kinetic energy before the impact). 

\begin{table}[h]
\caption{Overview of fractured glass layers after individual impacts complemented with average overall thicknesses, room temperatures during the experimental testing, and impact energies. With the left-hand side impacted by the hard body, symbols \textbar$\sbullet[.72]$\textbar indicate the previously fractured glass layers, empty spaces \textbar~\textbar the unfractured ones, and circles \textbar$\ccirc$\textbar the crack initiation in a glass layer.}
    \centering
    \footnotesize
    \begin{tabular}{llllllllll} \hline
         \multicolumn{2}{l}{Sample} & 5LG--1 & 5LG--2 & 5LG--3 & 5LG--4 & 7LG--1 & 7LG--2 & 7LG--3 & 7LG--4\\
         \hline
         \multicolumn{2}{l}{Thickness (mm)} 
         & 18.5 & 18.4 & 18.2 & 18.3 & 28.2 & 28.0 & 28.1 & 28.1
         \\
         \multicolumn{2}{l}{Temperature ($^\circ$C)} &
24.9
&
25.8 &
25.9 &
26.0 &
         24.9 &
25.6 &
25.7 &
25.8 
         \\
         Impact & \\
         Height (cm) & Energy (J)
         &
         \multicolumn{8}{c}{Position of fractured glass layers after individual impacts} \\
         5 & 24 & \textbar~\textbar~\textbar~\textbar & \textbar~\textbar~\textbar~\textbar & \textbar~\textbar~\textbar~\textbar & \textbar~\textbar~\textbar~\textbar & \textbar~\textbar~\textbar~\textbar$\ccirc$\textbar & \textbar~\textbar~\textbar~\textbar~\textbar & \textbar~\textbar~\textbar~\textbar~\textbar & \textbar~\textbar~\textbar~\textbar~\textbar \\ 
        10 & 47 & \textbar~\textbar~\textbar~\textbar & \textbar~\textbar~\textbar$\ccirc$\textbar & \textbar~\textbar~\textbar~\textbar & \textbar~\textbar~\textbar~\textbar & \textbar$\ccirc$\textbar~\textbar~\textbar$\sbullet[.72]$\textbar & \textbar~\textbar~\textbar~\textbar~\textbar & \textbar~\textbar~\textbar~\textbar~\textbar & \textbar~\textbar~\textbar~\textbar~\textbar \\ 
         15 & 71 & \textbar~\textbar~\textbar$\ccirc$\textbar & \textbar~\textbar~\textbar$\sbullet[.72]$\textbar & \textbar~\textbar~\textbar~\textbar & \textbar~\textbar~\textbar~\textbar & \textbar$\sbullet[.72]$\textbar~\textbar~\textbar$\sbullet[.72]$\textbar & \textbar~\textbar~\textbar~\textbar~\textbar & \textbar~\textbar~\textbar~\textbar~\textbar & \textbar$\ccirc$\textbar~\textbar~\textbar~\textbar \\ 
         20 & 95 & \textbar~\textbar~\textbar$\sbullet[.72]$\textbar & \textbar~\textbar~\textbar$\sbullet[.72]$\textbar & \textbar~\textbar~\textbar$\ccirc$\textbar & \textbar~\textbar~\textbar~\textbar & \textbar$\sbullet[.72]$\textbar~\textbar~\textbar$\sbullet[.72]$\textbar & \textbar~\textbar~\textbar~\textbar~\textbar & \textbar~\textbar~\textbar~\textbar~\textbar & \textbar$\sbullet[.72]$\textbar~\textbar~\textbar~\textbar \\ 
         25 & 118 & \textbar$\ccirc$\textbar~\textbar$\sbullet[.72]$\textbar & \textbar~\textbar$\ccirc$\textbar$\sbullet[.72]$\textbar & \textbar~\textbar$\ccirc$\textbar$\sbullet[.72]$\textbar & \textbar~\textbar~\textbar~\textbar & \textbar$\sbullet[.72]$\textbar~\textbar~\textbar$\sbullet[.72]$\textbar & \textbar~\textbar~\textbar~\textbar~\textbar & \textbar~\textbar~\textbar~\textbar~\textbar & \textbar$\sbullet[.72]$\textbar~\textbar~\textbar$\ccirc$\textbar \\ 
         30 & 142 & \textbar$\sbullet[.72]$\textbar~\textbar$\sbullet[.72]$\textbar & \textbar~\textbar$\sbullet[.72]$\textbar$\sbullet[.72]$\textbar & \textbar$\ccirc$\textbar$\sbullet[.72]$\textbar$\sbullet[.72]$\textbar & \textbar~\textbar~\textbar~\textbar & \textbar$\sbullet[.72]$\textbar~\textbar~\textbar$\sbullet[.72]$\textbar & \textbar~\textbar~\textbar~\textbar~\textbar & \textbar~\textbar~\textbar~\textbar~\textbar & \textbar$\sbullet[.72]$\textbar~\textbar~\textbar$\sbullet[.72]$\textbar \\ 
         35 & 165 & \textbar$\sbullet[.72]$\textbar~\textbar$\sbullet[.72]$\textbar & \textbar~\textbar$\sbullet[.72]$\textbar$\sbullet[.72]$\textbar & \textbar$\sbullet[.72]$\textbar$\sbullet[.72]$\textbar$\sbullet[.72]$\textbar & \textbar~\textbar~\textbar~\textbar & \textbar$\sbullet[.72]$\textbar~\textbar~\textbar$\sbullet[.72]$\textbar & \textbar~\textbar~\textbar~\textbar~\textbar & \textbar~\textbar~\textbar~\textbar~\textbar & \textbar$\sbullet[.72]$\textbar~\textbar~\textbar$\sbullet[.72]$\textbar \\ 
         40 & 189 & \textbar$\sbullet[.72]$\textbar~\textbar$\sbullet[.72]$\textbar & \textbar$\ccirc$\textbar$\sbullet[.72]$\textbar$\sbullet[.72]$\textbar & \textbar$\sbullet[.72]$\textbar$\sbullet[.72]$\textbar$\sbullet[.72]$\textbar & \textbar~\textbar~\textbar~\textbar & \textbar$\sbullet[.72]$\textbar~\textbar~\textbar$\sbullet[.72]$\textbar & \textbar~\textbar~\textbar~\textbar$\ccirc$\textbar & \textbar$\ccirc$\textbar~\textbar~\textbar$\ccirc$\textbar & \textbar$\sbullet[.72]$\textbar~\textbar~\textbar$\sbullet[.72]$\textbar \\ 
         45 & 213 & \textbar$\sbullet[.72]$\textbar$\ccirc$\textbar$\sbullet[.72]$\textbar & \textbar$\sbullet[.72]$\textbar$\sbullet[.72]$\textbar$\sbullet[.72]$\textbar & \textbar$\sbullet[.72]$\textbar$\sbullet[.72]$\textbar$\sbullet[.72]$\textbar & \textbar~\textbar~\textbar~\textbar & \textbar$\sbullet[.72]$\textbar~\textbar~\textbar$\sbullet[.72]$\textbar & \textbar~\textbar~\textbar$\ccirc$\textbar$\sbullet[.72]$\textbar & \textbar$\sbullet[.72]$\textbar~\textbar~\textbar$\sbullet[.72]$\textbar & \textbar$\sbullet[.72]$\textbar~\textbar~\textbar$\sbullet[.72]$\textbar \\ 
         50 & 236 & \textbar$\sbullet[.72]$\textbar$\sbullet[.72]$\textbar$\sbullet[.72]$\textbar & \textbar$\sbullet[.72]$\textbar$\sbullet[.72]$\textbar$\sbullet[.72]$\textbar & \textbar$\sbullet[.72]$\textbar$\sbullet[.72]$\textbar$\sbullet[.72]$\textbar & \textbar~\textbar~\textbar~\textbar & \textbar$\sbullet[.72]$\textbar~\textbar~\textbar$\sbullet[.72]$\textbar & \textbar$\ccirc$\textbar$\ccirc$\textbar$\sbullet[.72]$\textbar$\sbullet[.72]$\textbar & \textbar$\sbullet[.72]$\textbar~\textbar~\textbar$\sbullet[.72]$\textbar & \textbar$\sbullet[.72]$\textbar~\textbar~\textbar$\sbullet[.72]$\textbar \\ 
         55 & 260 &  &  &  & \textbar~\textbar~\textbar~\textbar &  &  &  &  \\ 
         60 & 284 &  &  &  & \textbar~\textbar~\textbar~\textbar &  &  &  &  \\ 
         65 & 307 &  &  &  & \textbar~\textbar~\textbar~\textbar &  &  &  &  \\  
         70 & 331 &  &  &  & \textbar~\textbar~\textbar$\ccirc$\textbar &  &  &  &  \\ 
         75 & 355 &  &  &  & \textbar~\textbar~\textbar$\sbullet[.72]$\textbar &  &  &  &  \\  
         80 & 378 &  &  &  & \textbar$\ccirc$\textbar$\ccirc$\textbar$\sbullet[.72]$\textbar &  &  &  & \\  
         \hline
    \end{tabular}
    \label{tab:EXP_fracture}
\end{table}

For 5LG-samples, the first cracks appeared for the impact energies 47--331~J consistently in the back glass layer (right position in the scheme), i.e., the farthest one from the contact area. 
The samples were often able to resist additional more powerful impact(s) without a fracture of other glass layer(s). Then, the sequence of gradual collapse differed for individual samples: the fracture of the back glass layer was followed by the middle glass layer in two cases, by the front glass (the impacted one) in one case, and both remaining layers fractured also in one case for the highest impact energy. The complete fracture of all layers corresponds to the impact energies from 189 to 378~J. 

A certain level of differences in forces leading to glass fracture is not unusual for glass components and indicates the stochastic nature of tensile strength of glass described in many publications, e.g.,~\citep{nie2009effect, datsiou2018weibull, osnes2020rate}. This stochastic fracture behaviour is caused by initial surface microflaws. Moreover, the glass strength is also rate-dependent~\citep{nie2009effect} and it seems that a different tensile strength should be assumed for the free surfaces compared to the inner (laminated) glass surfaces, which are protected by the interlayer against mechanical scratches after the lamination process and also against the access of water to the crack tip~\citep{alter2017enhanced}. Finally, a possible reason for the differences could also be a different level of residual stresses that might be present even in annealed  float glass panels~\citep{zaccaria2016thermal}.

For 7LG-samples, the difference in thicknesses of inner and outer glass layers was designed more pronounced than for the 5LG-laminates (\Tref{tab:samples}).
The first fracture in these 7LG-samples corresponds to three different failure cases, i.e., to the fracture of the farthest glass layer (twice), the impacted layer fracture (once), or the fracture of both outer glass layers during an impact event (once). The range of impact energies leading to the first damage, 24--189~J, was again wide. For three out of four samples, the fracture of an outer glass layer was followed after several impacts by the fracture of the opposite outer layer. Only for one sample, the cracks formed in the middle glass layer in the second stage, which was followed by the fracture of the rest intact glass layers after an additional impact.
Inner glass layers stayed intact for three out of four tested 7LG-samples for the impact energies up to 236~J (impact height of 50~cm). 

\subsection{Non-destructive tests and the prefracture response of laminated glass}
\label{S:NOND}

\begin{figure}[ht]
    \centering
    \begin{tabular}{cc}
         \includegraphics[width=0.475\textwidth]{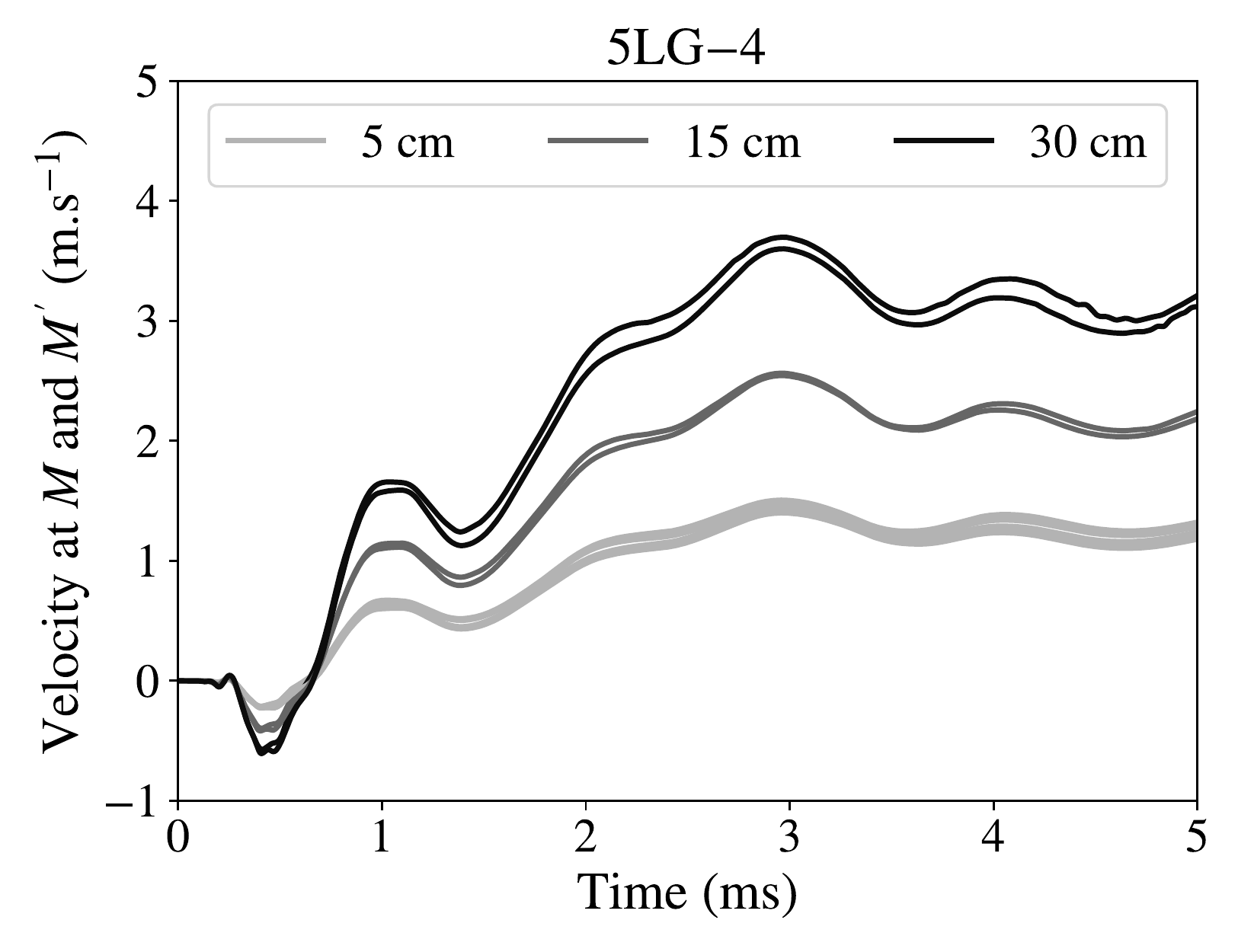} 
&  
\includegraphics[width=0.475\textwidth]{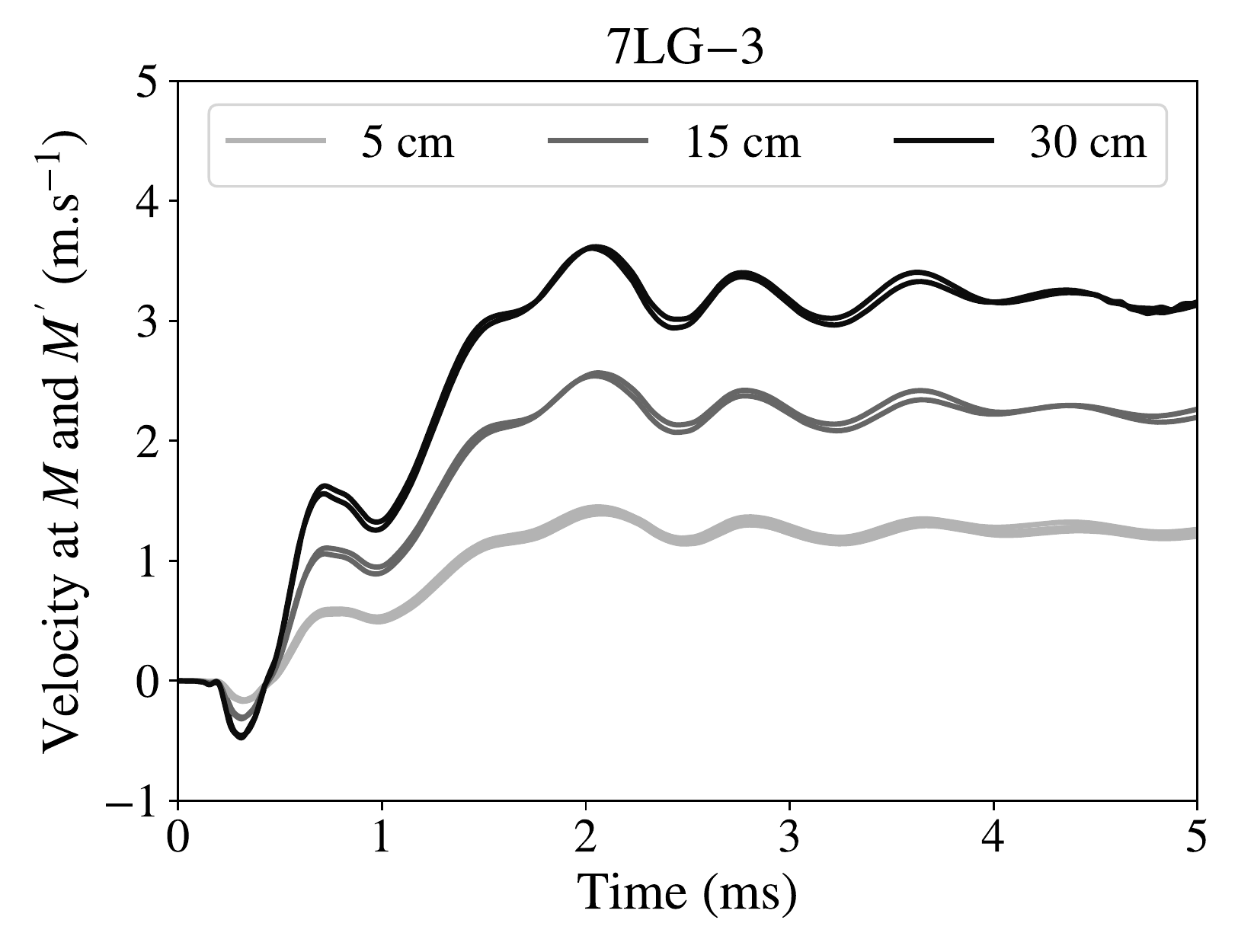}
\end{tabular}
    \
    \caption{Comparison of velocities at positions $M$ (0, 250, 0)~mm and $M'$ (250, 0, 0)~mm derived from the corresponding sensors located in the middle of bottom and left edge (\Fref{fig:exp_photo}) for a 5LG-sample and a 7LG-sample and three impact heights (5~cm, 15~cm, and 30~cm). 
}
    \label{fig:vel_sym}
\end{figure}

\begin{figure}[hp]
    \centering
    \begin{tabular}{cc}
    \includegraphics[width=0.475\textwidth]{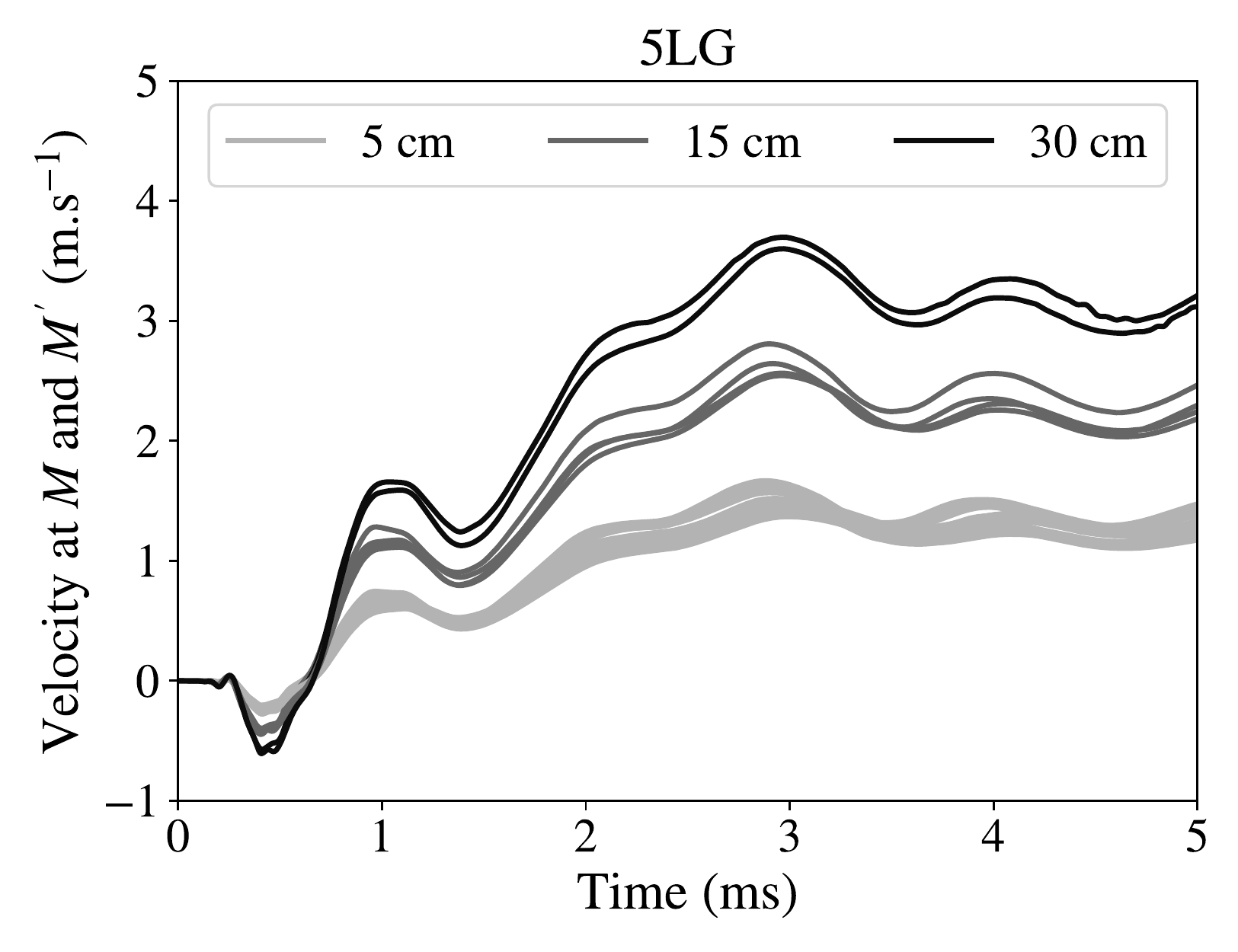}
& 
         \includegraphics[width=0.475\textwidth]{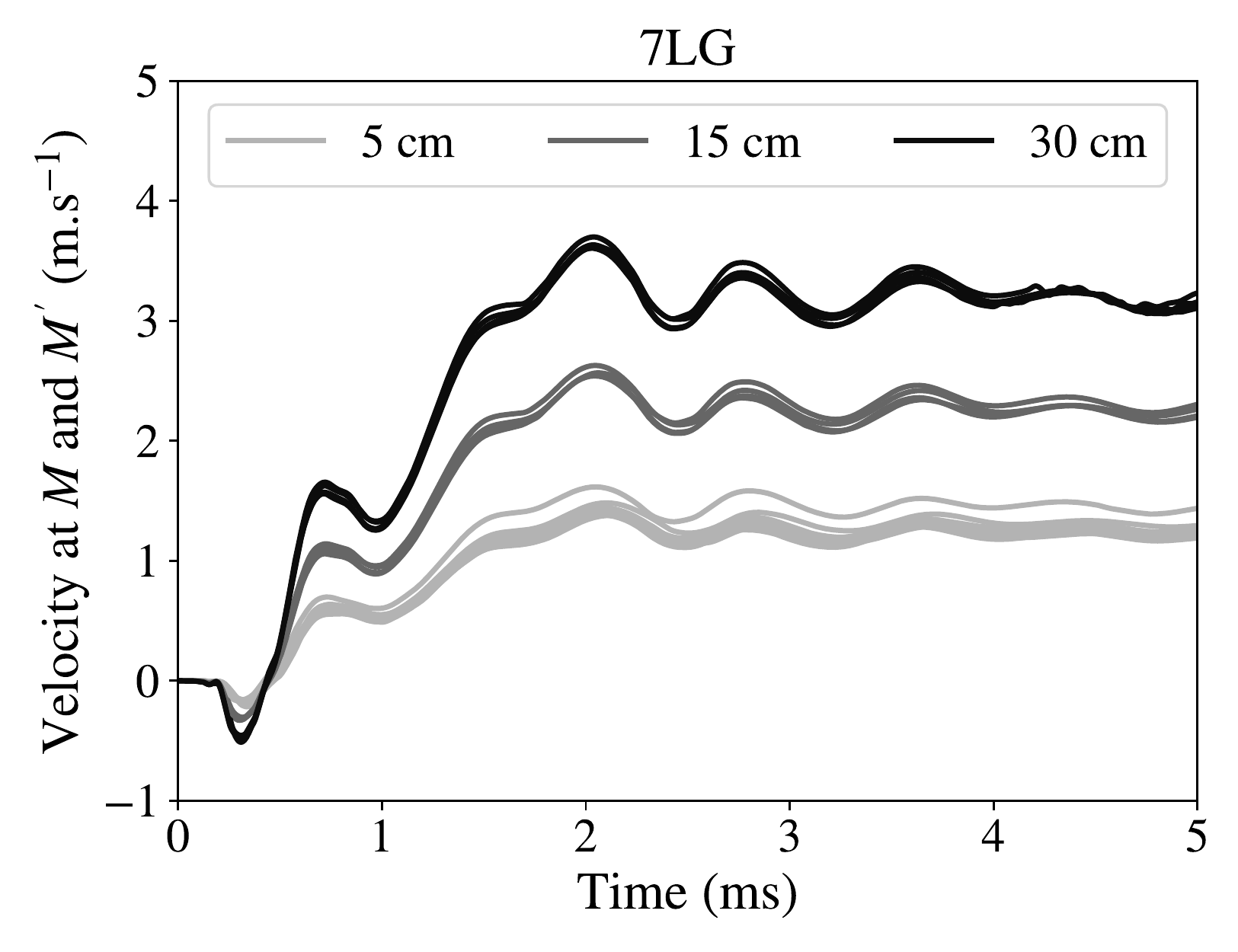}
\\
         \includegraphics[width=0.475\textwidth]{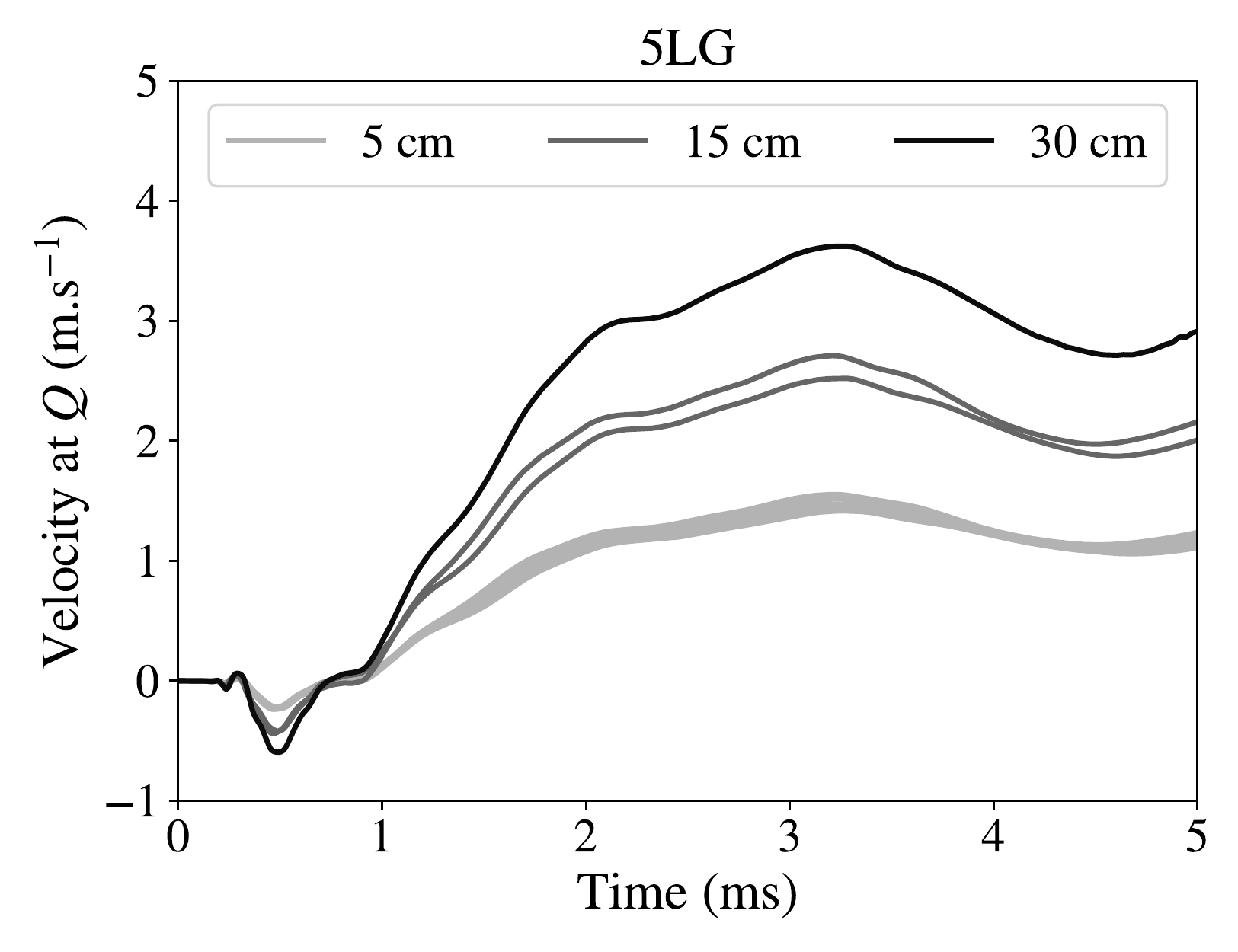}
         & 
         \includegraphics[width=0.475\textwidth]{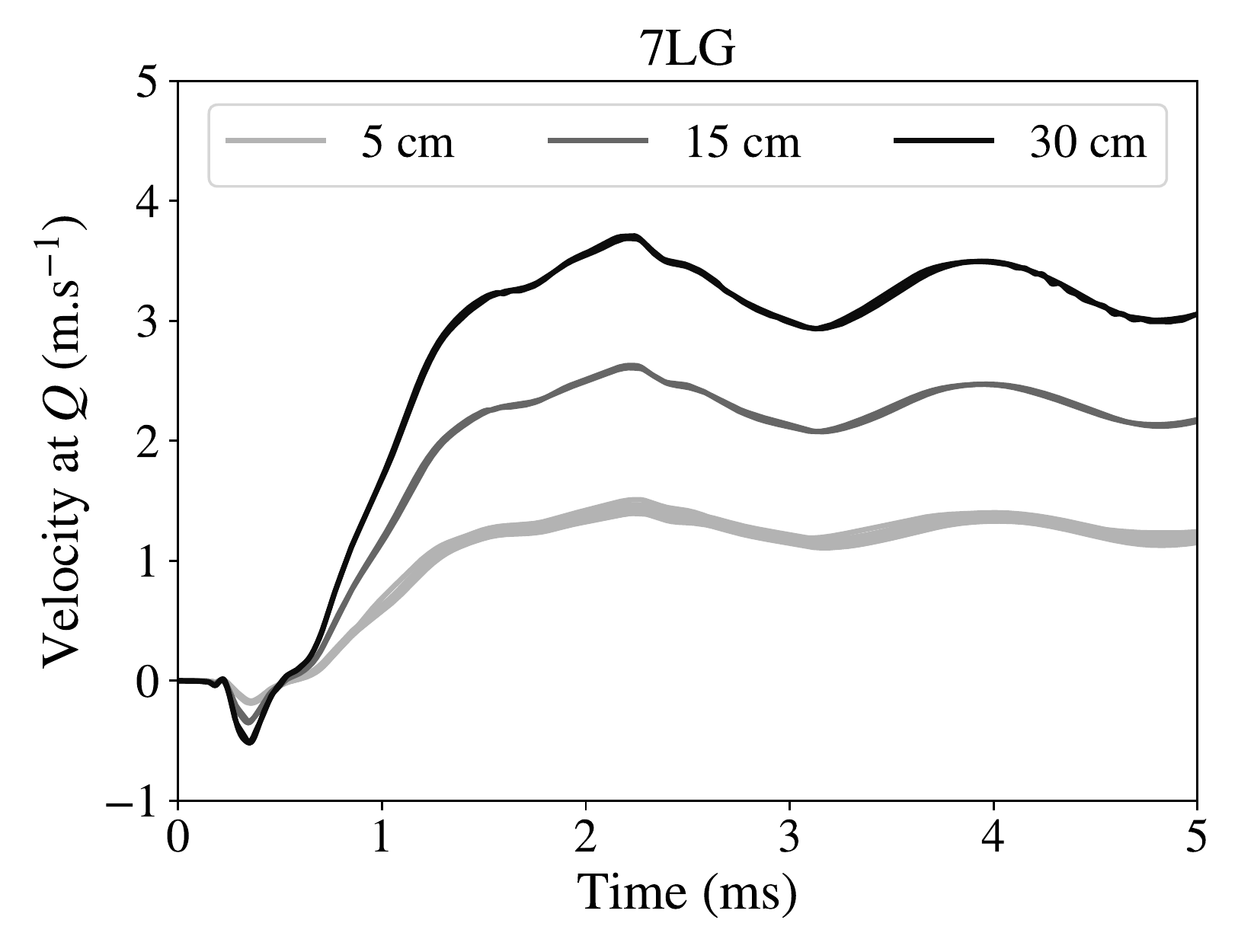}
         \\
         \includegraphics[width=0.475\textwidth]{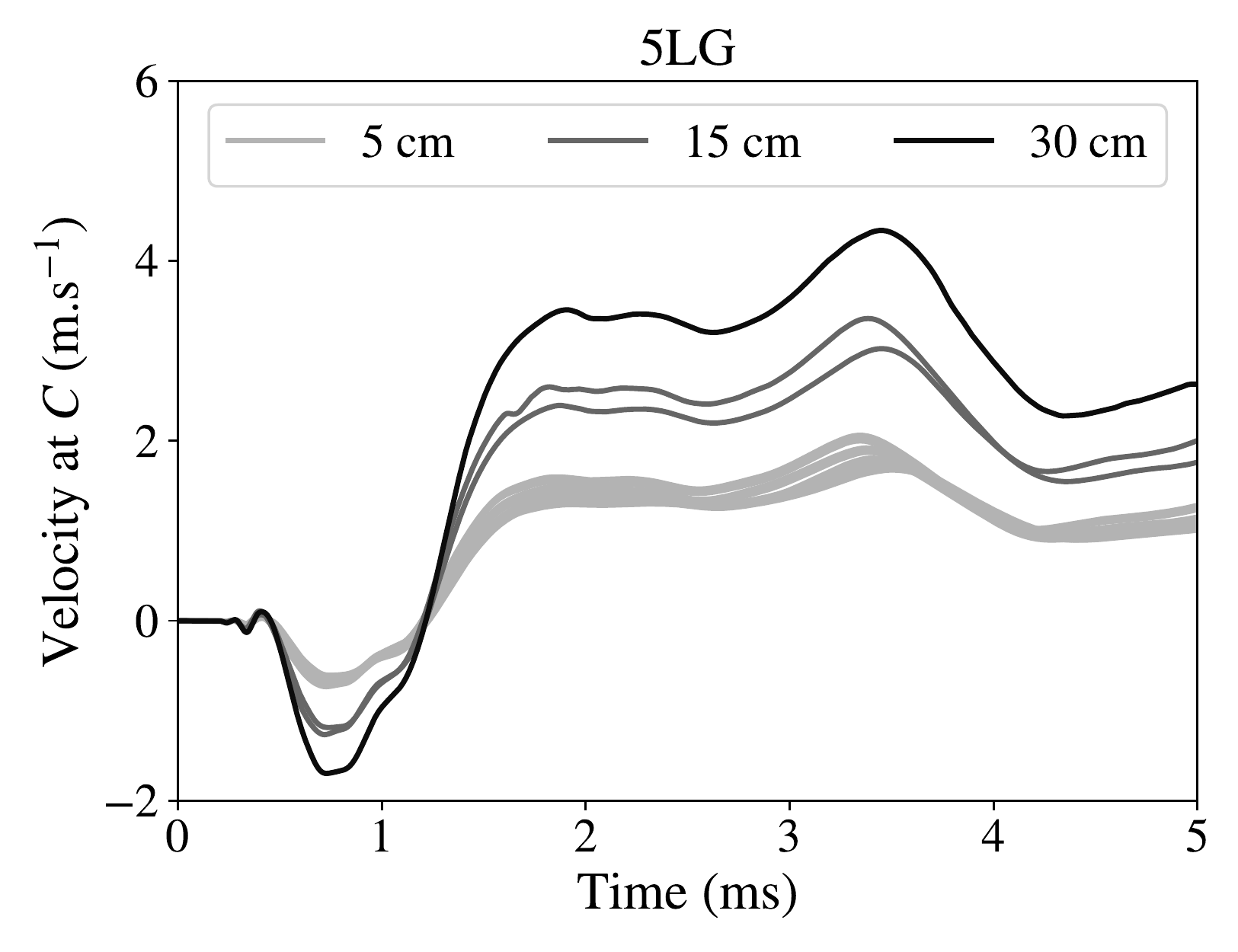}
         & 
         \includegraphics[width=0.475\textwidth]{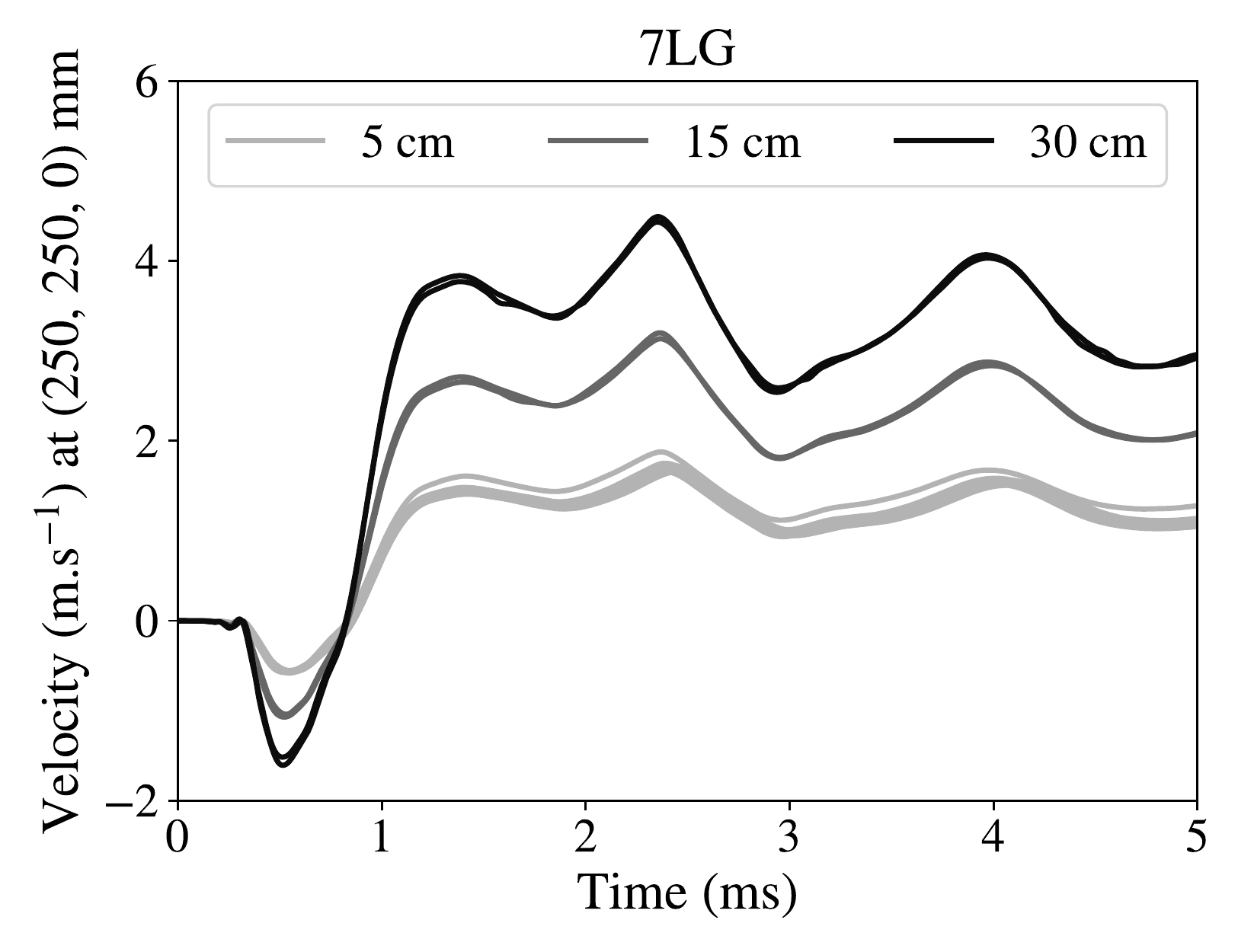}
    \end{tabular}
    \caption{Assessment of the impact test repeatability by comparison of velocities for all non-destructive tests, samples, measured locations (\Fref{fig:exp_photo}), and three impact heights (5~cm, 15~cm, and 30~cm). 
}
    \label{fig:vel_rep}
\end{figure}

\begin{figure}[ht]
    \centering
    \footnotesize
    \begin{tabular}{cc}
    \multicolumn{2}{c}{Analysed part of the signal (between red lines)}
    \\
        \includegraphics[width=0.475\textwidth]{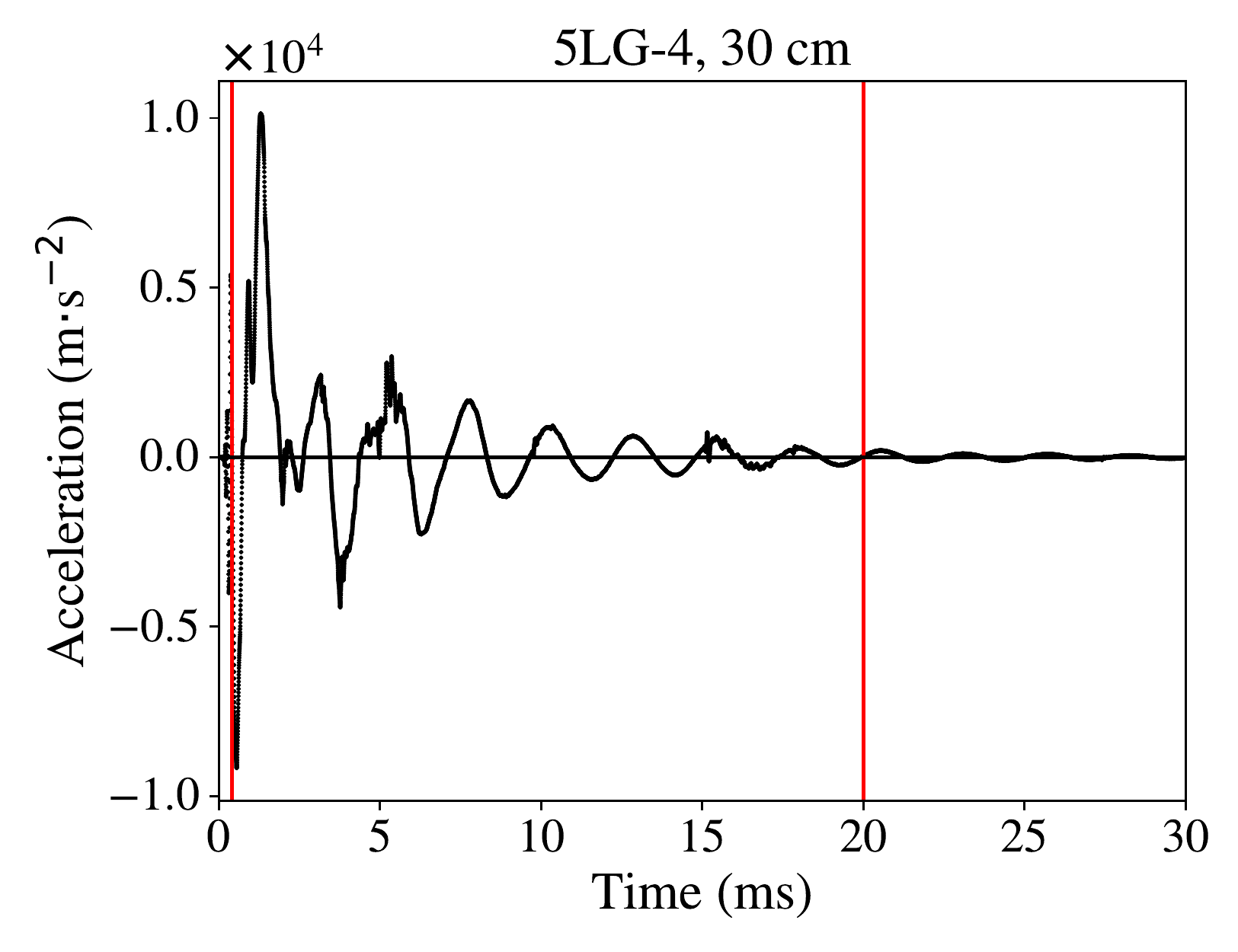}
         &  
         \includegraphics[width=0.475\textwidth]{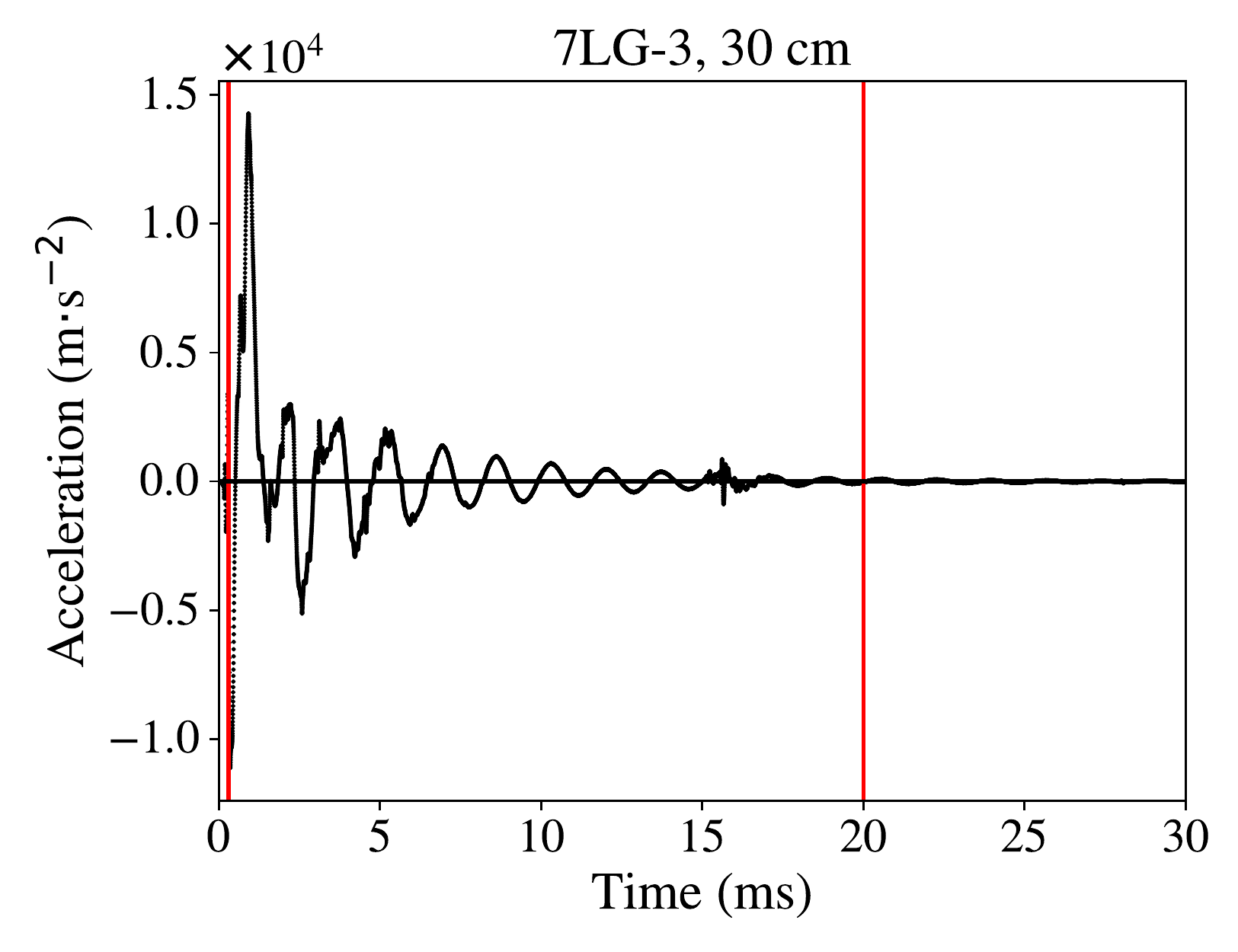} \\
         \multicolumn{2}{c}{Detail of the initial part of the signal}
\\
         \includegraphics[width=0.475\textwidth]{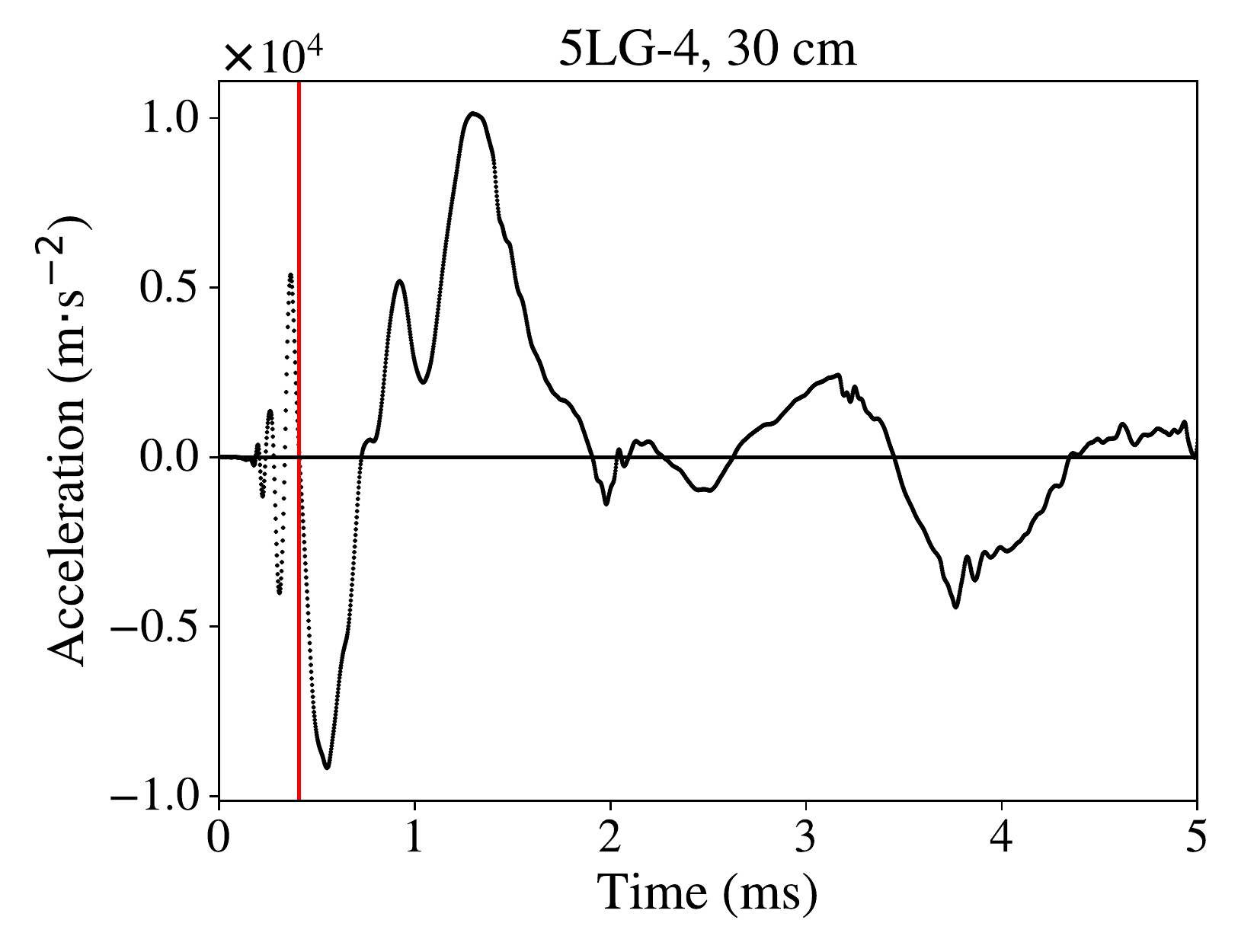}
         &  
         \includegraphics[width=0.475\textwidth]{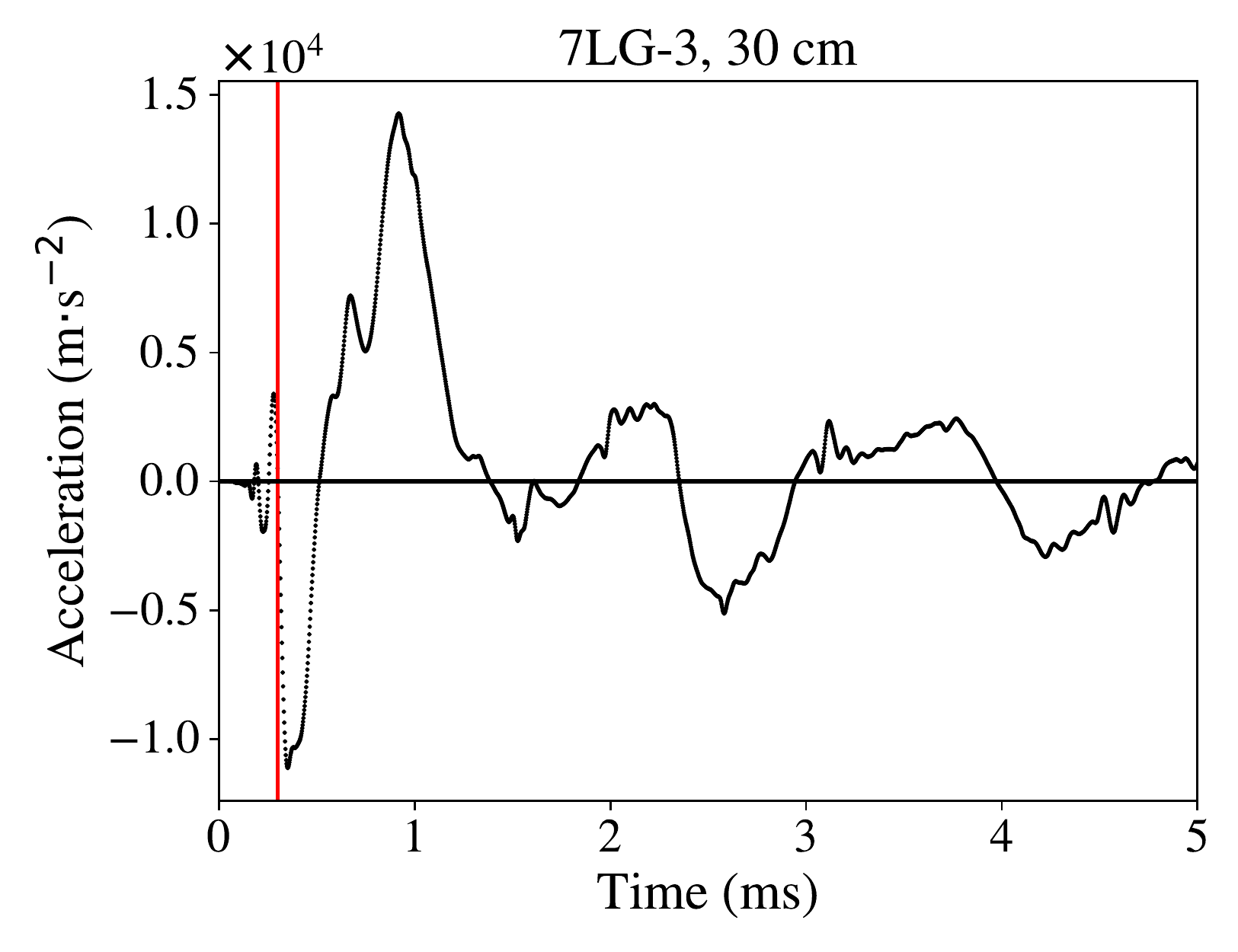}
    \end{tabular}
    \caption{Accelerations recorded by the corner sensors $C$ attached to the laminated glass samples; red lines define the part of the signals employed in the FFT analysis.}
    \label{fig:ACC_corner}
\end{figure}

\begin{figure}[hp]
    \centering
    \footnotesize
    \begin{tabular}{cc}
        \multicolumn{2}{c}{Sensor attached to the corner $C$}\\
        \includegraphics[width=0.475\textwidth]{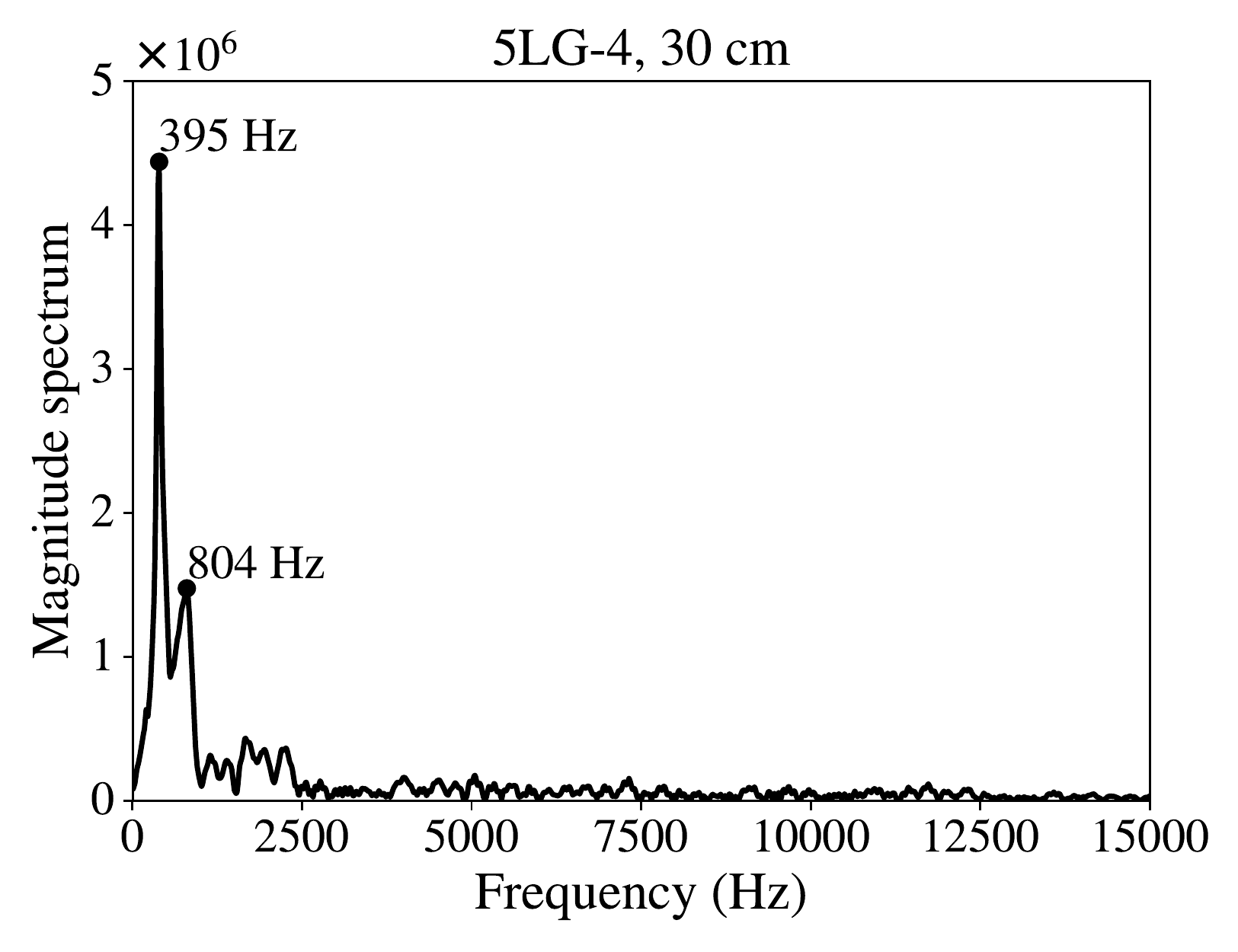}
         &  
         \includegraphics[width=0.475\textwidth]{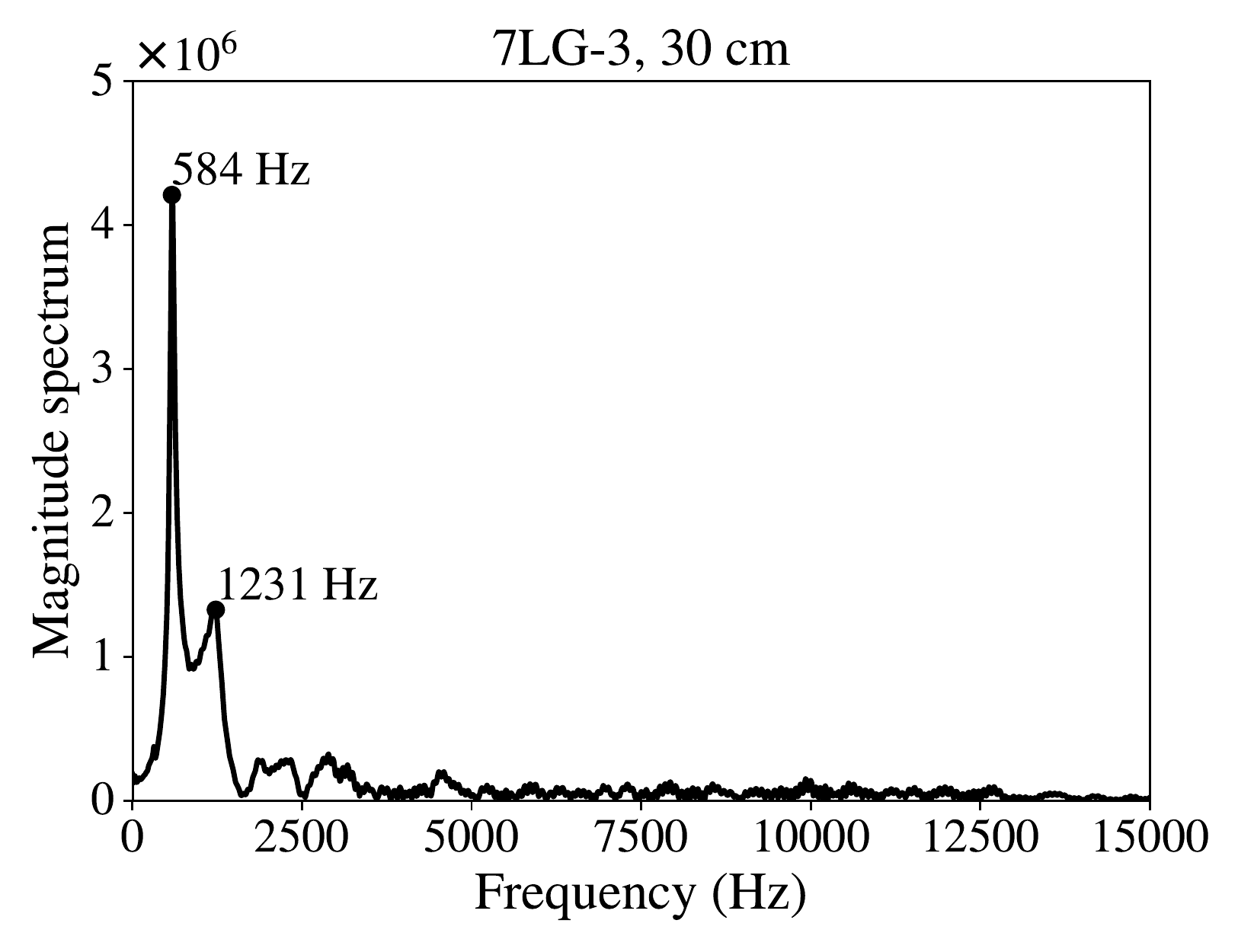}
         \\
        \multicolumn{2}{c}{Sensor attached to the impactor $I$}\\
     \includegraphics[width=0.475\textwidth]{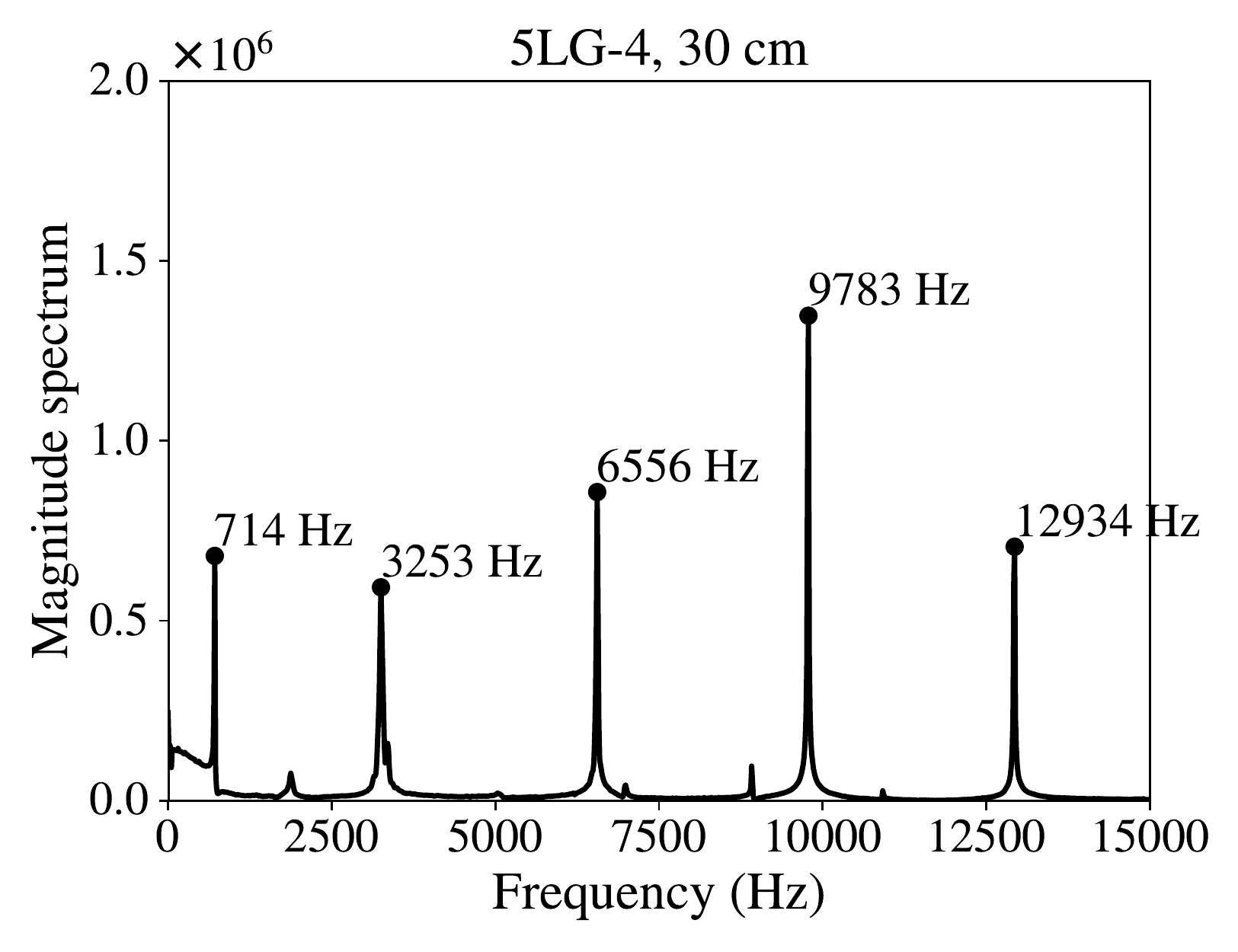}
         &  
         \includegraphics[width=0.475\textwidth]{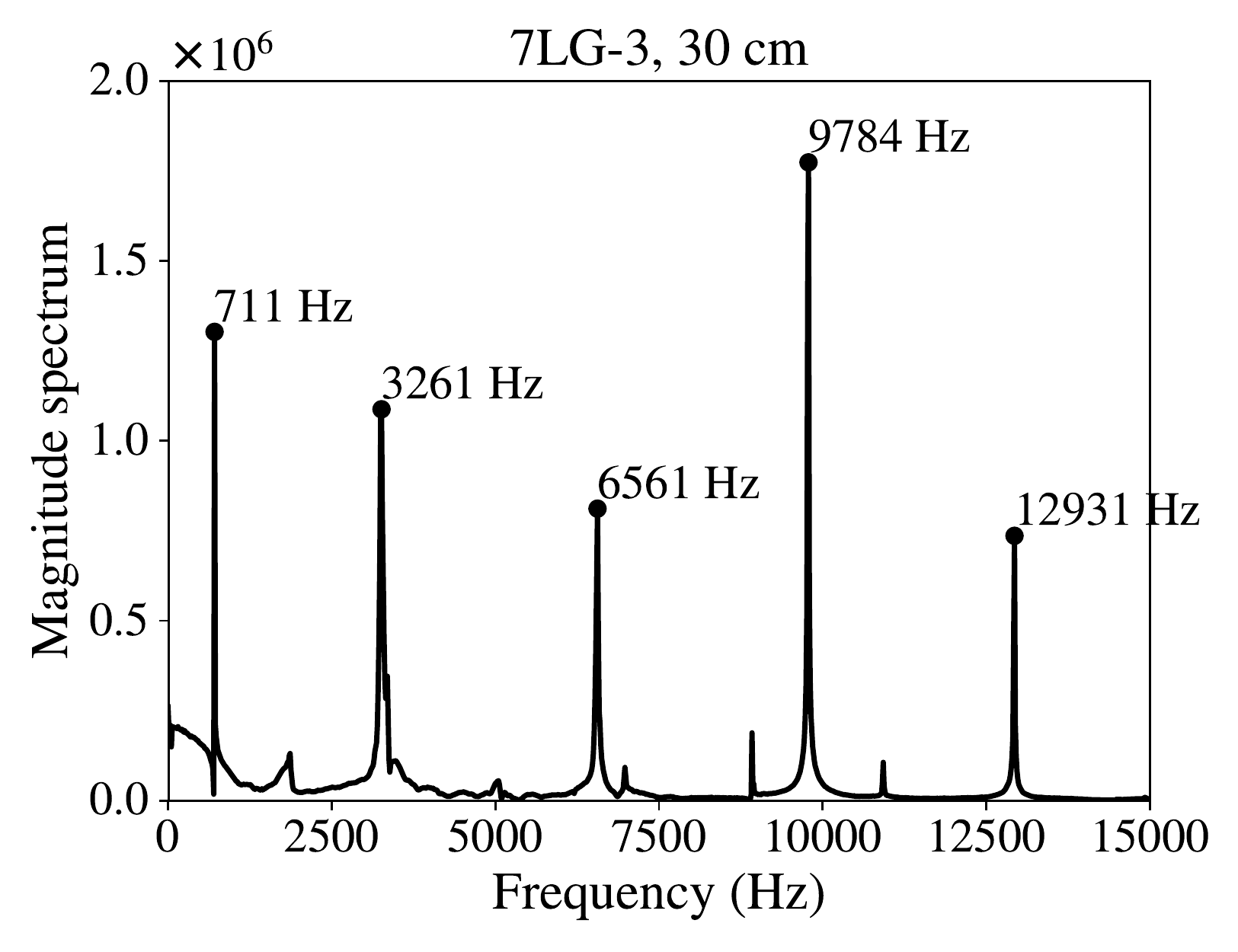}
    \end{tabular}
    \begin{tabular}{cccc}
        \multicolumn{2}{c}{Vibration modes of a laminated glass plate}
         &  
         \multicolumn{2}{c}{Vibration modes of the impactor}
         \\
         \includegraphics[width=0.21\textwidth]{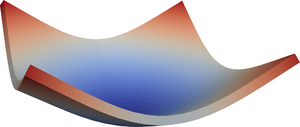}
         &
         \includegraphics[width=0.2\textwidth]{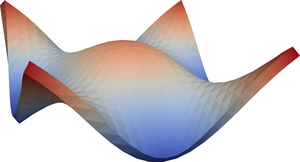}
         &
         \includegraphics[width=0.23\textwidth]{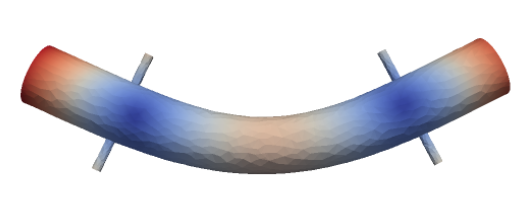}
         & 
         \includegraphics[width=0.25\textwidth]{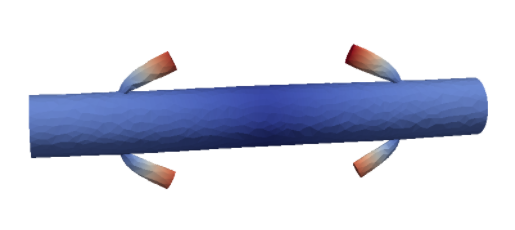}
         \\
         1. bi-sym. bending & 2. bi-sym. bending & 1. bending & 1. longitudinal\\
         395/584~Hz (5LG/7LG) &  804/1,231~Hz (5LG/7LG) & $\approx$ 700~Hz & $\approx$ 3,300~Hz
    \end{tabular}
    \caption{FFT-based signal analysis of accelerations recorded by sensors attached to laminated glass samples ($C$) and to the impactor ($I$) complemented with relevant vibration modes.}
    \label{fig:FFT_corner}
\end{figure}

\begin{figure}[hp]
    \centering
    \footnotesize
    \begin{tabular}{cc}
\includegraphics[width=0.475\textwidth]{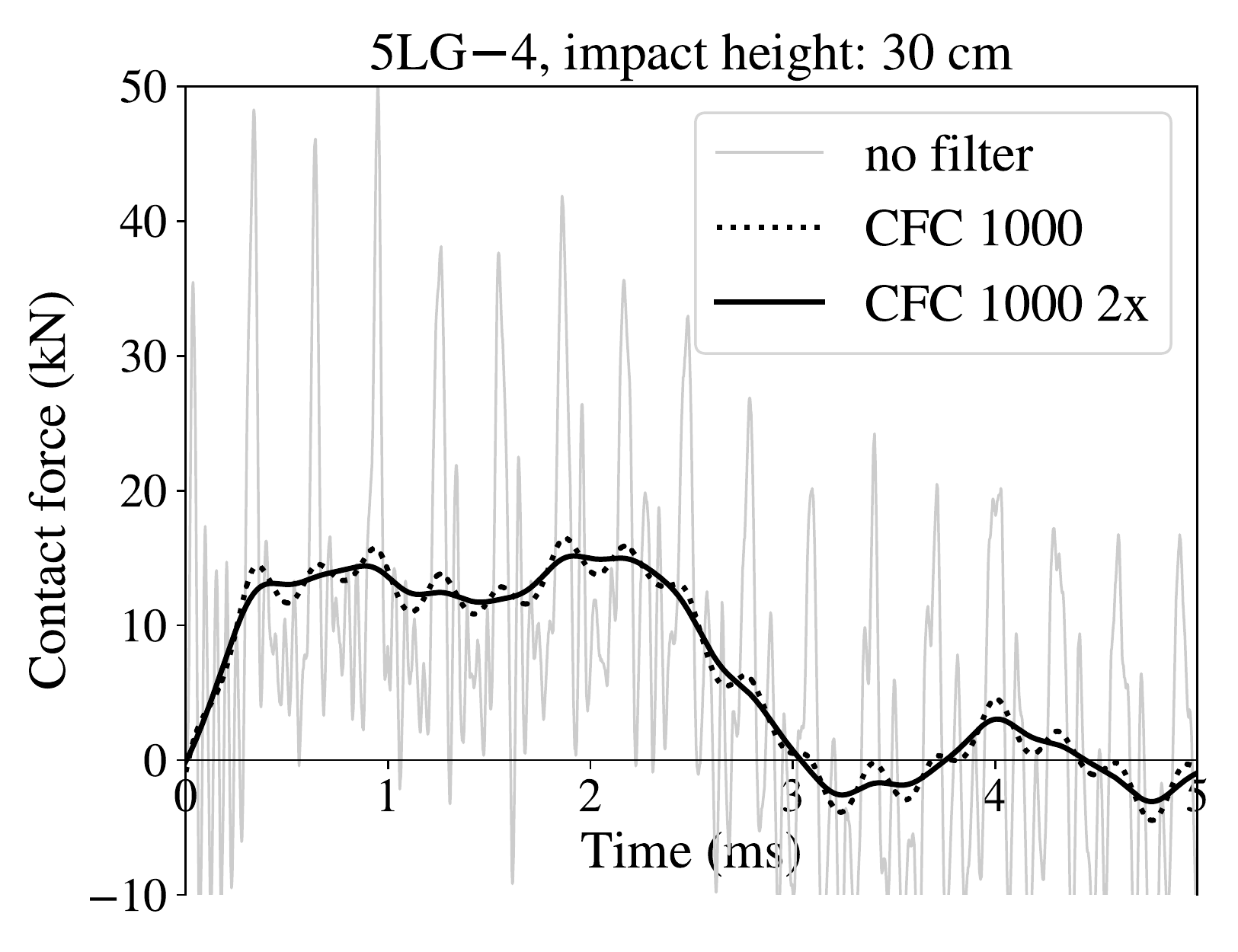}
         &
         \includegraphics[width=0.475\textwidth]{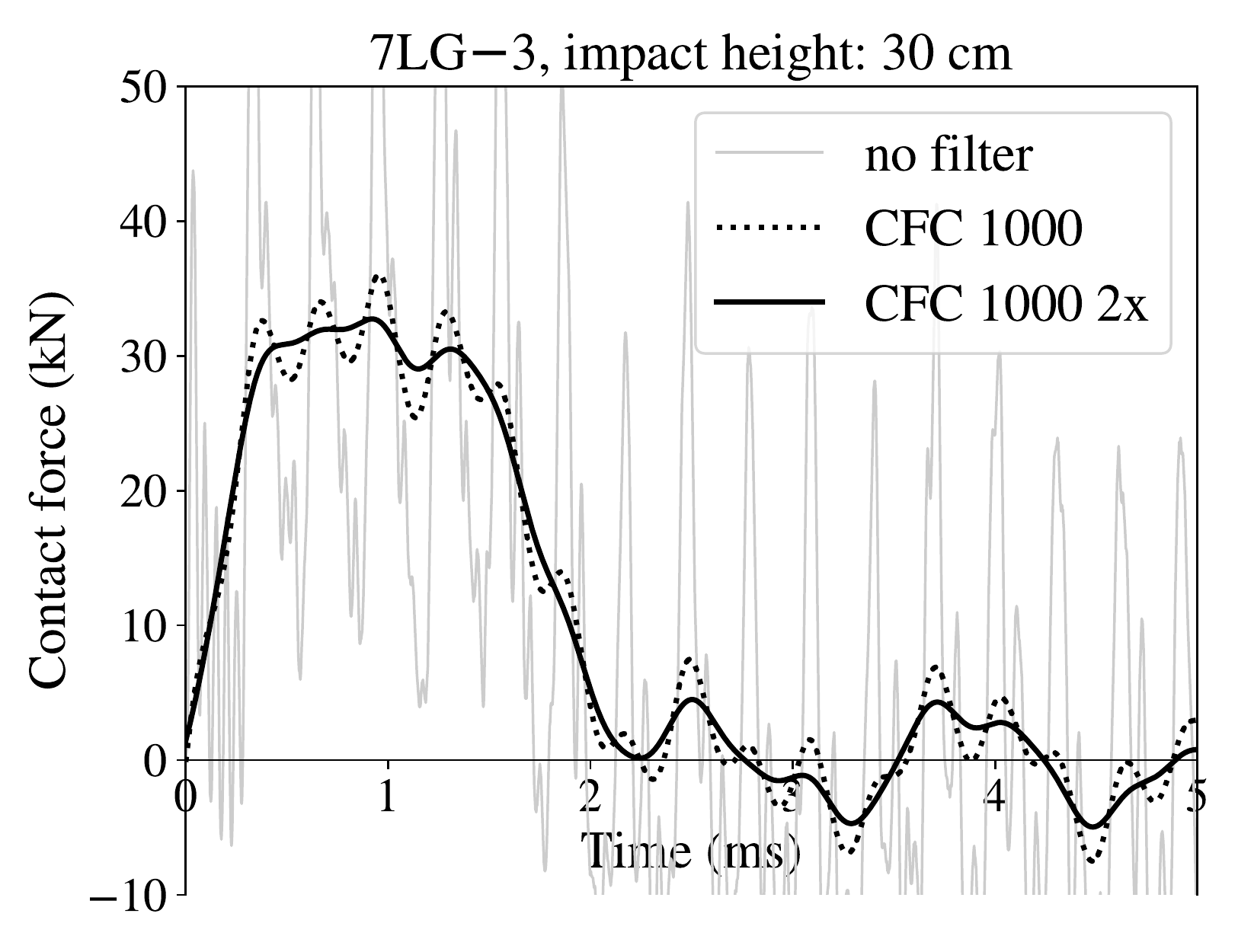}
         \\
\end{tabular}
    \caption{Comparison of contact forces derived from non-filtered and filtered accelerations  for two non-destructive tests representing both types of samples.}
    \label{fig:cof_filter}
\end{figure}

\begin{figure}[hp]
    \centering
    \begin{tabular}{cc}
         \includegraphics[width=0.475\textwidth]{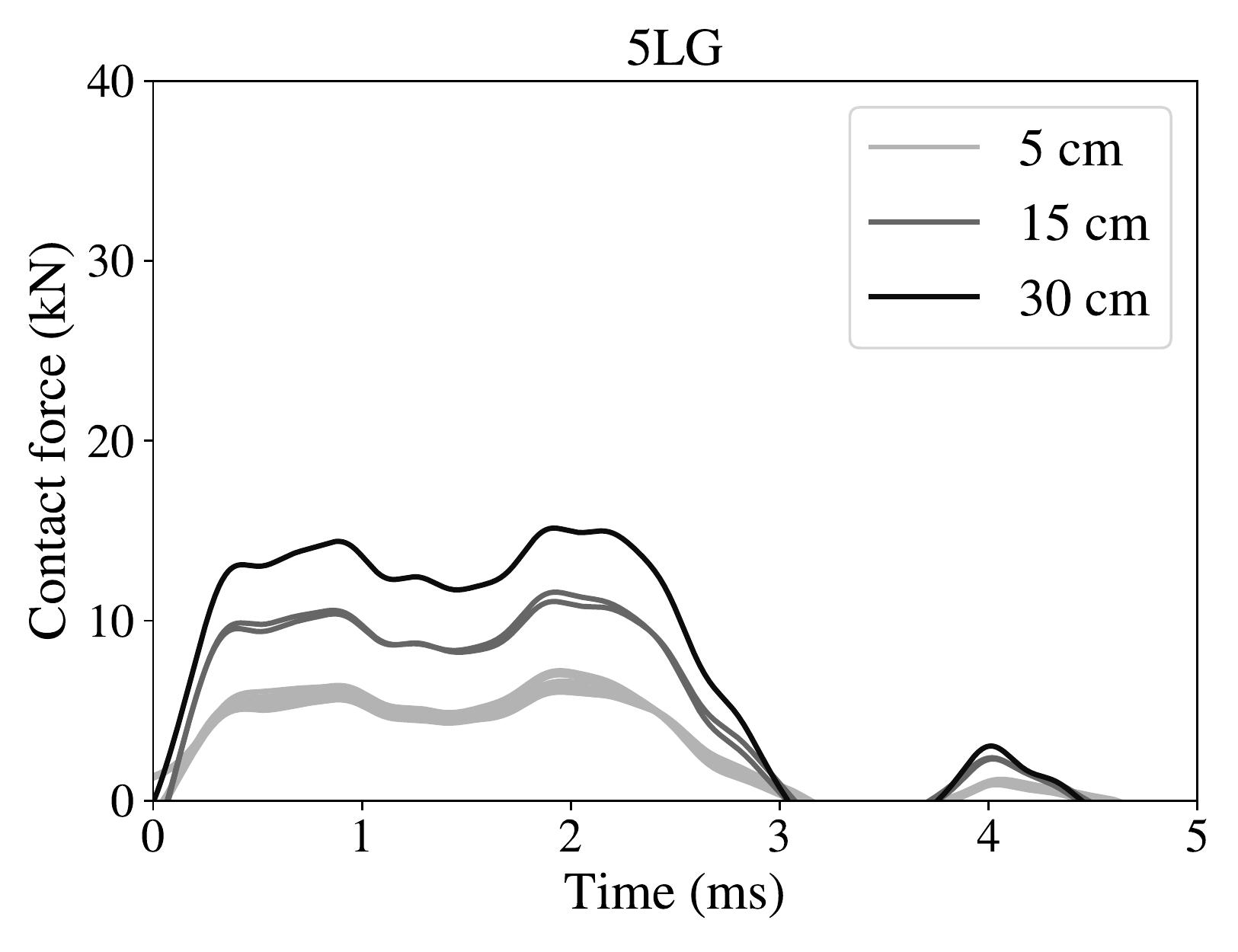}
         &  
         \includegraphics[width=0.475\textwidth]{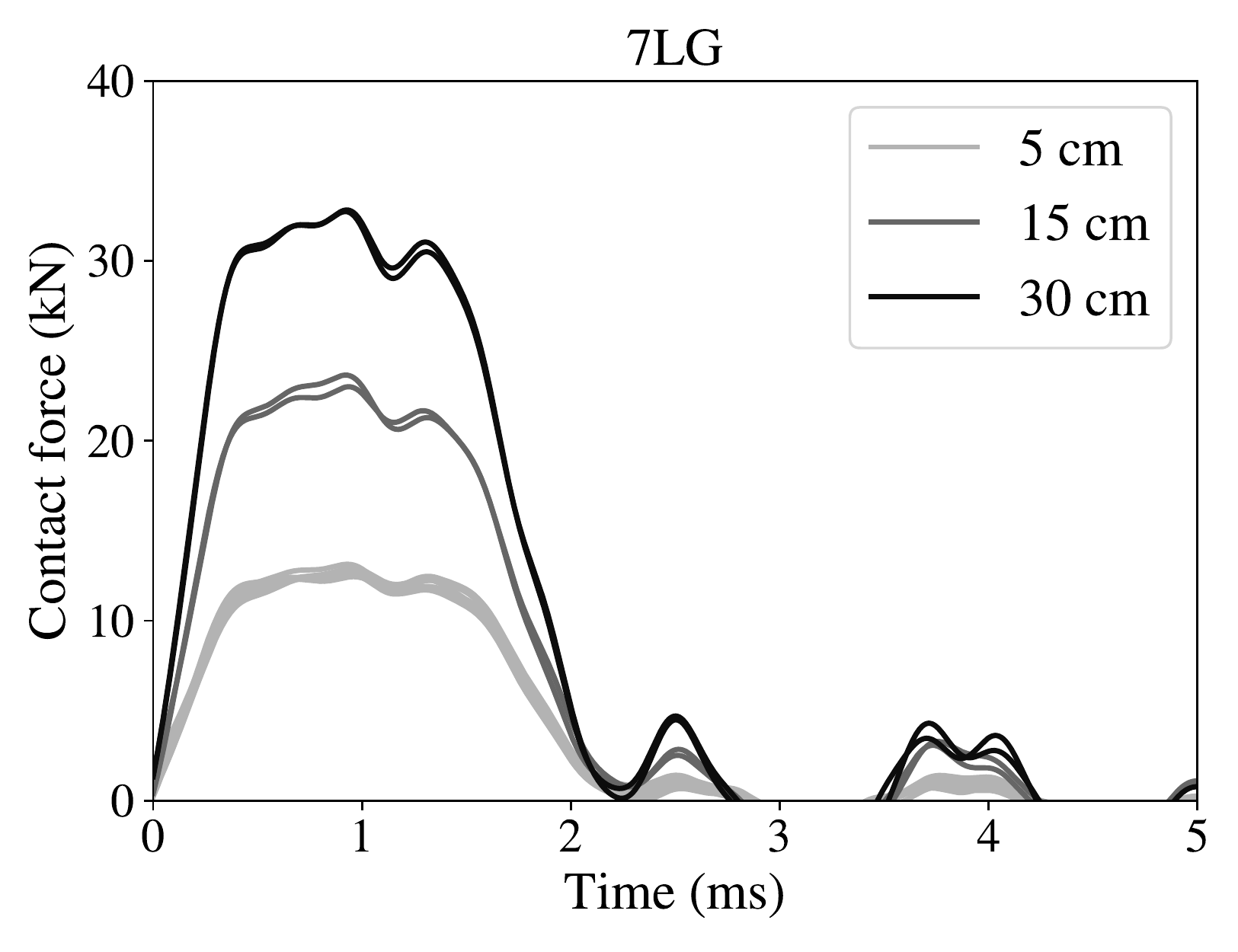}
\end{tabular}
    \caption{Comparison of contact forces (2$\times$ filter CFC~1000) for all non-destructive tests, samples, and three impact heights.}
    \label{fig:cof_rep}
\end{figure}

In this section, we start from the evaluation of the non-destructive impact tests to assess the effect of boundary conditions and the repeatability of experiments (\Fref{fig:vel_sym} and \Fref{fig:vel_rep}). Further, we analyse the vibrations of samples and impactor to set the filter parameters (\Fref{fig:ACC_corner} and \Fref{fig:FFT_corner}) and compare the derived contact forces leading to glass fracture for both cross-section geometries (\Fref{fig:cof_filter} and \Fref{fig:cof_rep}).

To analyse the response of samples, we derived velocities from measured accelerations using their time integration. For this purpose, we employed the classical Newmark scheme~\cite{newmark1959method} in the form corresponding to the constant average acceleration method. 
\Fref{fig:vel_sym} compares the velocities at positions $M$ and $M'$ (\Fref{fig:exp_photo}) that are located symmetrically about the diagonal axis. The response is plotted for three different impact heights (5~cm, 15~cm, and 30~cm) and two samples (with no visible failure for the selected impact heights) representing both geometries. The time axis range is set to cover the contact duration. The comparison of velocities from both sensors indicates that the steel ropes did not effect significantly the response during the contact because the differences were small and 
can more likely be attributed to other experimental uncertainties.  

\Fref{fig:vel_rep} summarises the velocities for all samples and non-destructive tests at three positions (midpoints of edges $M$ and $M'$ were plotted together) and three impact heights of 5, 15, and 30~cm.
As can be seen, the testing method provides consistent responses for the same conditions. The uncertainties related to the experimental measurements are caused by a slightly eccentric impact (see ahead to \Fref{fig:cracks_details_7LG}) and partially by imperfections in sample dimensions or positions of sensors. 
The oscillations of velocities, evident from Figures~\ref{fig:vel_sym} and~ \ref{fig:vel_rep}, correspond to vibrations of samples induced by the impact.
Their frequencies were obtained from the fast Fourier transform (FFT) analyses of signals recorded from the impact test similarly to~\citep{wang2020post}.
Before the FFT~analysis, the initial and final part of the data from accelerometers on glass samples were eliminated in accordance with the instructions in~\citep{lenci2015experimental} to have a sufficiently regular signal.  We eliminated the initial data before the largest amplitudes (approximately 0.3--0.4~ms of a signal) and the final part (after 20~ms) with small oscillations affected significantly by the experimental noise (\Fref{fig:ACC_corner}). 

The natural frequencies of laminated glass plates were derived from the corner sensor and were compared with those of the impactor (\Fref{fig:FFT_corner}). The two dominant frequencies of samples correspond to the first bi-symmetric vibration modes excited by the impact. The numerical simulation of modal response of the impactor explained that the modes excited by the impact corresponded to the first and higher bending and longitudinal modes. We employed a standard material parameters of steel (Young's modulus of 210~GPa, Poisson ratio of~0.3, and density of 7,800~kg$\cdot$m$^{-3}$).  

For the evaluation of contact forces, a sensor was attached to the impactor (\Fref{fig:exp_photo}), and the recorded 
accelerations were multiplied by its mass (48.2~kg).
The FFT analysis combined with the modal analysis for the impactor revealed that the signals included also the impactor's vibrations, and thus the corresponding frequencies had to be eliminated.
\Fref{fig:cof_filter} shows the comparison of contact forces derived from raw and filtered data.
The frequency corresponding to the most significant amplitudes observed for the non-filtered data matches that of the longitudinal vibration of the impactor (approximately 3,300~Hz), 
and the motion presumably corresponds to local oscillations of the sensor 
and not to the contact force (\Fref{fig:FFT_corner}).
Thus, the signal was filtered with an 8th-pole Butterworth low-pass filter 
(2$\times$ CFC~1000 filter with the cut-off frequency of 1,650~Hz) to completely eliminate these undesired vibrations. The selected cut-off frequency is above dominant natural frequencies detected for glass samples (\Fref{fig:FFT_corner}).

As visible from the filtered evolution of contact forces (\Fref{fig:cof_filter}), the acceleration of the impactor did not drop to zero after the contact, but some level of oscillations can be seen even after the impact. These vibrations correspond to the acceleration of the sensor due to the impactor's bending and the sensor's off-axis position (\Fref{fig:exp_photo}). If we omit this additional vibration, the impacts resulted in 3~ms contact for 5LG-samples and 2~ms contact for the 7LG-samples.
However, the first bending frequency of the impactor (approximately 700 Hz) is in the range of frequencies excited on laminated plates, and so their elimination is undesired. This impactor ringing must be kept in mind when validating numerical simulations against the result presented here, and differences corresponding to the amplitudes seen in the plots after the contact are expected. 

\Fref{fig:cof_rep} compares the contact forces for all unfractured samples (both geometries) and three impact heights. The contact duration was consistent and longer for the 5LG-samples, whereas the impact from the same height resulted in more than twice higher contact force for 7LG-samples. This observation highlights that an assessment of the impact resistance for different types of laminated glass based on the level of impact energy or height can be misleading for this type of boundary conditions, and the resultant contact force has to be evaluated as well.

\subsection{Destructive tests and the post-fracture response of laminated glass samples}

\begin{figure}[hp]
    \centering
    \begin{tabular}[t]{cc}
         \includegraphics[width=0.475\textwidth]{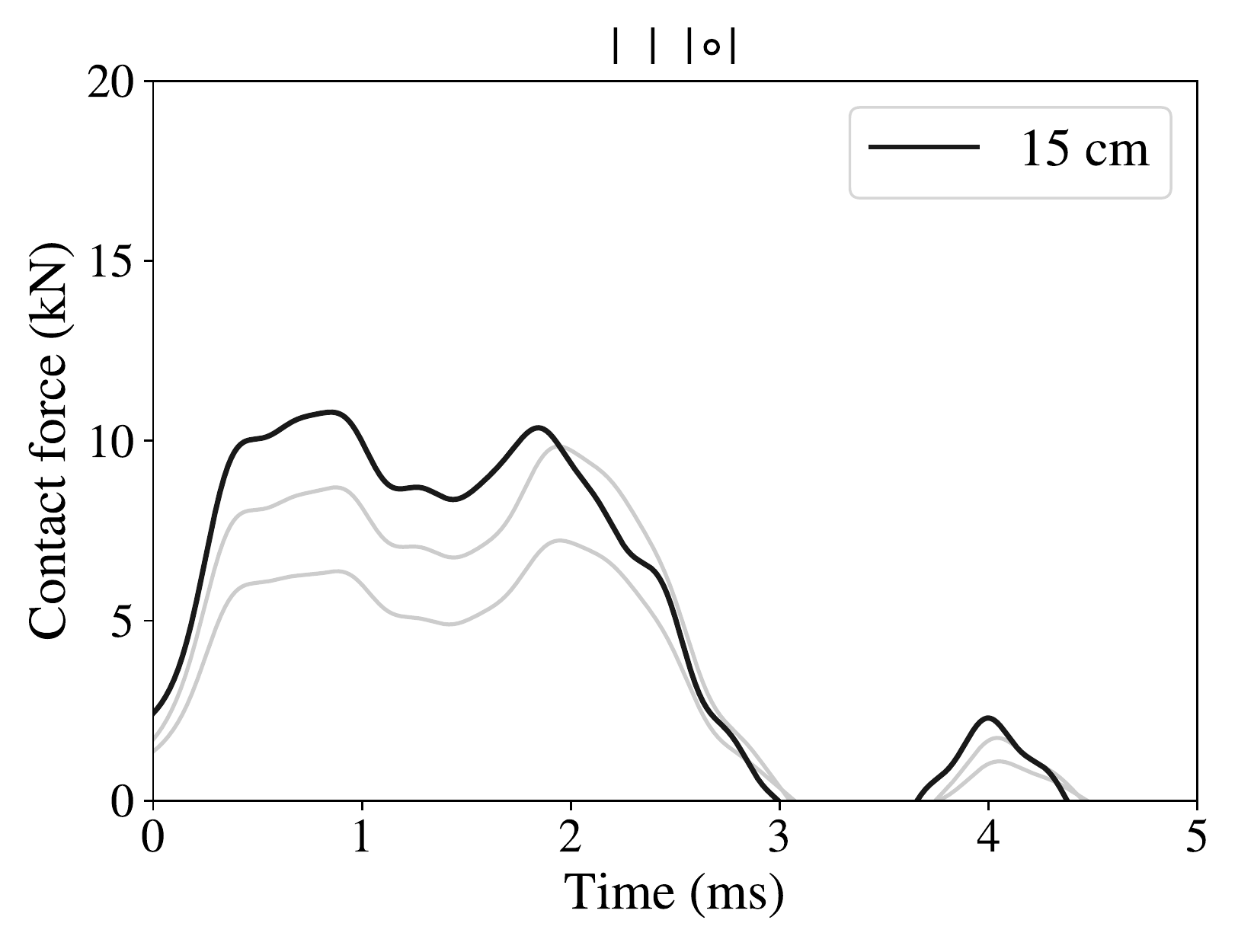}
         &  
         \includegraphics[width=0.35\textwidth]{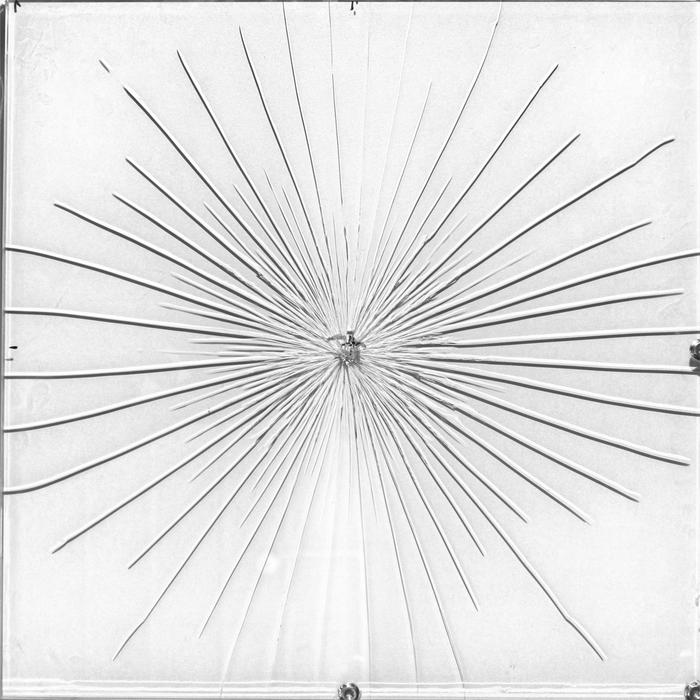}
         \\
         \\
         \includegraphics[width=0.475\textwidth]{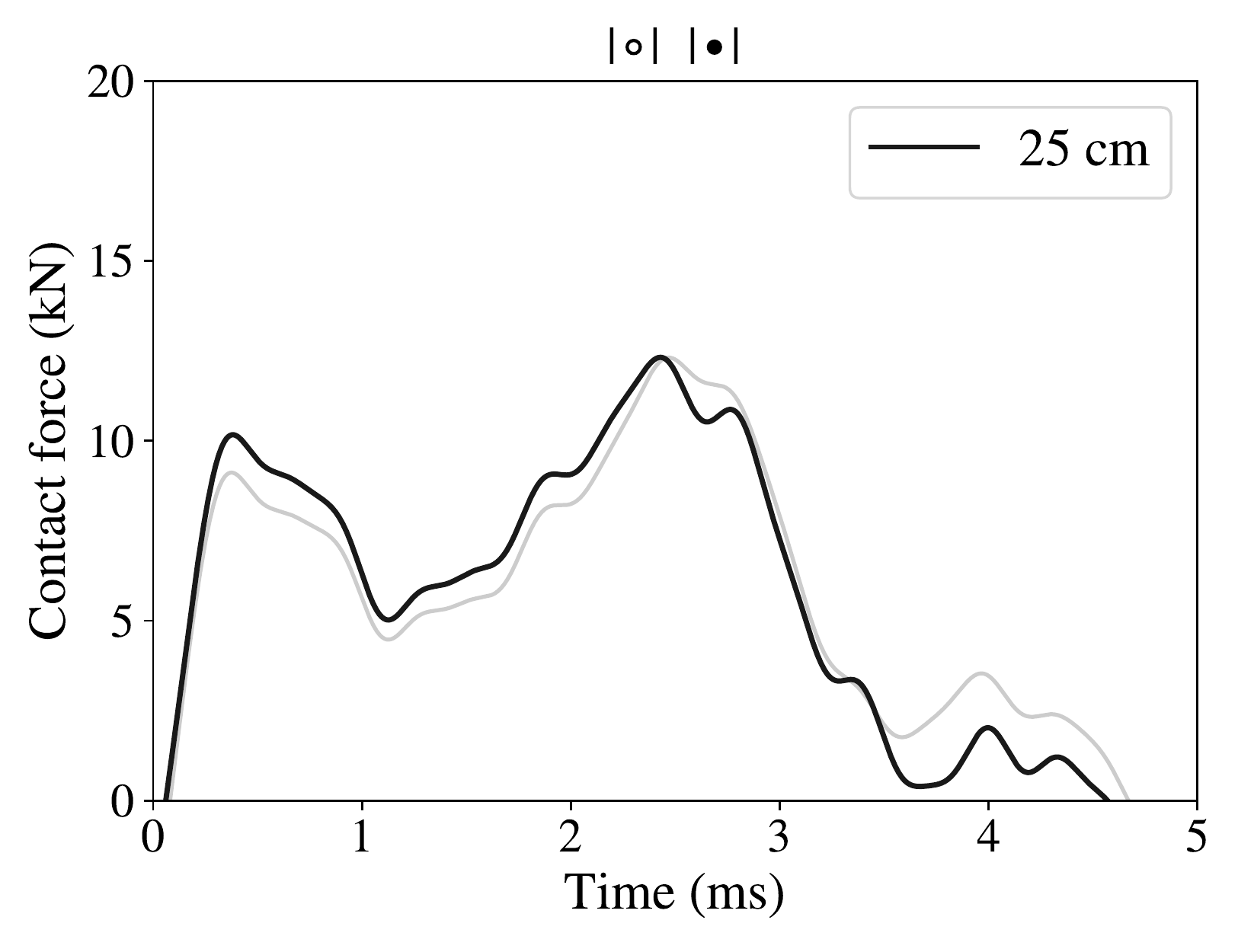}
         & 
         \includegraphics[width=0.35\textwidth]{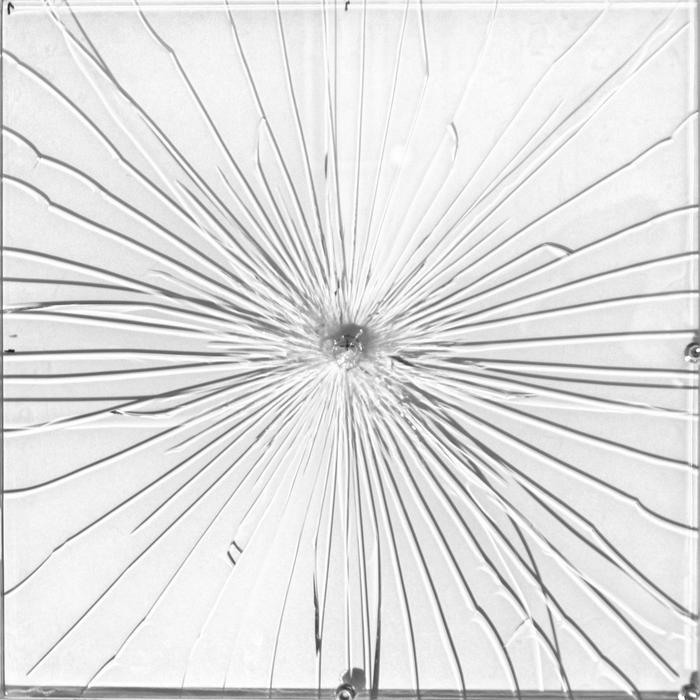}
         \\
         \\
         \includegraphics[width=0.475\textwidth]{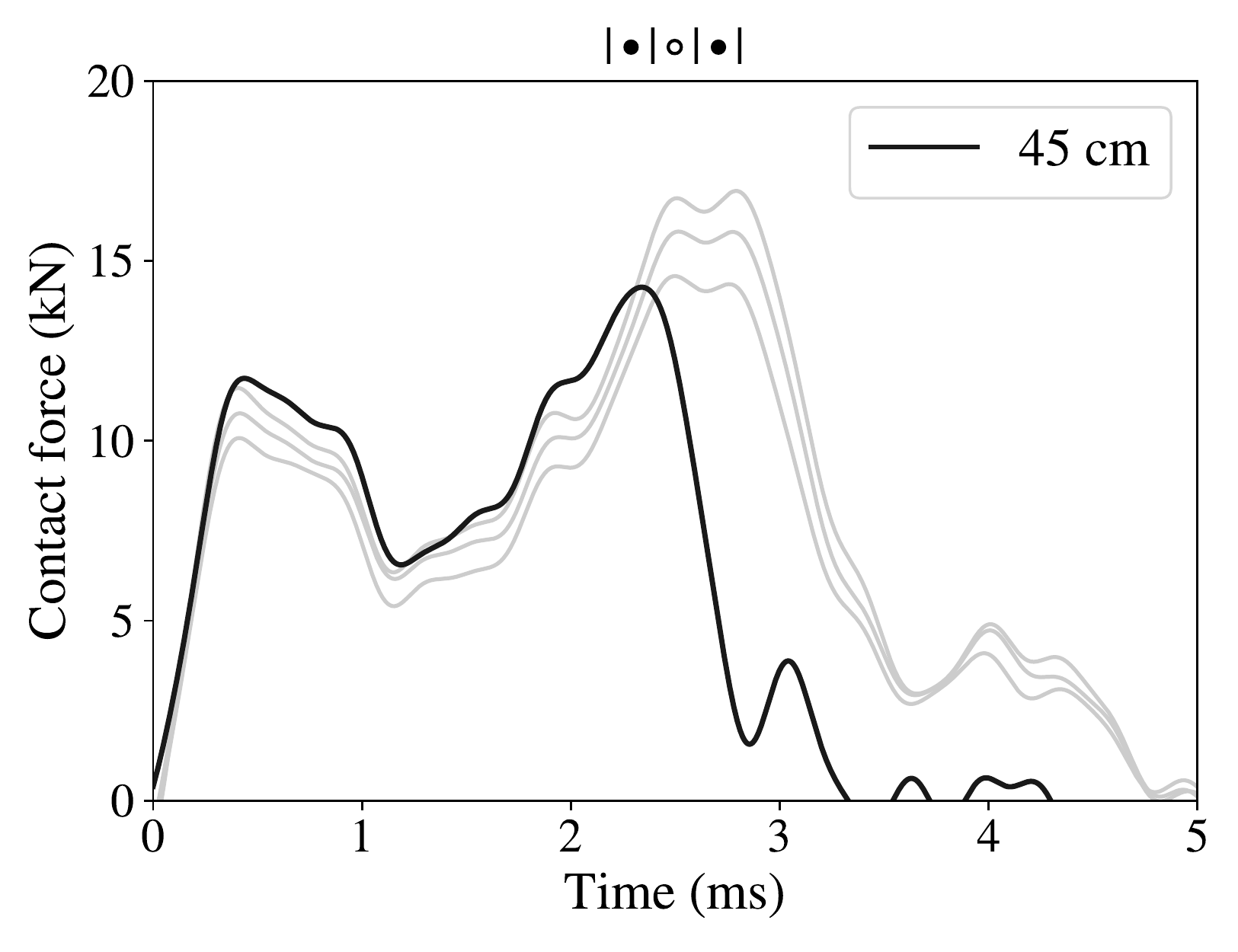}
         &
         \includegraphics[width=0.35\textwidth]{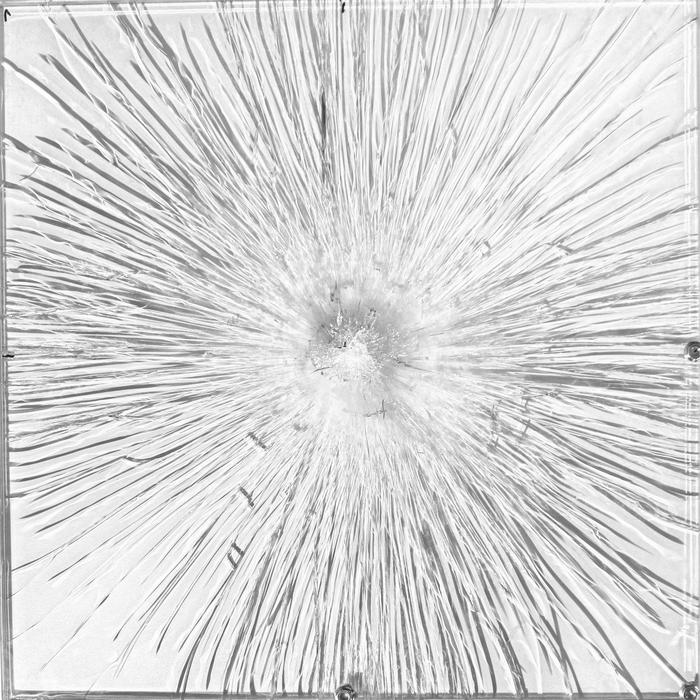}
    \end{tabular}
    \caption{Evolution of contact forces (2$\times$ filter CFC~1000) and corresponding fracture patterns for 5LG--1. Black lines denote the contact forces during the destructive tests, whereas the grey ones correspond to the previous sequence of impact events at lower impact heights (increased by 5~cm for each following impact event).}
    \label{fig:cf_5LG-1}
\end{figure}

\begin{figure}[hp]
    \centering
    \begin{tabular}[t]{cc}
         \includegraphics[width=0.475\textwidth]{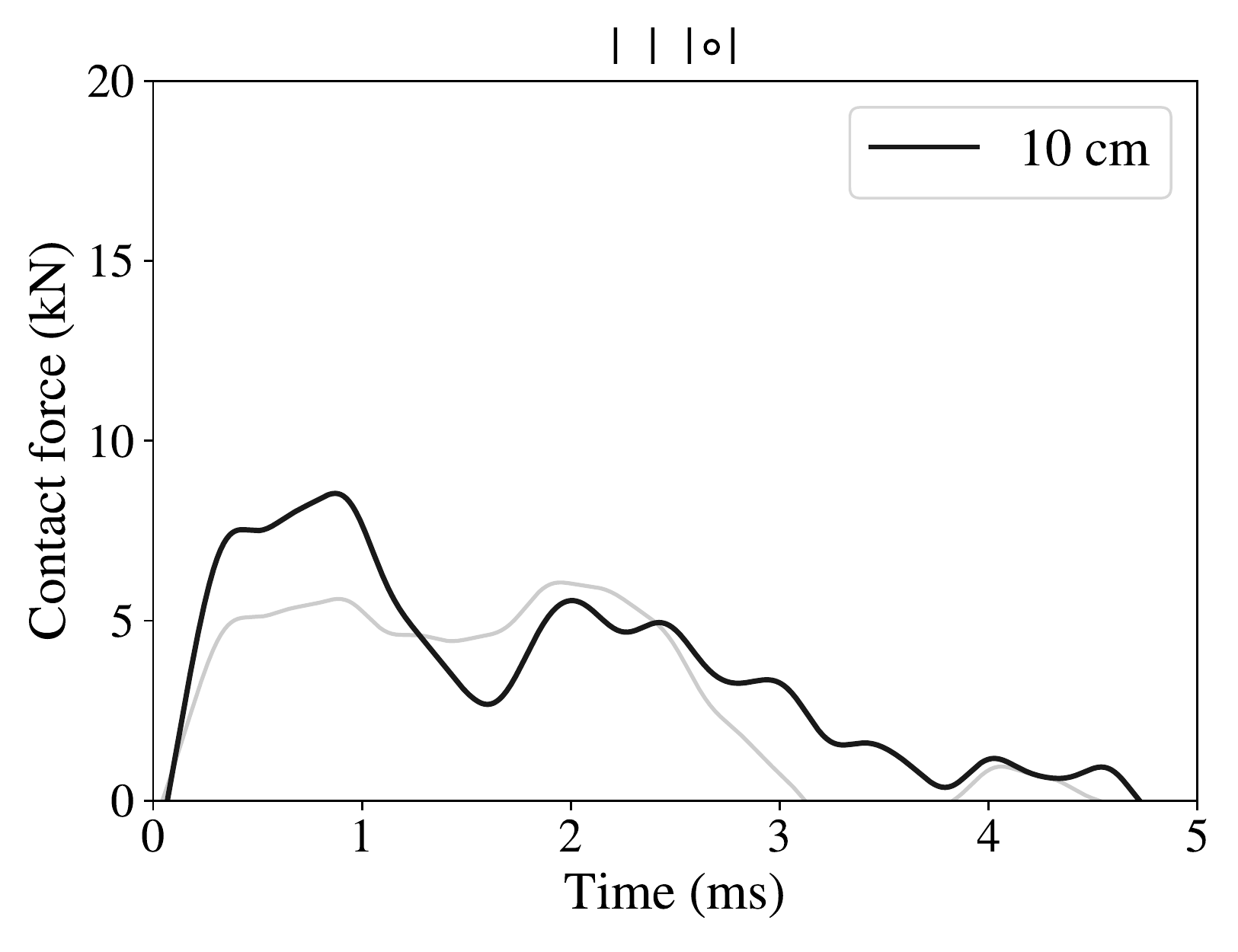}
         &  
         \includegraphics[width=0.35\textwidth]{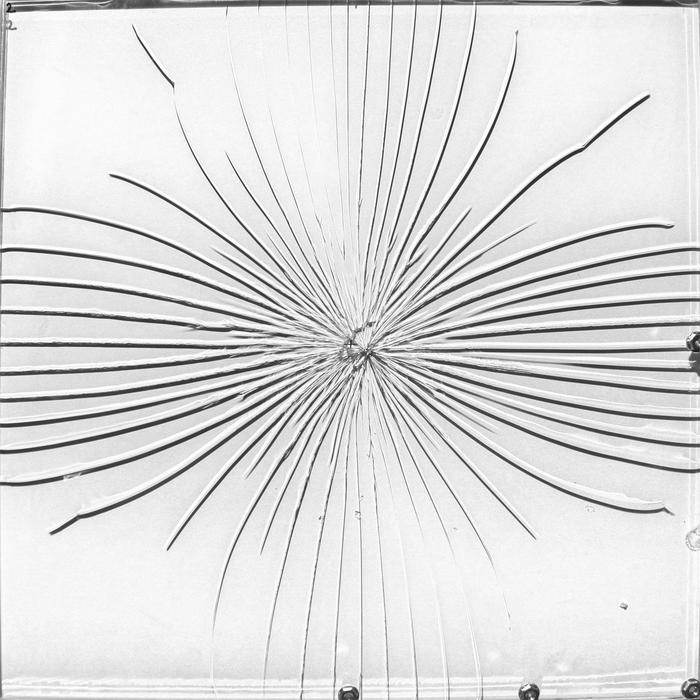}
         \\
         \\
         \includegraphics[width=0.475\textwidth]{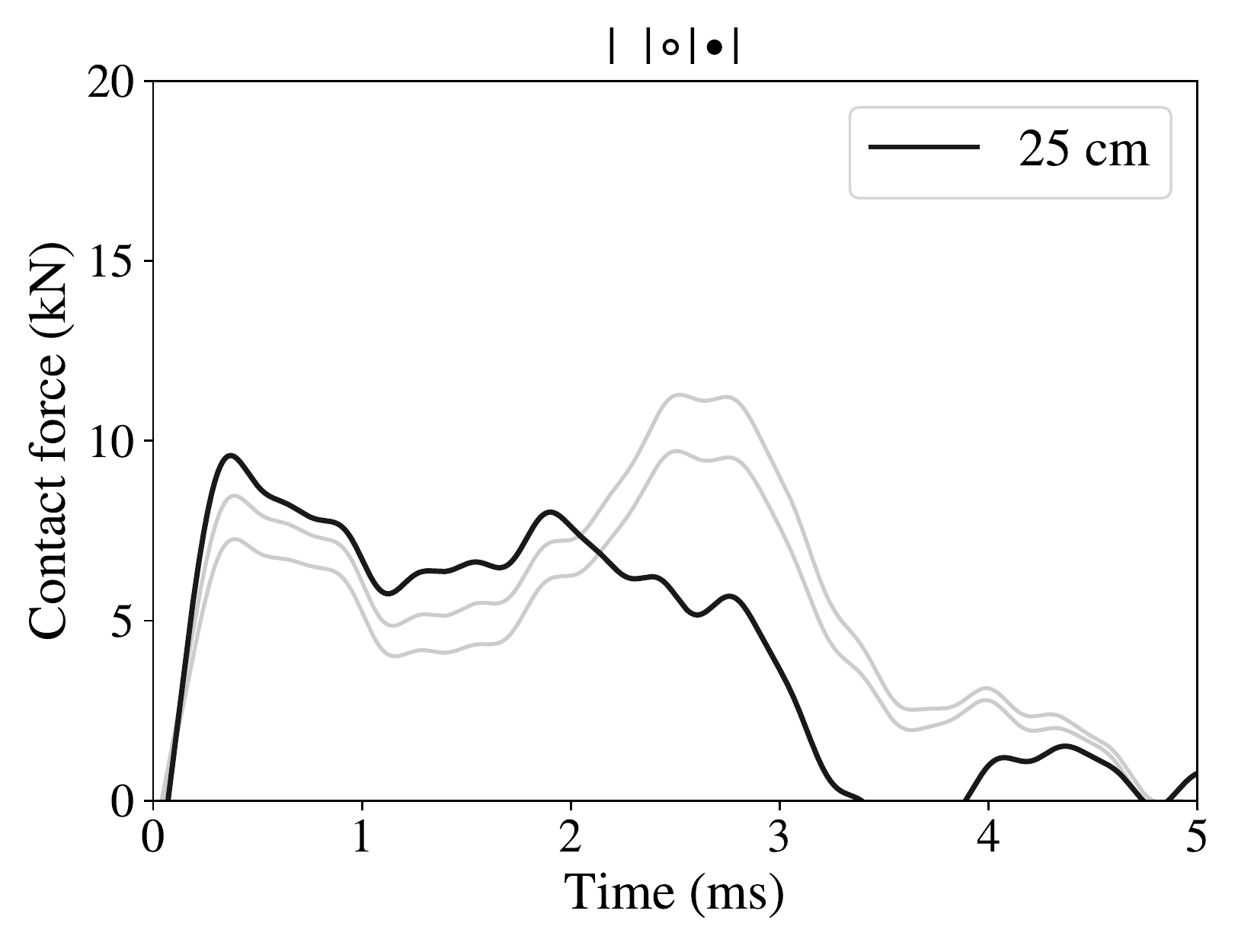}
         & 
         \includegraphics[width=0.35\textwidth]{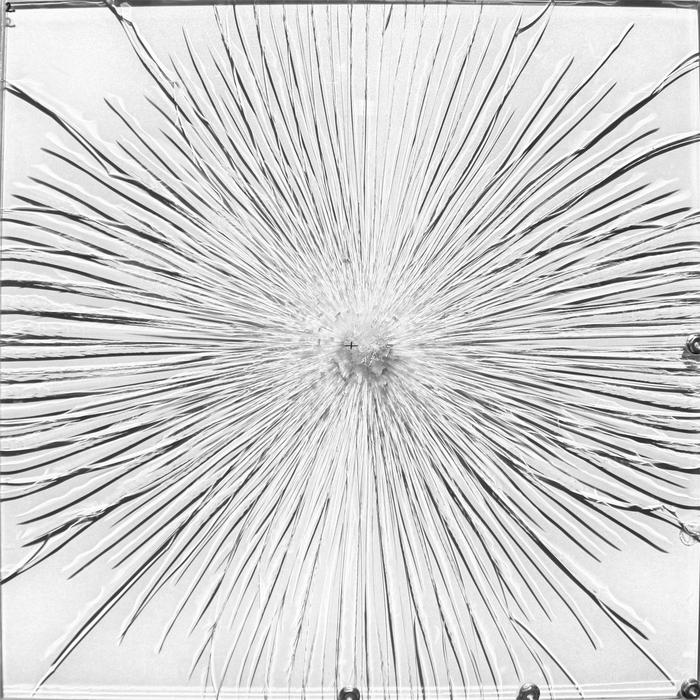}
         \\
         \\
         \includegraphics[width=0.475\textwidth]{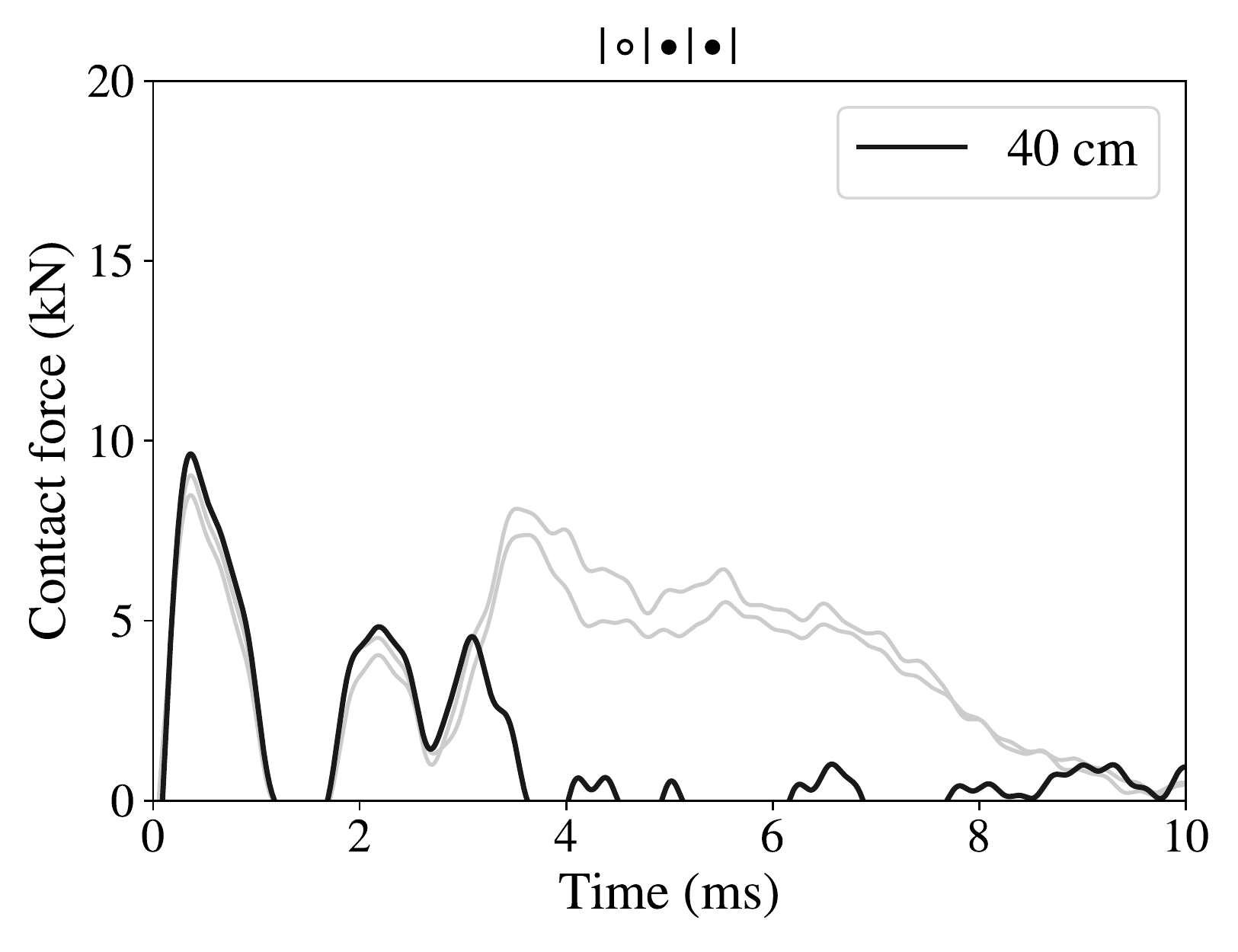}
         &
         \includegraphics[width=0.35\textwidth]{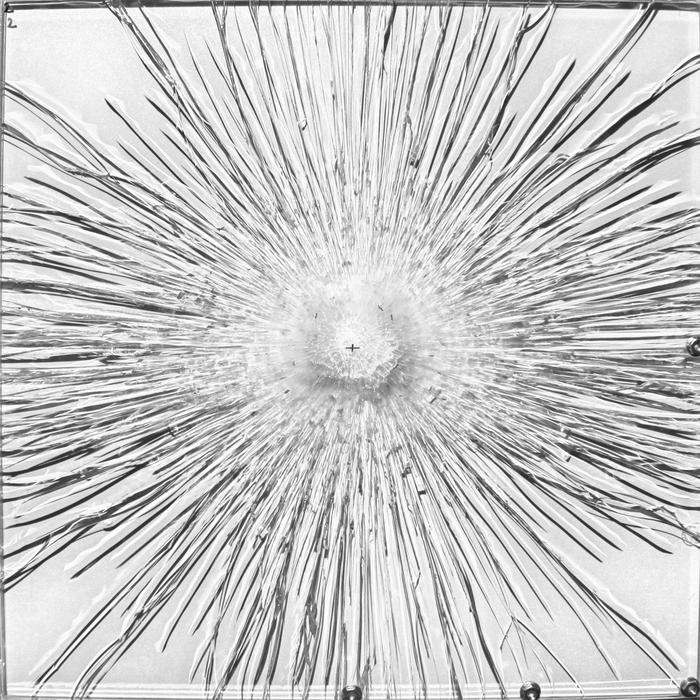}
    \end{tabular}
    \caption{Evolution of contact forces (2$\times$ filter CFC~1000) and corresponding fracture patterns for 5LG--2. Black lines denote the contact forces during the destructive tests, whereas the grey ones correspond to the previous sequence of impact events at lower impact heights (increased by 5~cm for each following impact event).}
    \label{fig:cf_5LG-2}
\end{figure}

\begin{figure}[hp]
    \centering
\begin{tabular}{ccc}
    \includegraphics[width=0.3\textwidth]{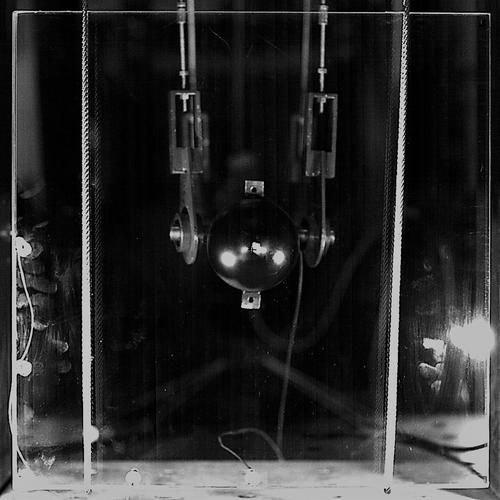}
     &  
    \includegraphics[width=0.3\textwidth]{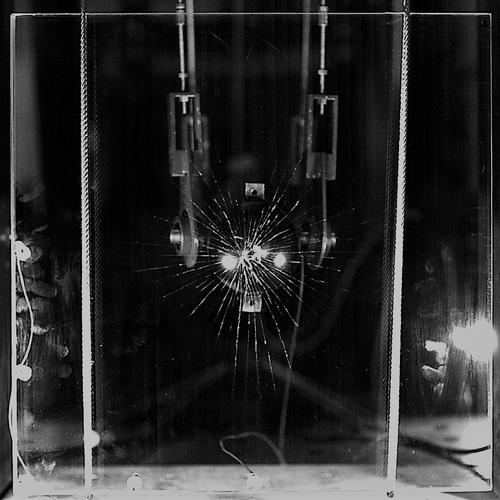}
    &
    \includegraphics[width=0.3\textwidth]{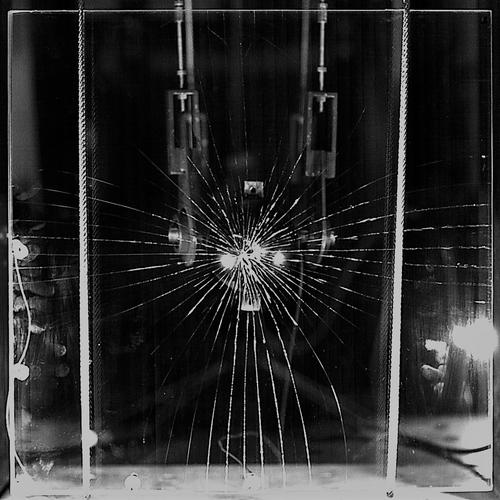}
     \\
& 
    \footnotesize 10~cm 
\textbar~\textbar~\textbar$\ccirc$\textbar 
    &
\\
     \\
    \includegraphics[width=0.3\textwidth]{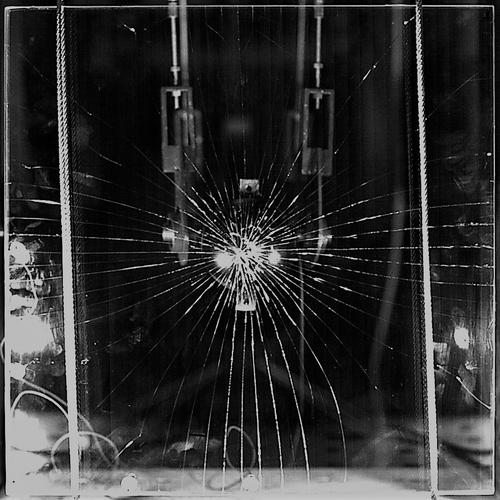}
    &
    \includegraphics[width=0.3\textwidth]{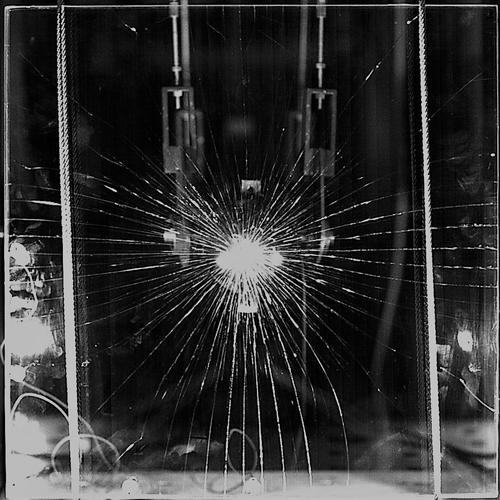}
    &
    \includegraphics[width=0.3\textwidth]{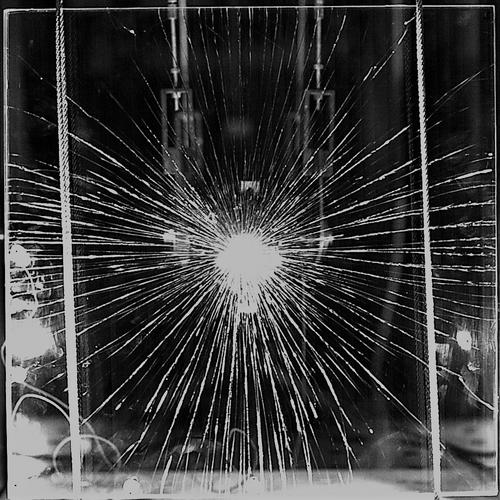}
    \\
& 
    \footnotesize 25~cm 
\textbar~\textbar$\ccirc$\textbar$\sbullet[.72]$\textbar 
    &
\\
    \\
    \includegraphics[width=0.3\textwidth]{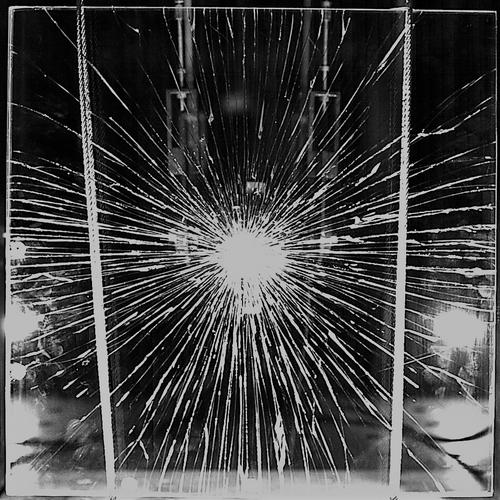}
    &
    \includegraphics[width=0.3\textwidth]{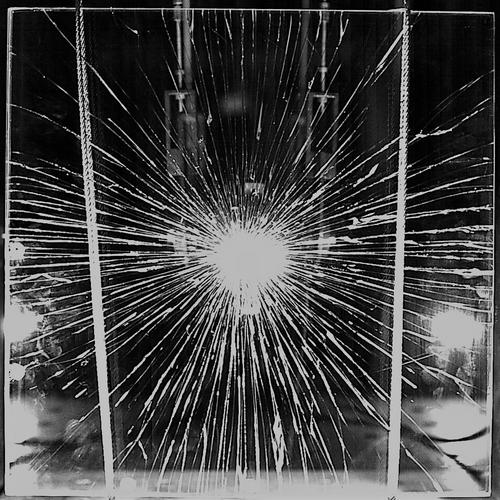}
    &
    \includegraphics[width=0.3\textwidth]{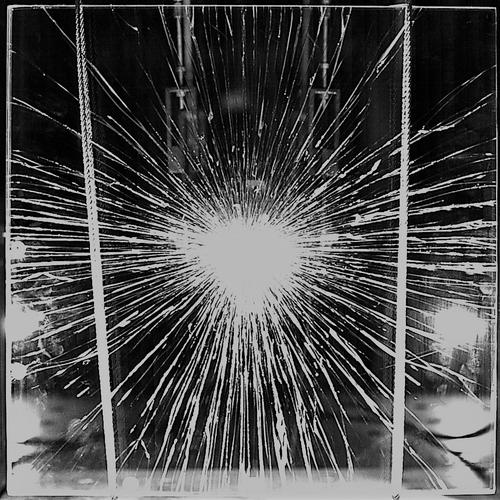}
    \\
&
    \footnotesize 40~cm 
\textbar$\ccirc$\textbar$\sbullet[.72]$\textbar$\sbullet[.72]$\textbar
    &
\end{tabular}
    \caption{Evolution of cracks in 5LG-2 sample (frame rate 5,000 frames per second, i.e., time step 0.2~ms between two adjacent photos in a row).}
    \label{fig:crack_ev_5LG2}
\end{figure}

\begin{figure}[hp]
    \centering
    \begin{tabular}[t]{cc}
         \includegraphics[width=0.475\textwidth]{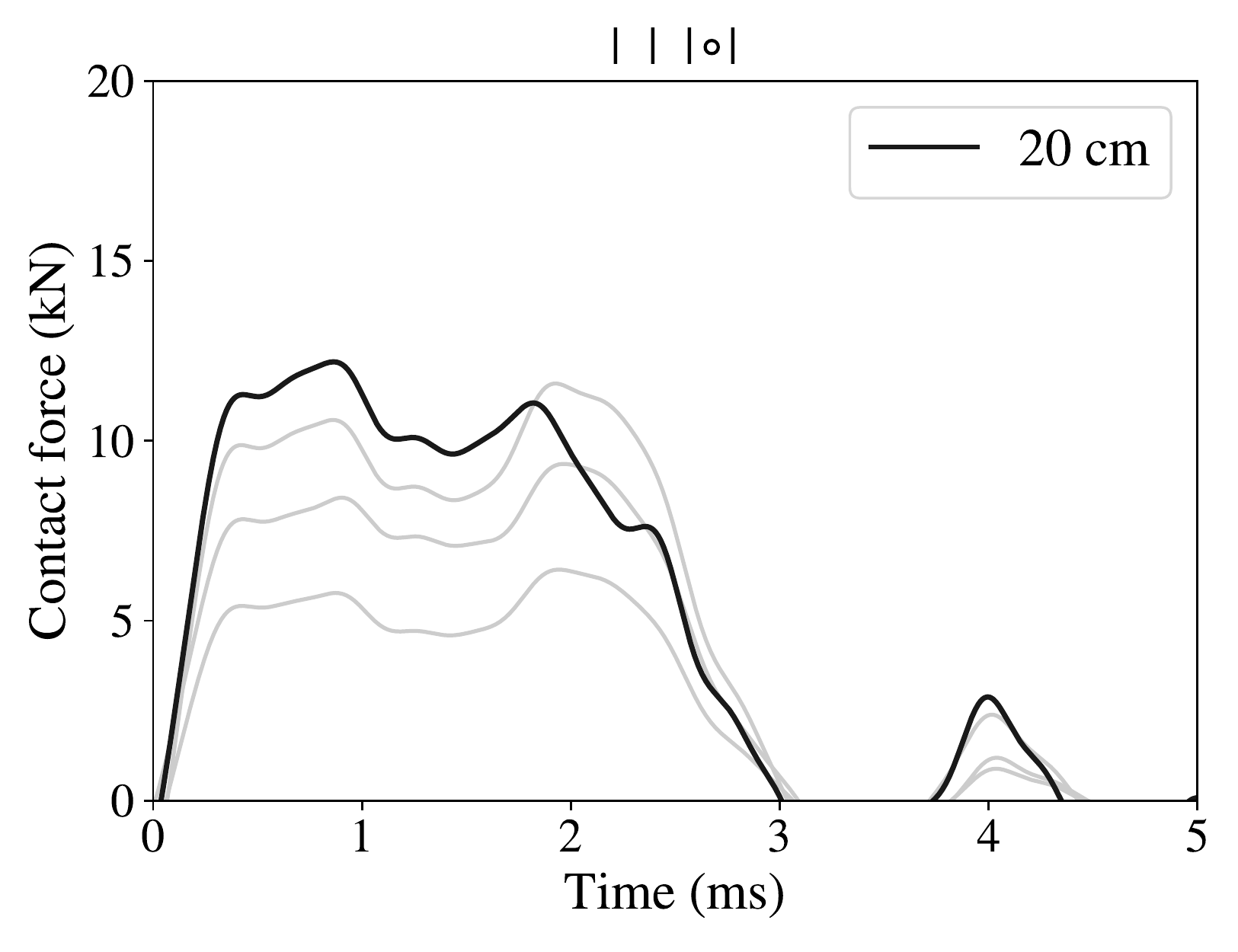}
         &  
         \includegraphics[width=0.35\textwidth]{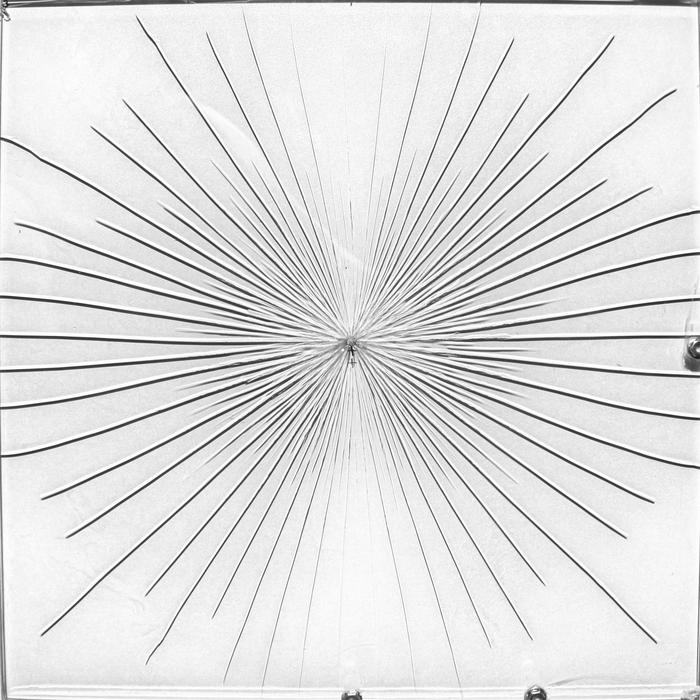}
         \\
         \\
         \includegraphics[width=0.475\textwidth]{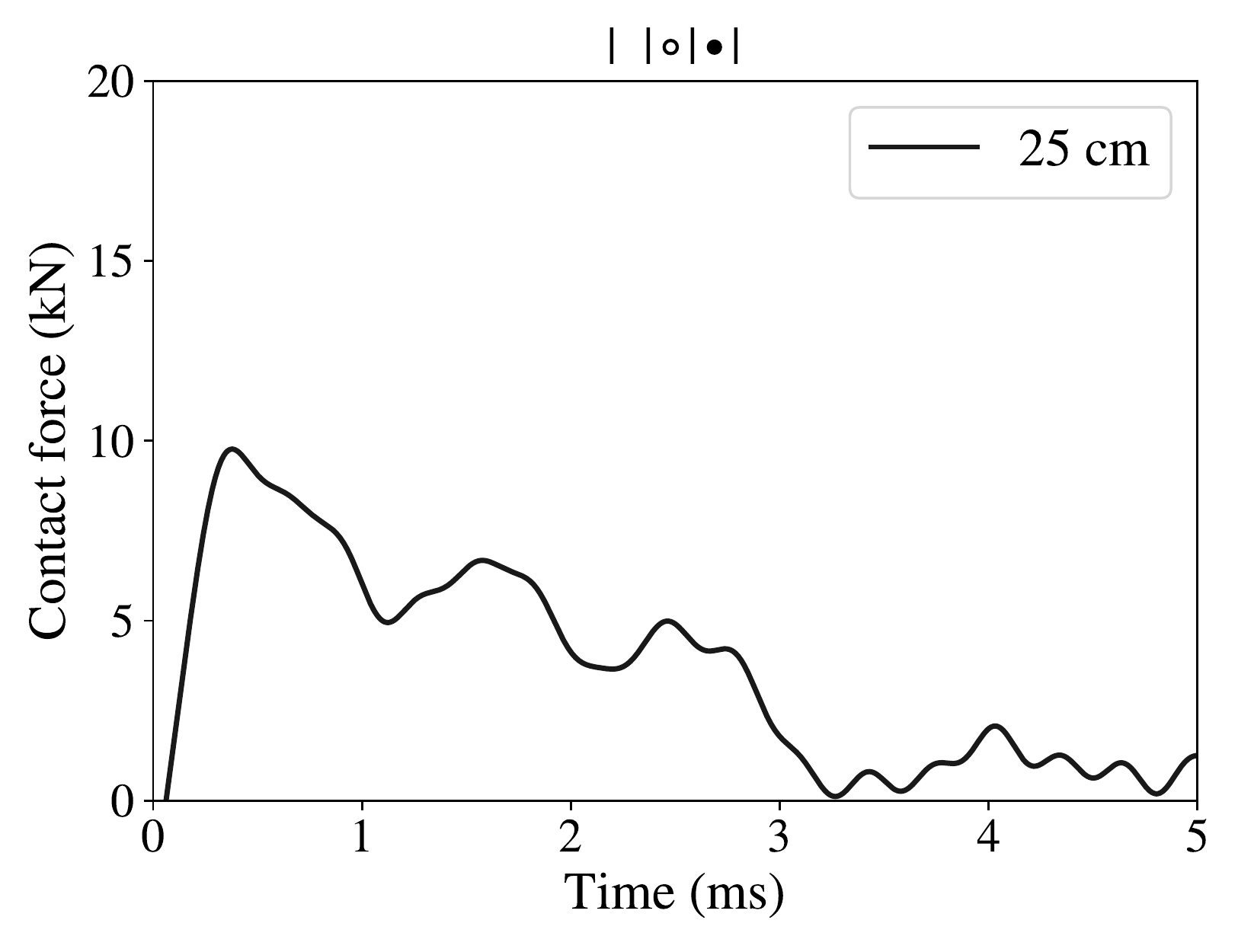}
         & 
         \includegraphics[width=0.35\textwidth]{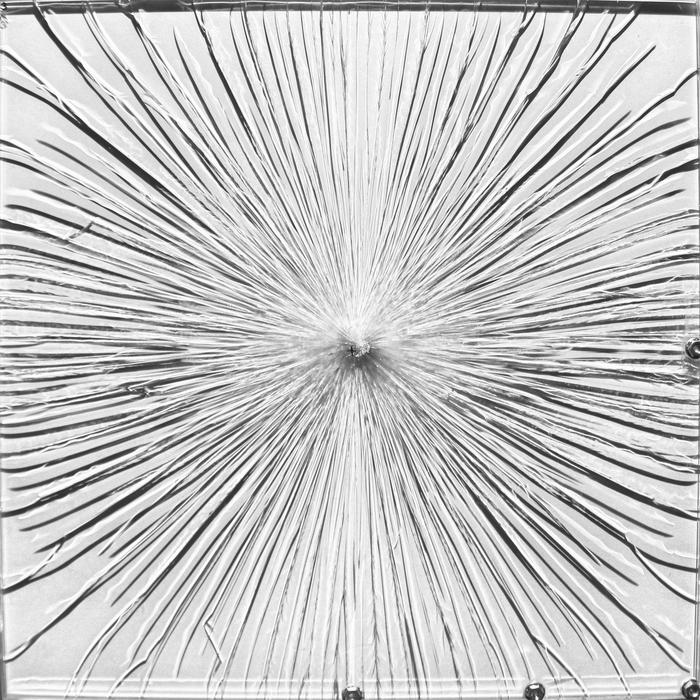}
         \\
         \\
         \includegraphics[width=0.475\textwidth]{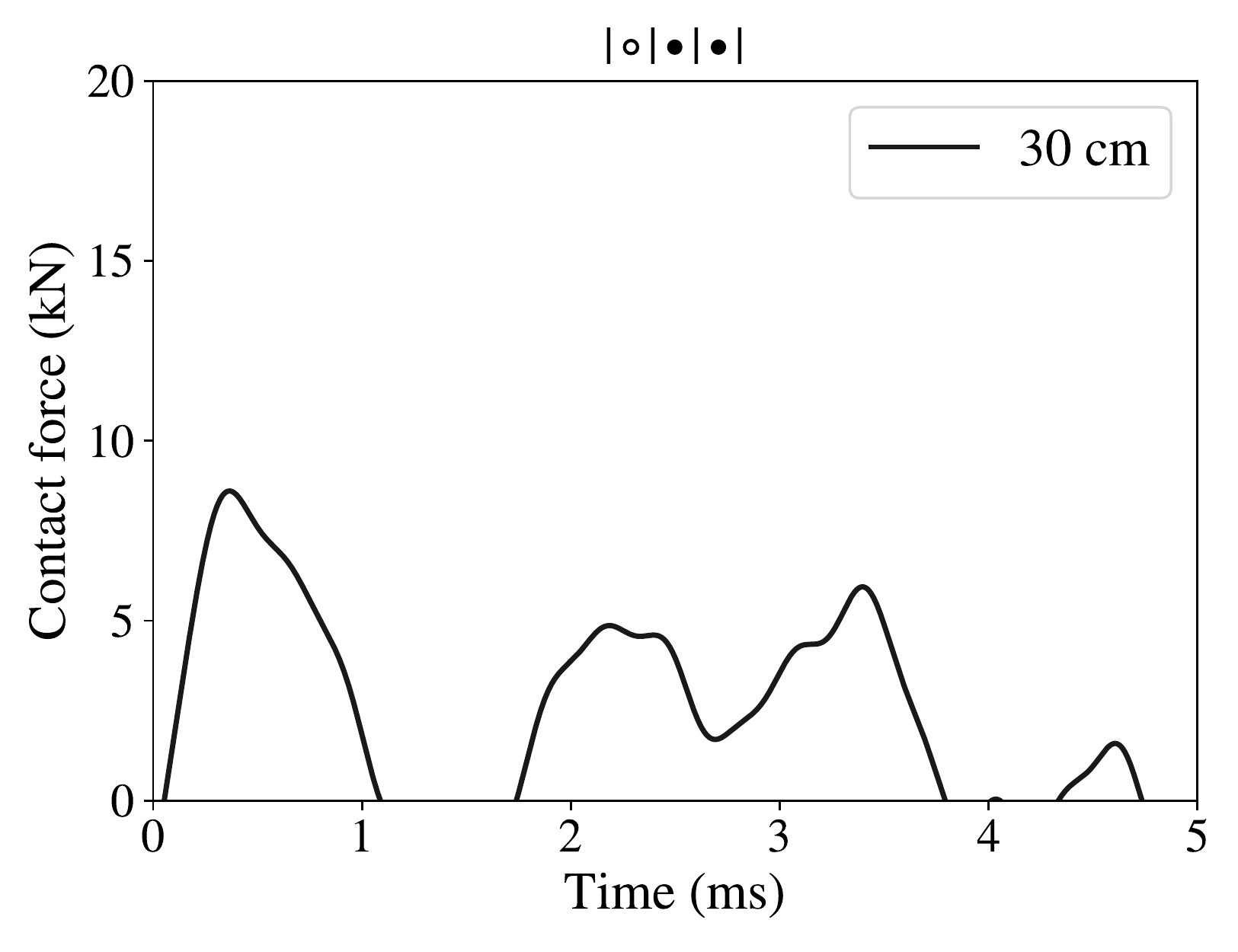}
         &
         \includegraphics[width=0.35\textwidth]{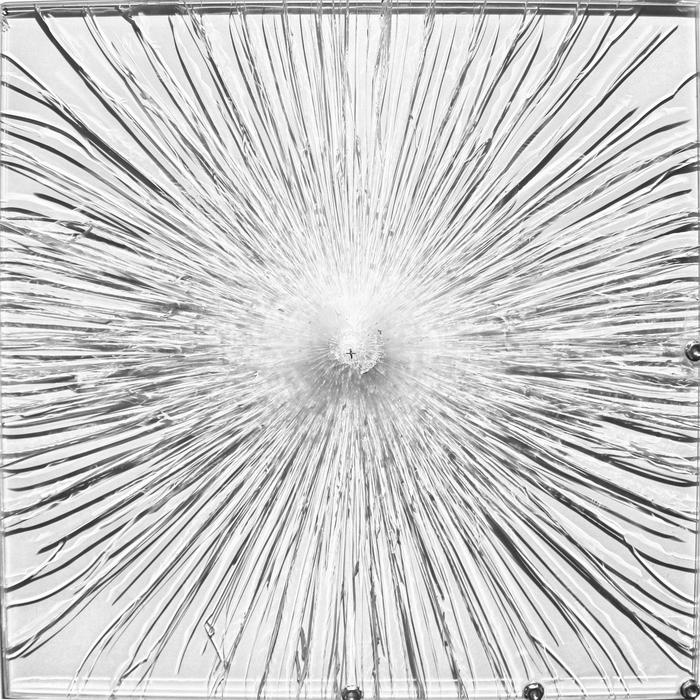}
    \end{tabular}
    \caption{Evolution of contact forces (2$\times$ filter CFC~1000) and corresponding fracture patterns for 5LG--3. Black lines denote the contact forces during the destructive tests, whereas the grey ones correspond to the previous sequence of impact events at lower impact heights (increased by 5~cm for each following impact event).}
    \label{fig:cf_5LG-3}
\end{figure}

\begin{figure}[hp]
    \centering
\begin{tabular}{ccc}
    \includegraphics[width=0.3\textwidth]{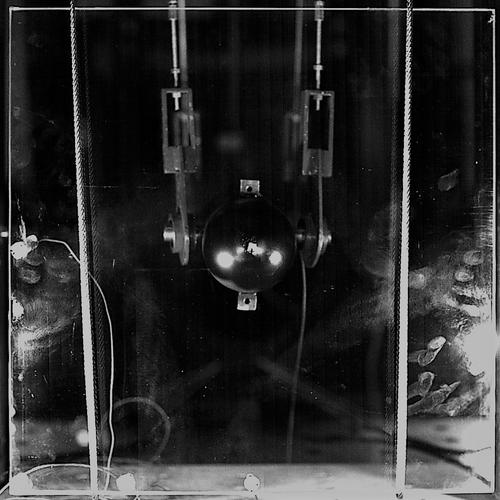}
     &  
    \includegraphics[width=0.3\textwidth]{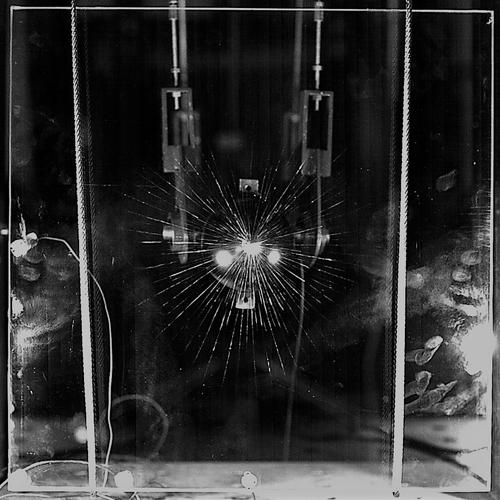}
    &
    \includegraphics[width=0.3\textwidth]{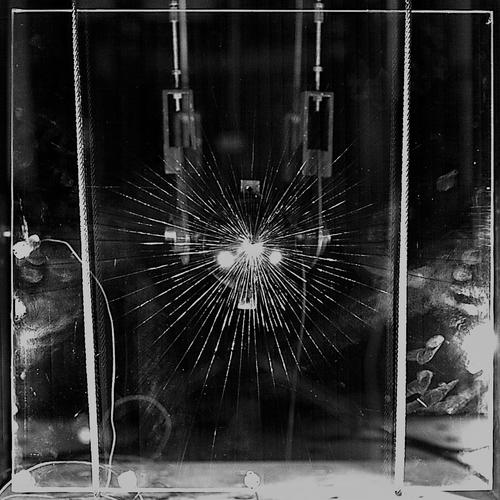}
     \\
& 
    \footnotesize 20~cm 
\textbar~\textbar~\textbar$\ccirc$\textbar 
    &
\\
     \\
    \includegraphics[width=0.3\textwidth]{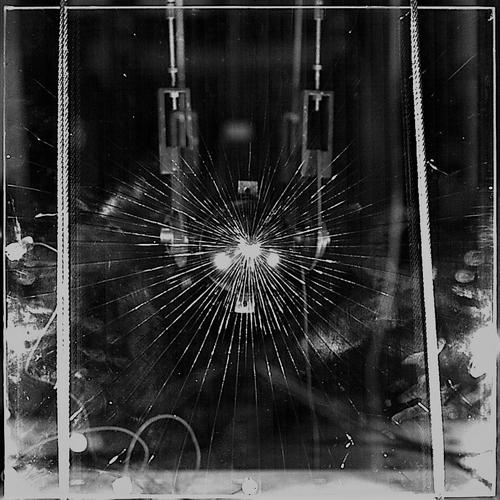}
    &
    \includegraphics[width=0.3\textwidth]{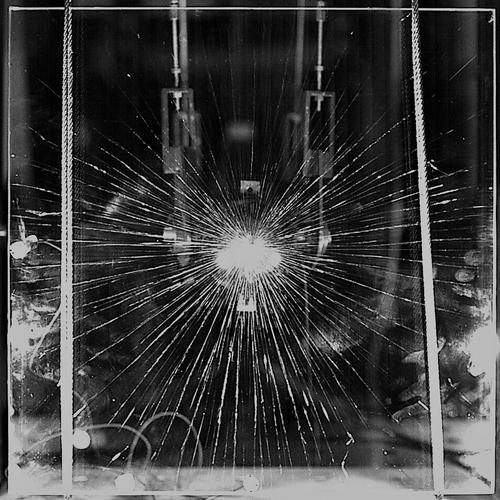}
    &
    \includegraphics[width=0.3\textwidth]{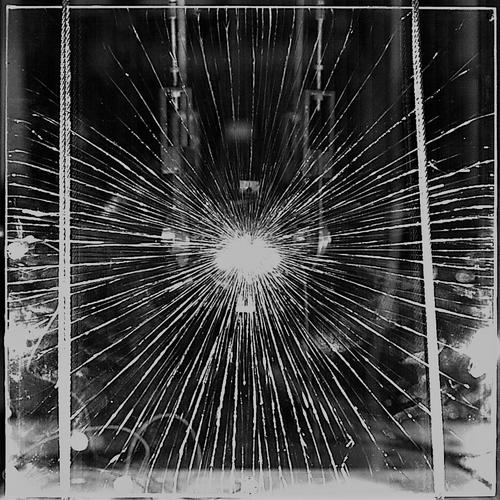}
    \\
& 
    \footnotesize 25~cm 
\textbar~\textbar$\ccirc$\textbar$\sbullet[.72]$\textbar 
    &
\\
    \\
    \includegraphics[width=0.3\textwidth]{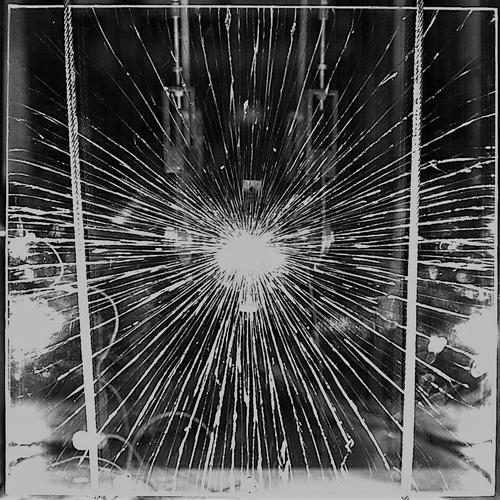}
    &
    \includegraphics[width=0.3\textwidth]{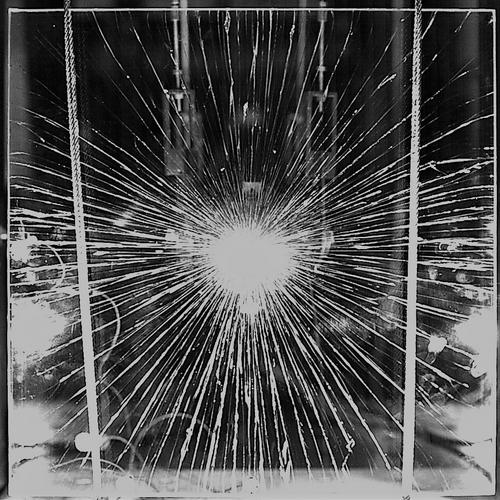}
    &
    \includegraphics[width=0.3\textwidth]{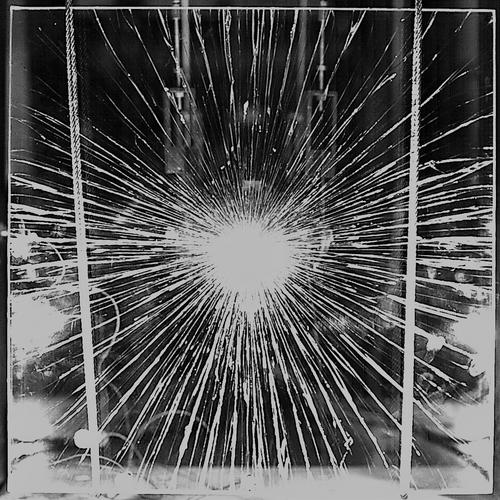}
    \\
&
    \footnotesize 30~cm 
\textbar$\ccirc$\textbar$\sbullet[.72]$\textbar$\sbullet[.72]$\textbar
    &
\end{tabular}
    \caption{Evolution of cracks in 5LG-3 sample (frame rate 5,000 frames per second, i.e., time step 0.2~ms between two adjacent photos in a row).}
    \label{fig:crack_ev_5LG3}
\end{figure}

\begin{figure}[hp]
    \centering
    \begin{tabular}[t]{cc}
         \includegraphics[width=0.475\textwidth]{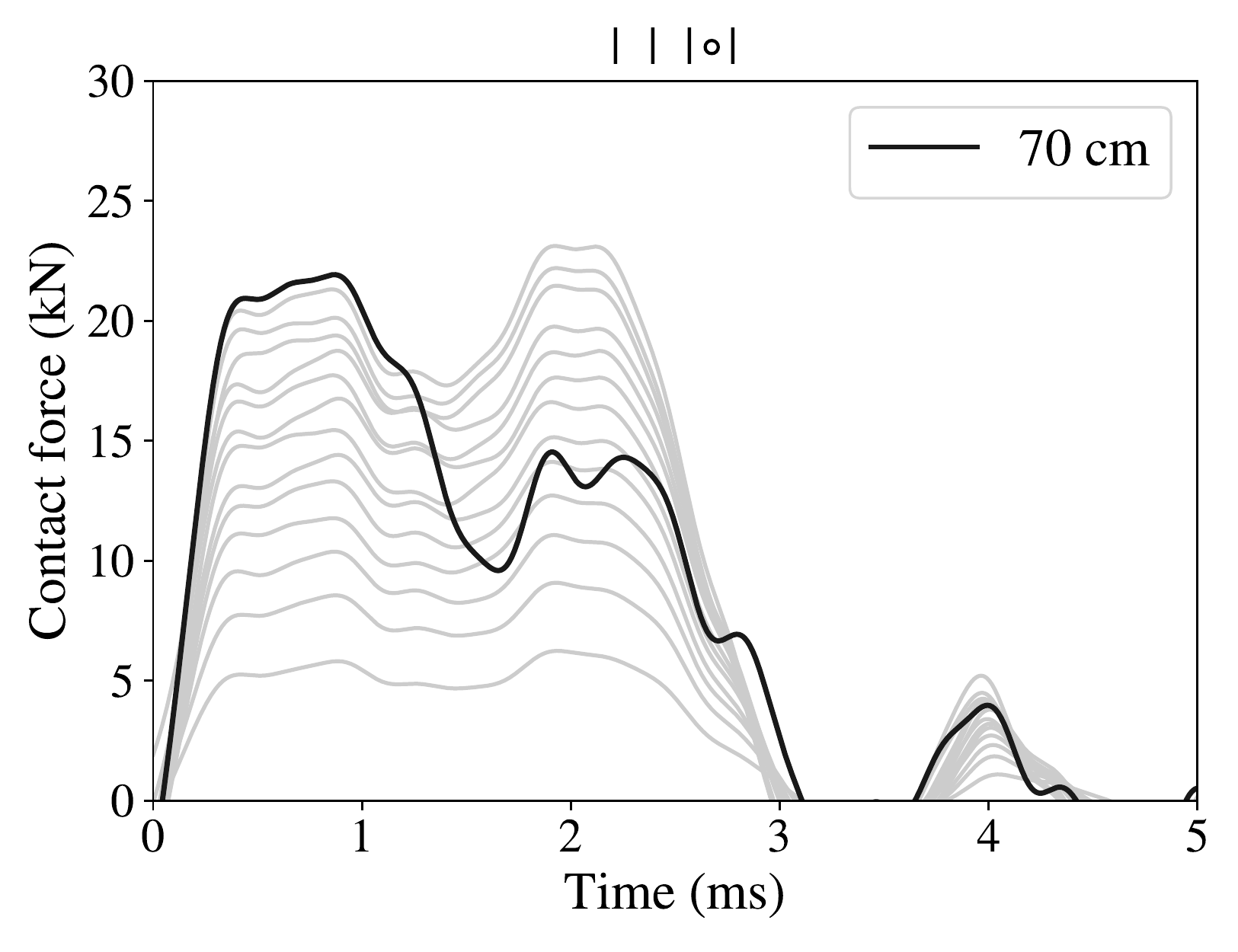}
         &  
         \includegraphics[width=0.35\textwidth]{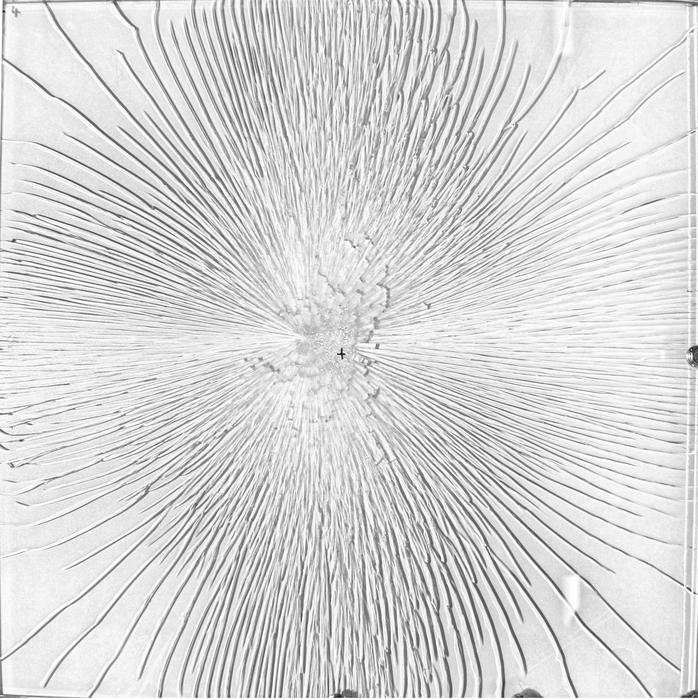}
         \\
         \\
         \includegraphics[width=0.475\textwidth]{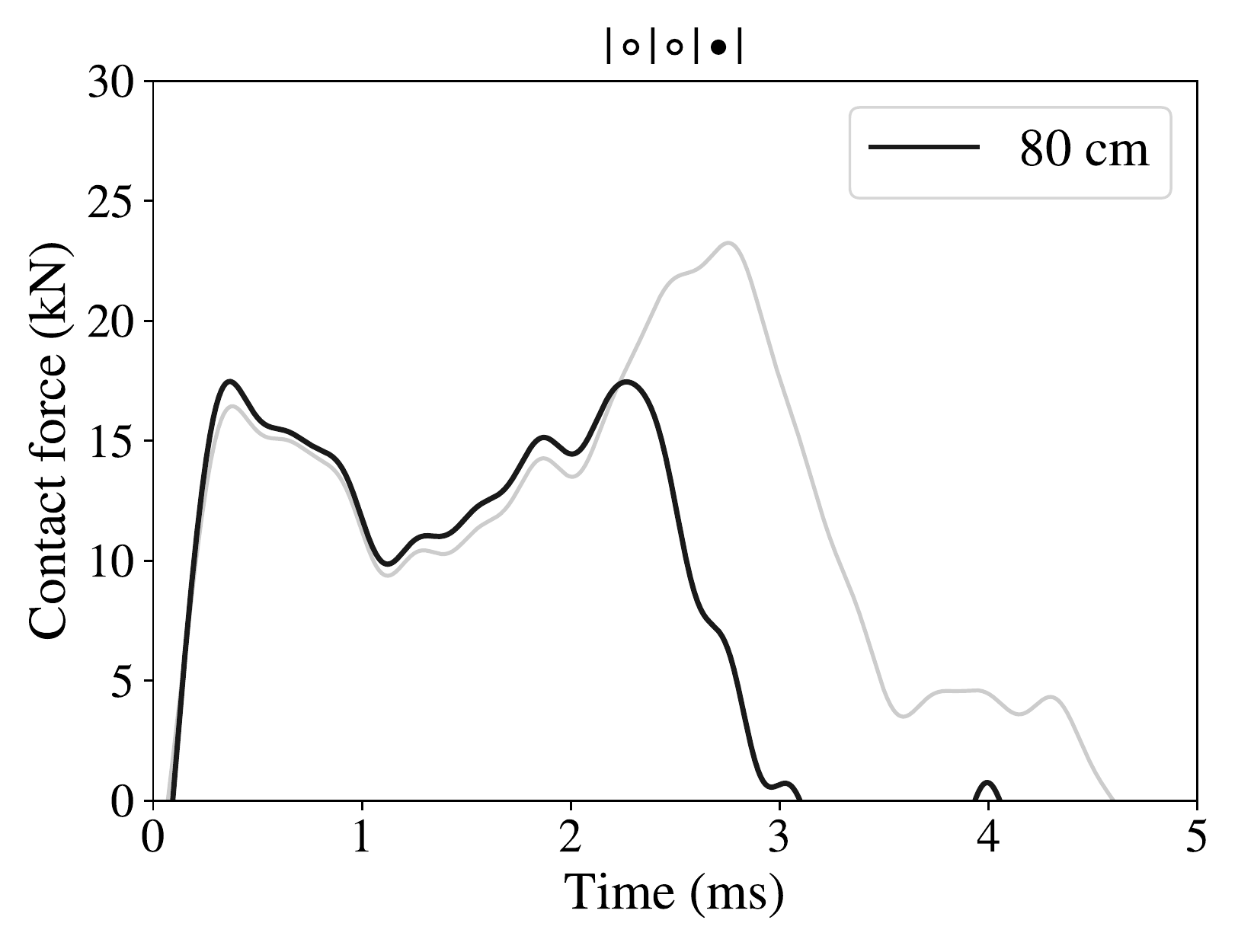}
         & 
         \includegraphics[width=0.35\textwidth]{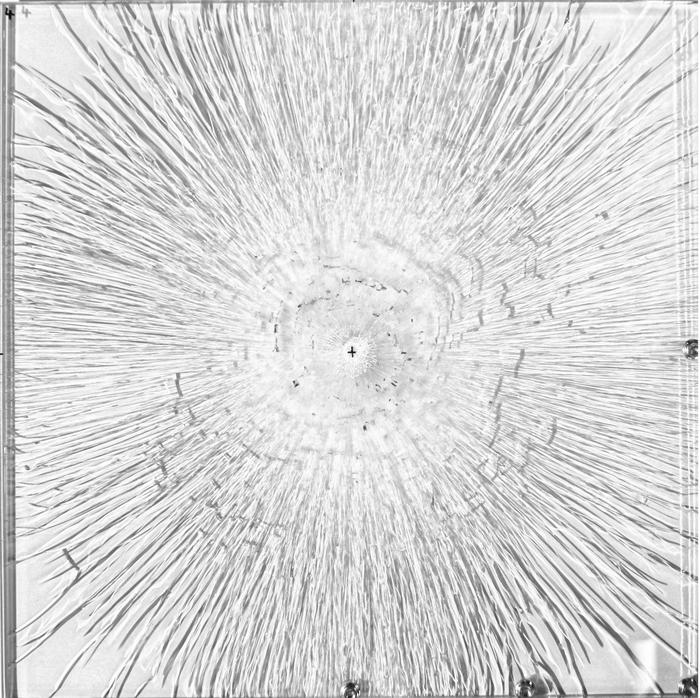}
    \end{tabular}
    \caption{Evolution of contact forces (2$\times$ filter CFC~1000) and corresponding fracture patterns for 5LG--4. Black lines denote the contact forces during the destructive tests, whereas the grey ones correspond to the previous sequence of impact events at lower impact heights (increased by 5~cm for each following impact event).}
    \label{fig:cf_5LG-4}
\end{figure}

\begin{figure}[hp]
    \centering
    \begin{tabular}[t]{cc}
         \includegraphics[width=0.475\textwidth]{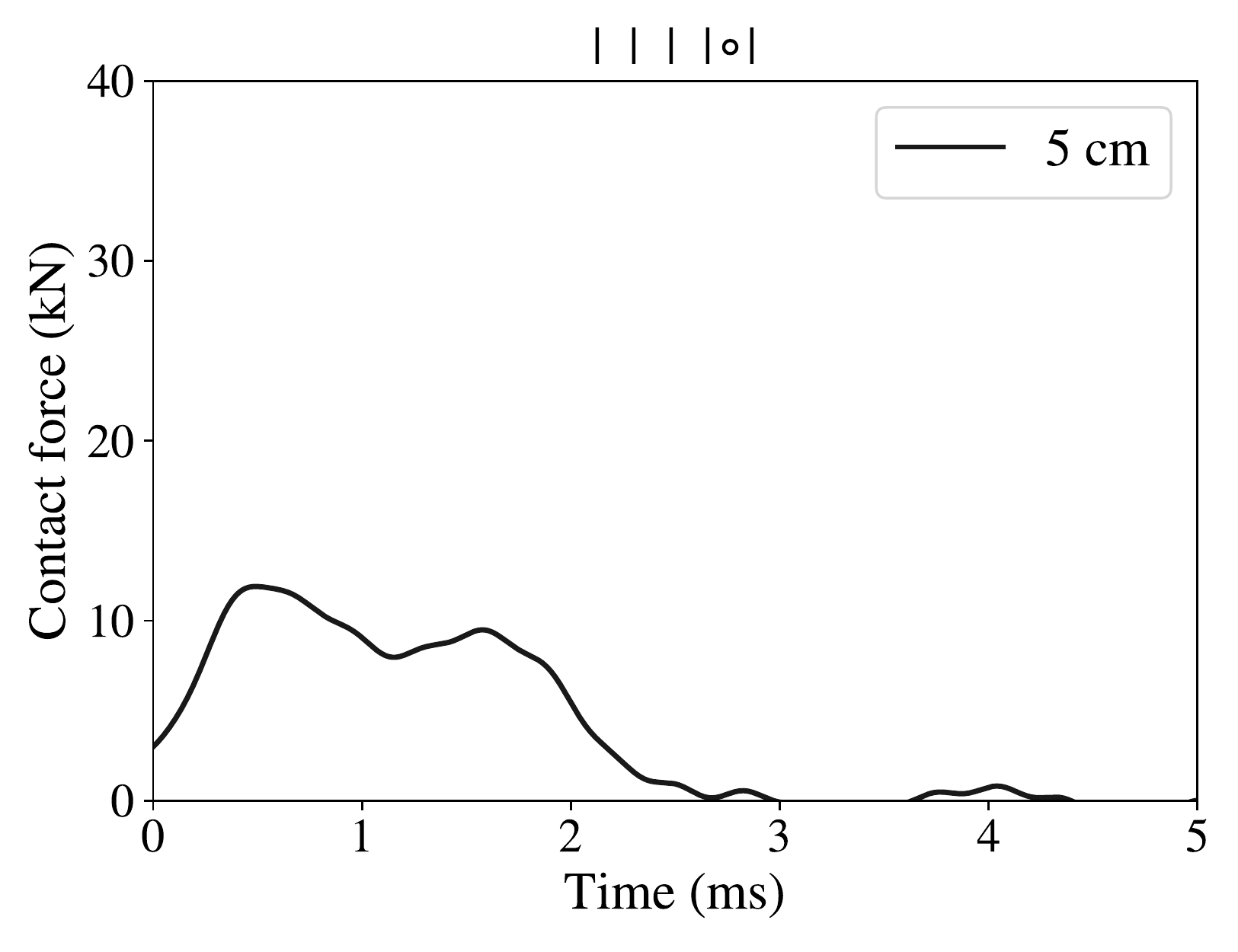}
         &  
         \includegraphics[width=0.35\textwidth]{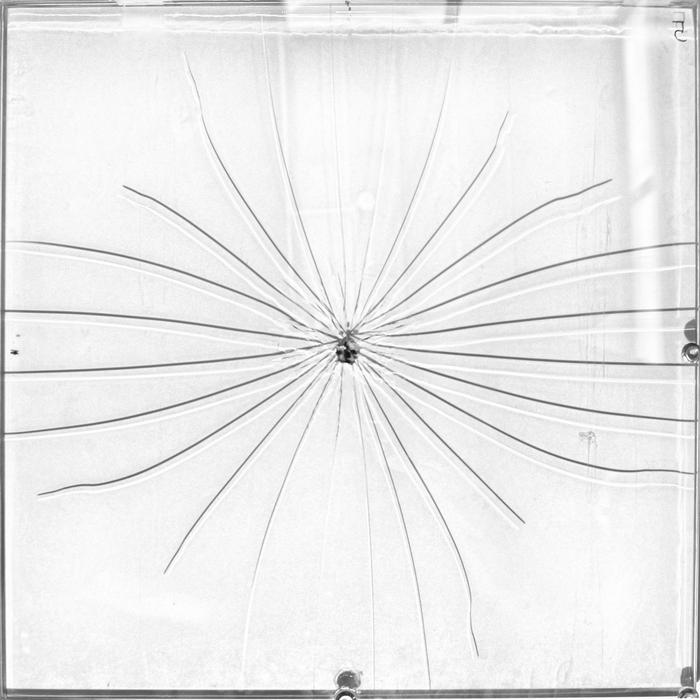}
         \\
         \\
         \includegraphics[width=0.475\textwidth]{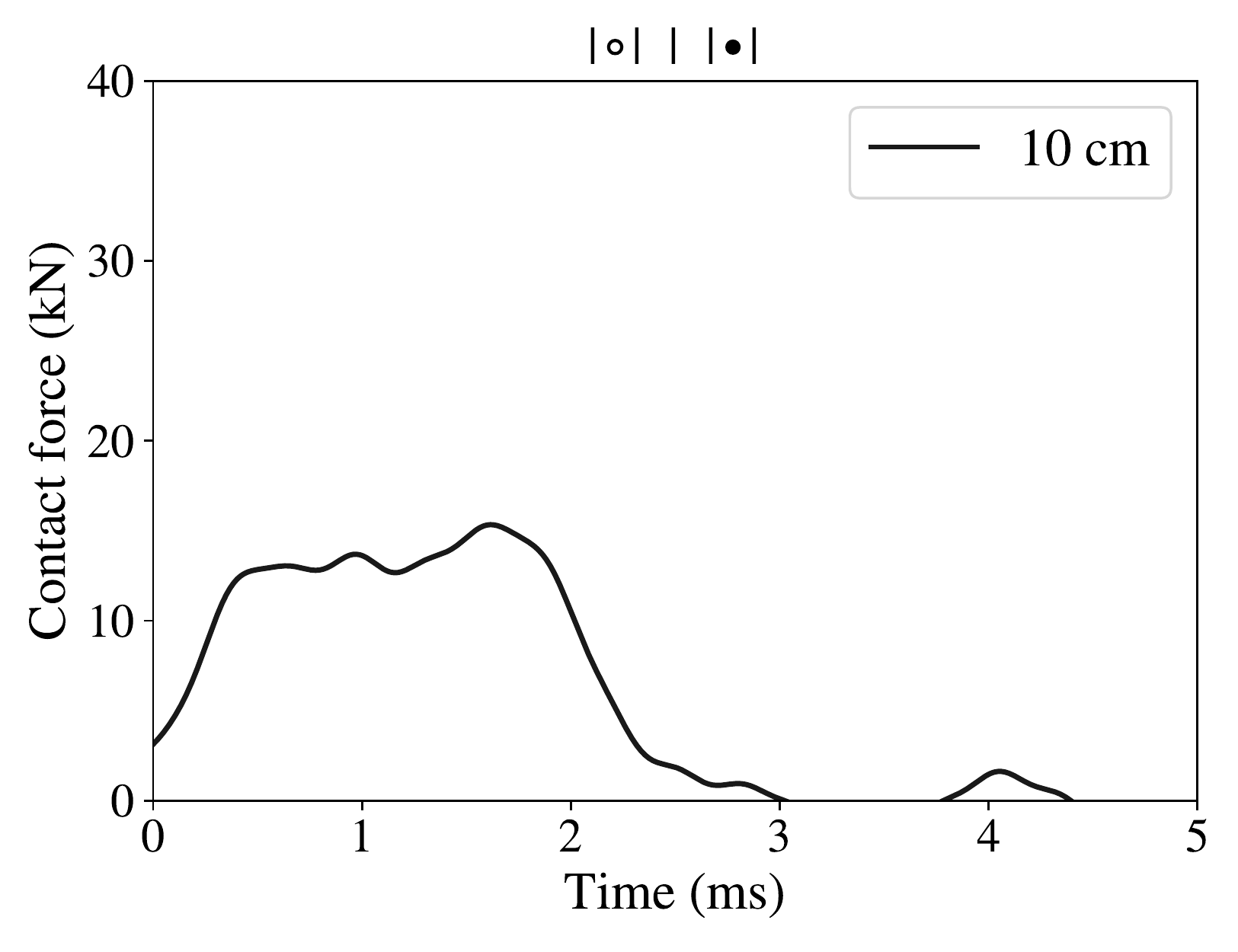}
         & 
         \includegraphics[width=0.35\textwidth]{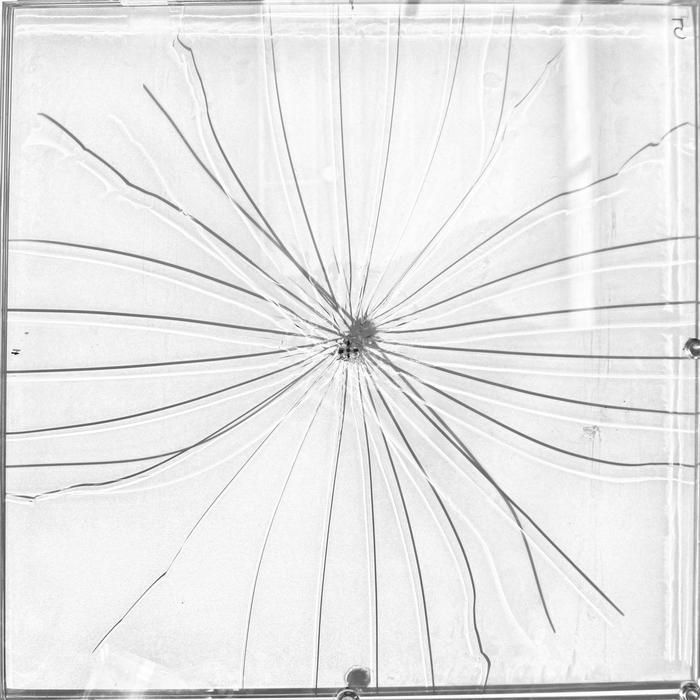}
         \\
         \\
         \includegraphics[width=0.475\textwidth]{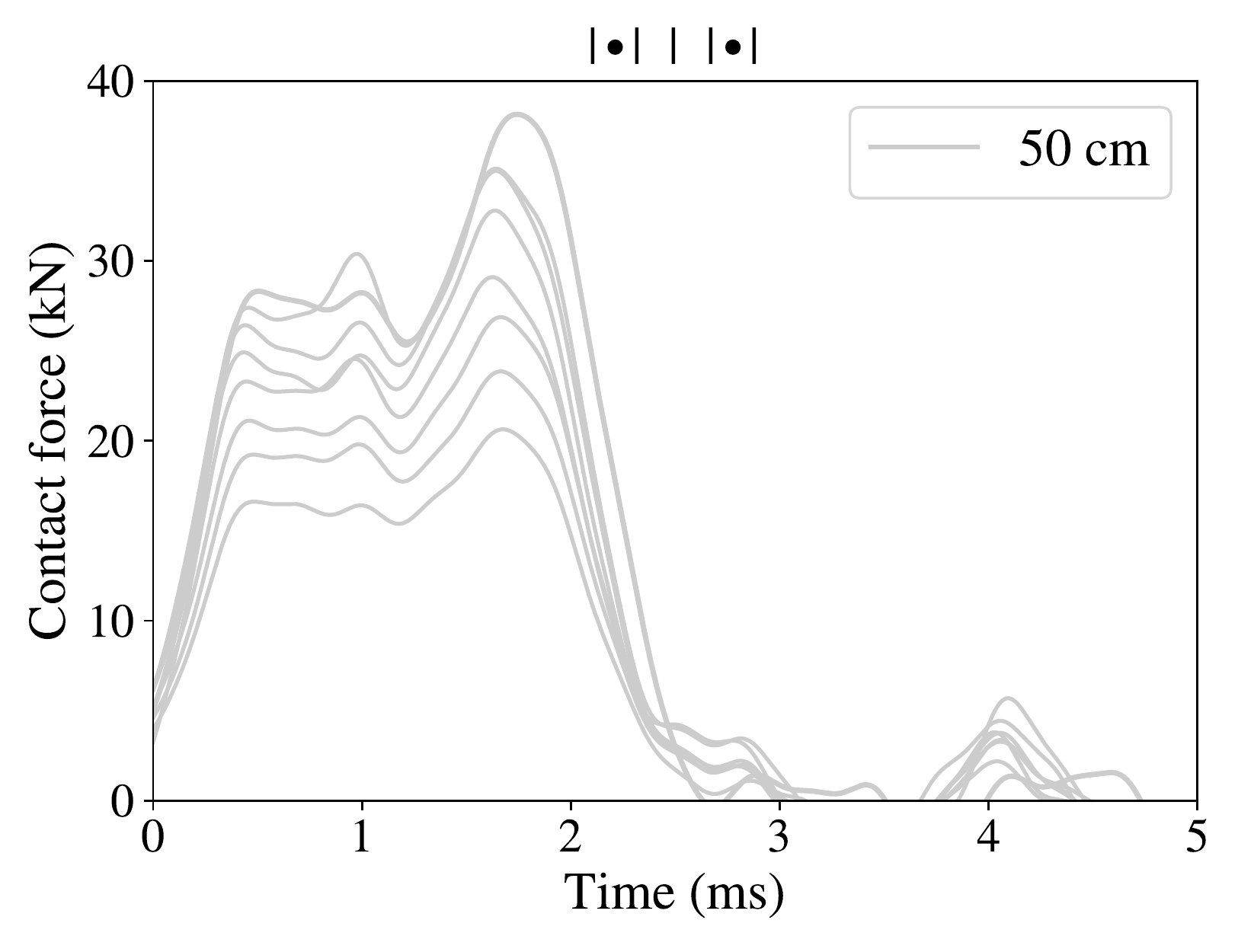}
         &
         \includegraphics[width=0.35\textwidth]{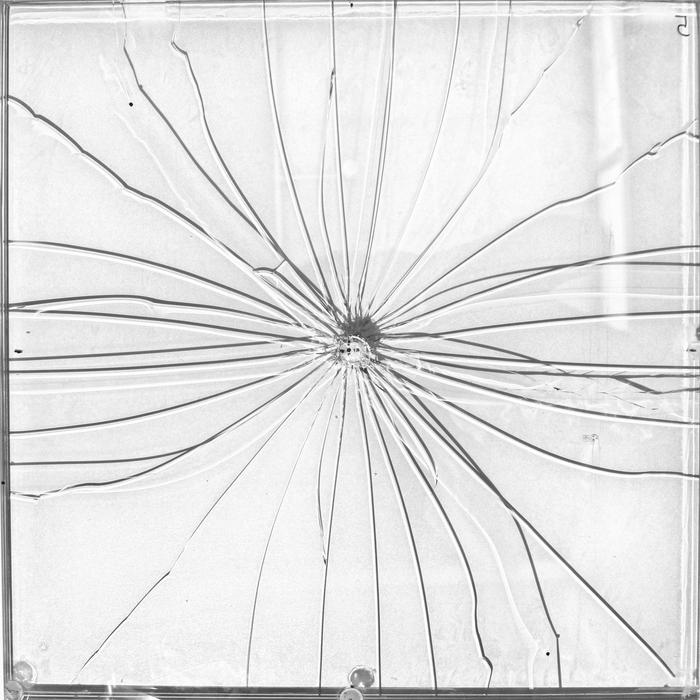}
    \end{tabular}
    \caption{Evolution of contact forces (2$\times$ filter CFC~1000) and corresponding fracture patterns for 7LG--1. Black lines denote the contact forces during the destructive tests, whereas the grey ones correspond to the previous sequence of impact events at lower impact heights (increased by 5~cm for each following impact event).}
    \label{fig:cf_7LG-1}
\end{figure}

\begin{figure}[hp]
    \centering
    \begin{tabular}[t]{cc}
         \includegraphics[width=0.475\textwidth]{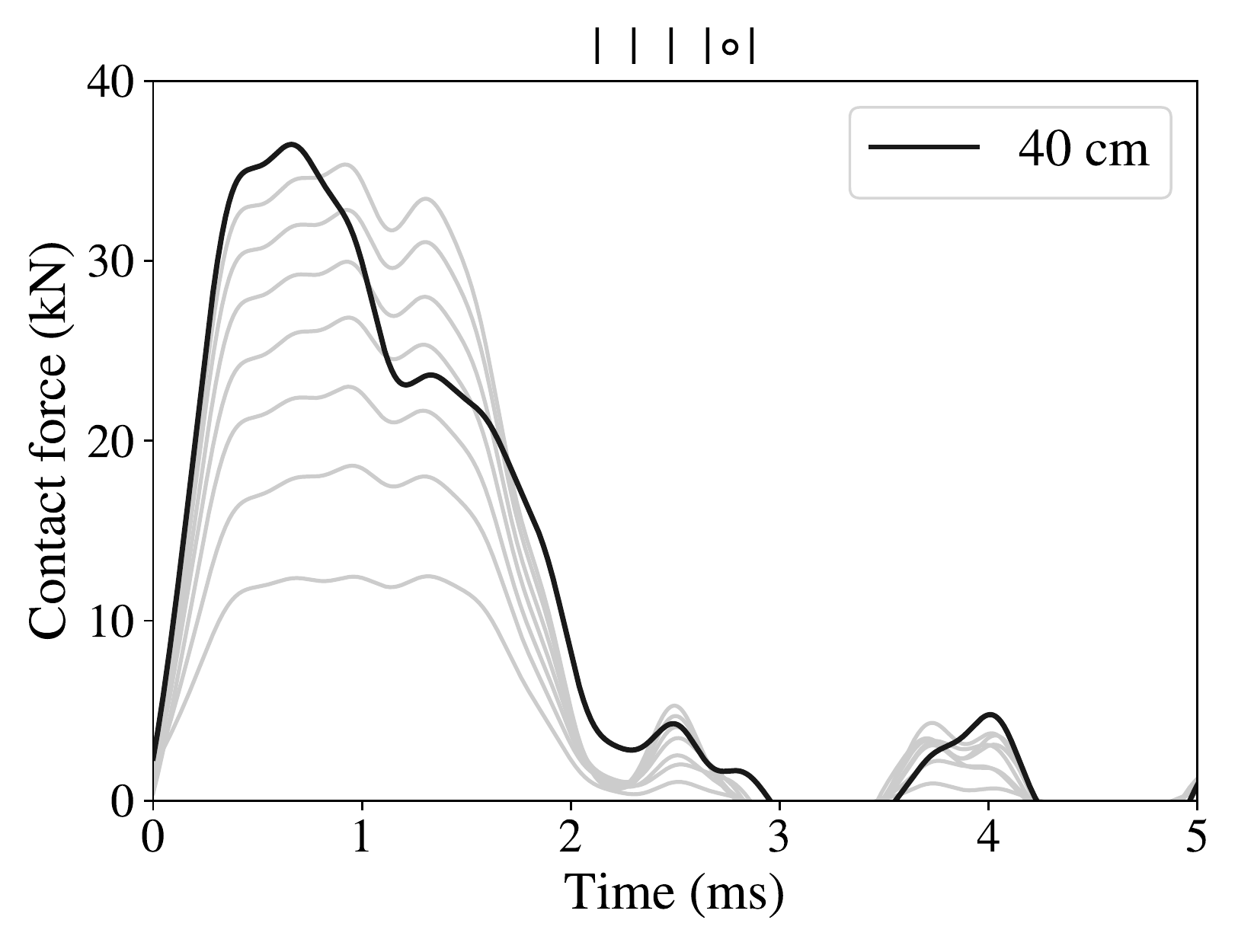}
         &  
         \includegraphics[width=0.35\textwidth]{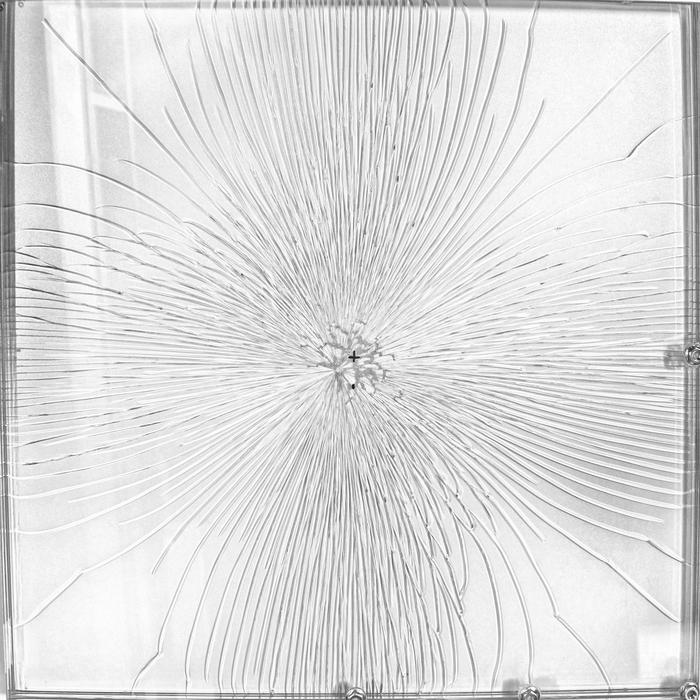}
         \\
         \\
         \includegraphics[width=0.475\textwidth]{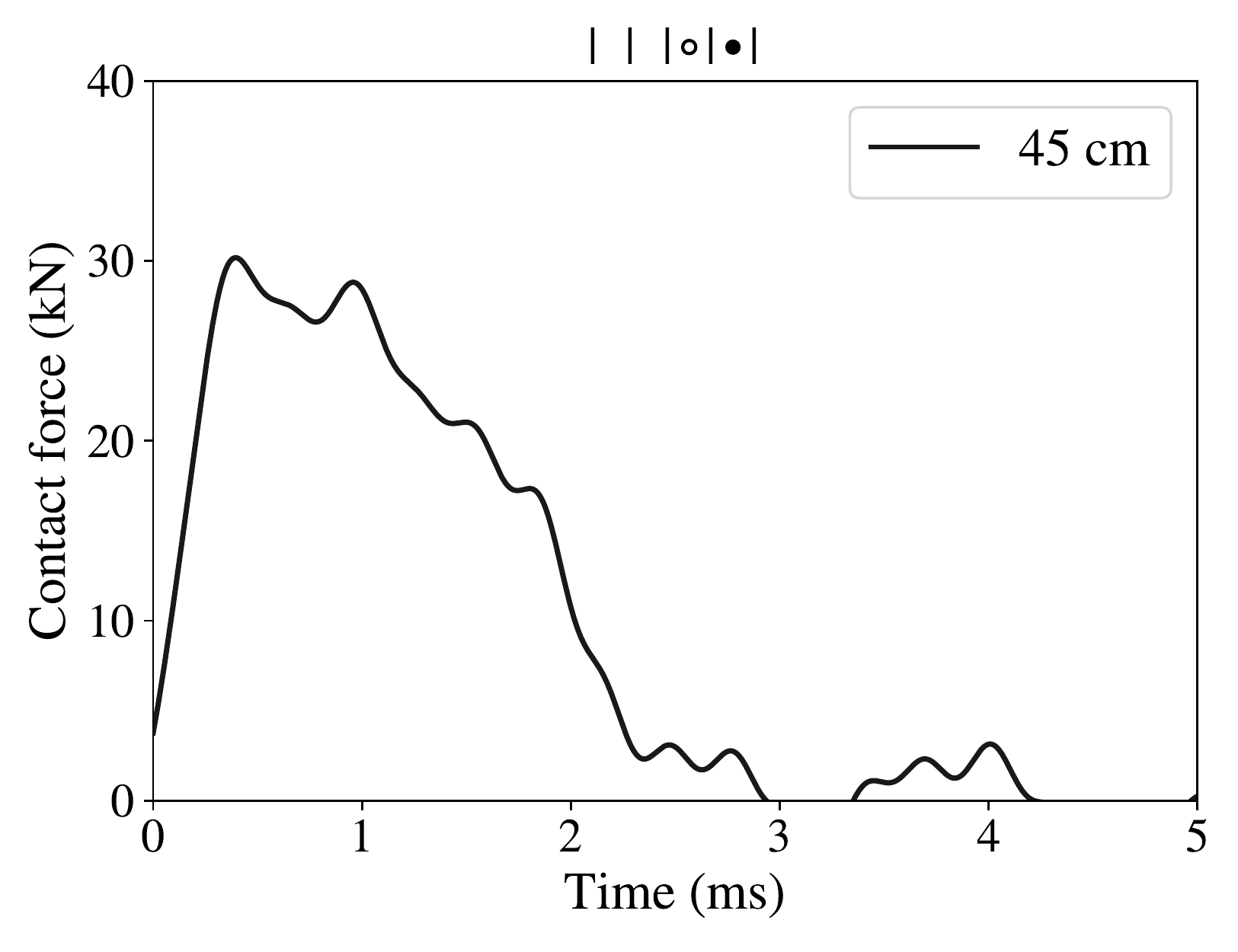}
         & 
         \includegraphics[width=0.35\textwidth]{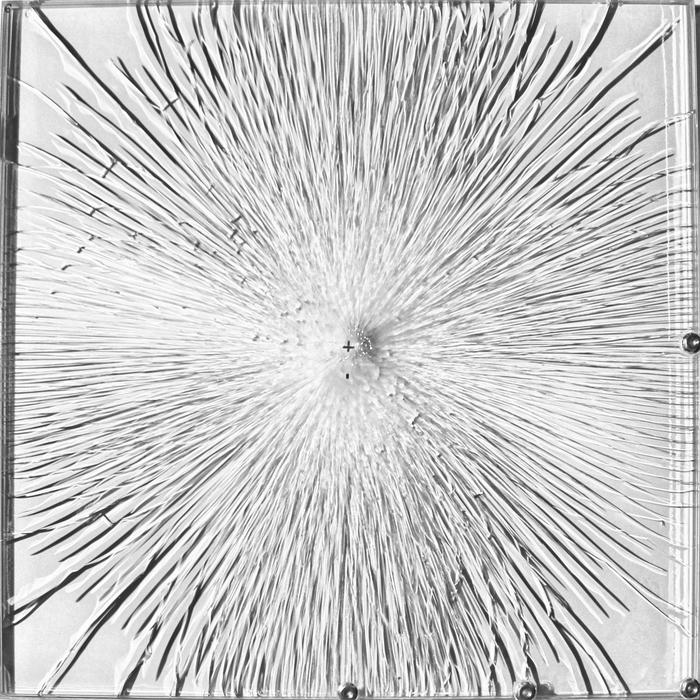}
         \\
         \\
         \includegraphics[width=0.475\textwidth]{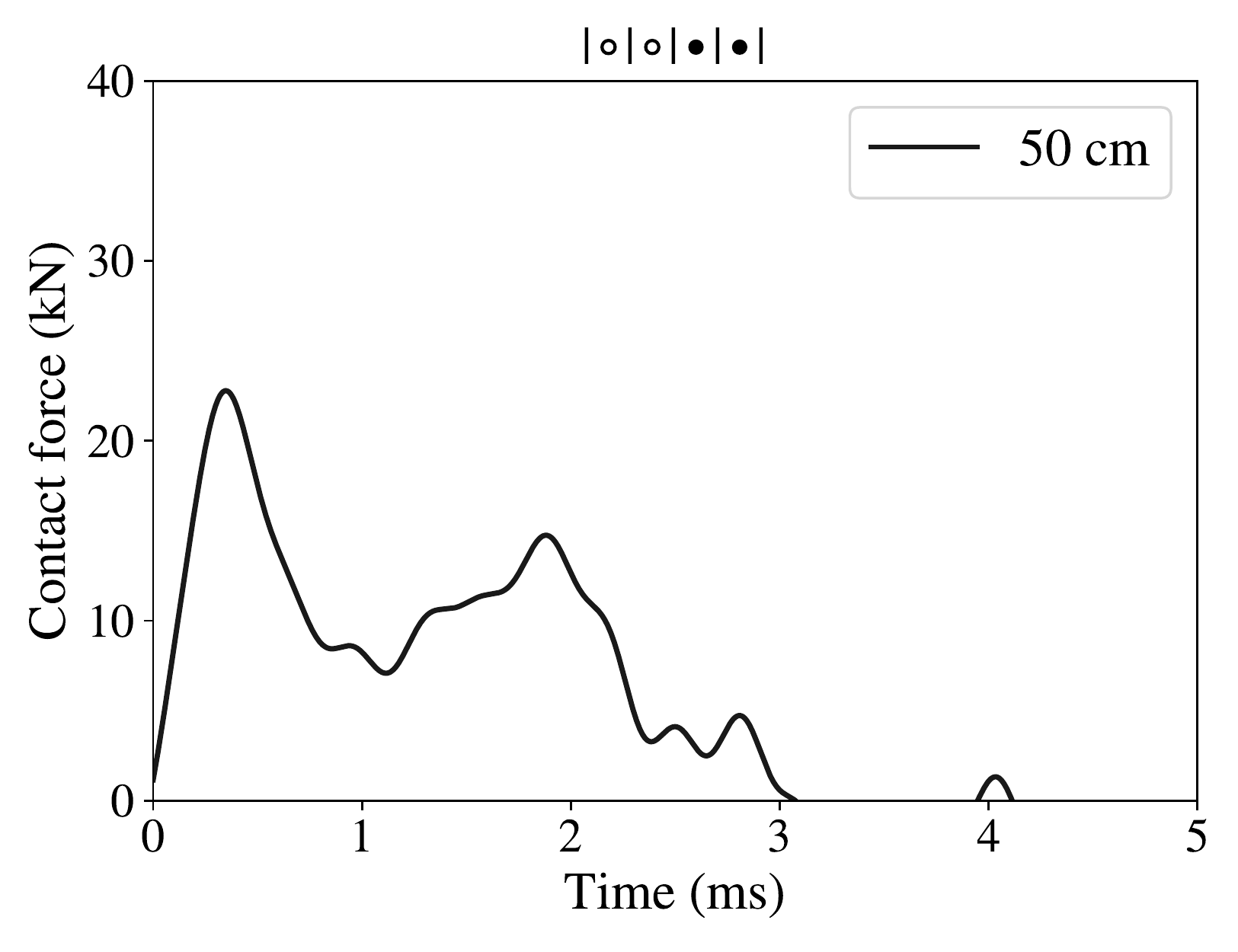}
         &
         \includegraphics[width=0.35\textwidth]{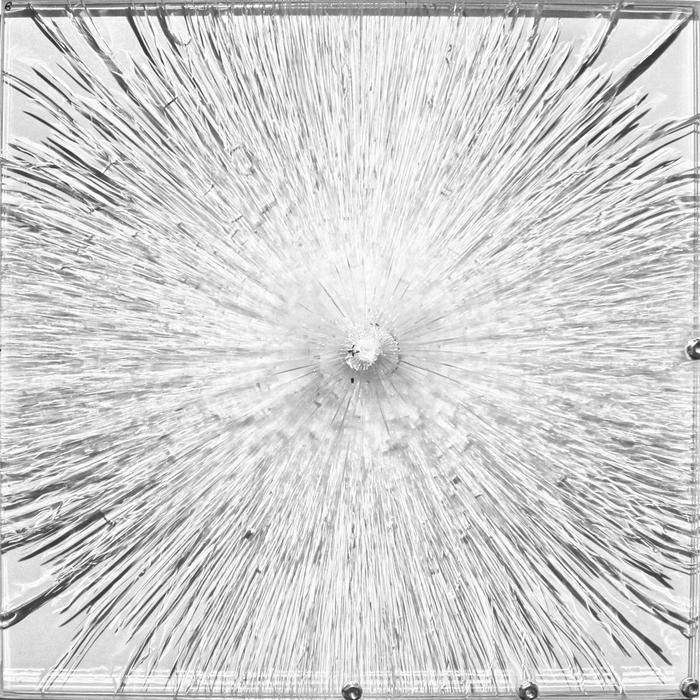}
    \end{tabular}
    \caption{Evolution of contact forces (2$\times$ filter CFC~1000) and corresponding fracture patterns for 7LG--2. Black lines denote the contact forces during the destructive tests, whereas the grey ones correspond to the previous sequence of impact events at lower impact heights (increased by 5~cm for each following impact event).}
    \label{fig:cf_7LG-2}
\end{figure}

\begin{figure}[hp]
    \centering
    \begin{tabular}[t]{cc}
         \includegraphics[width=0.475\textwidth]{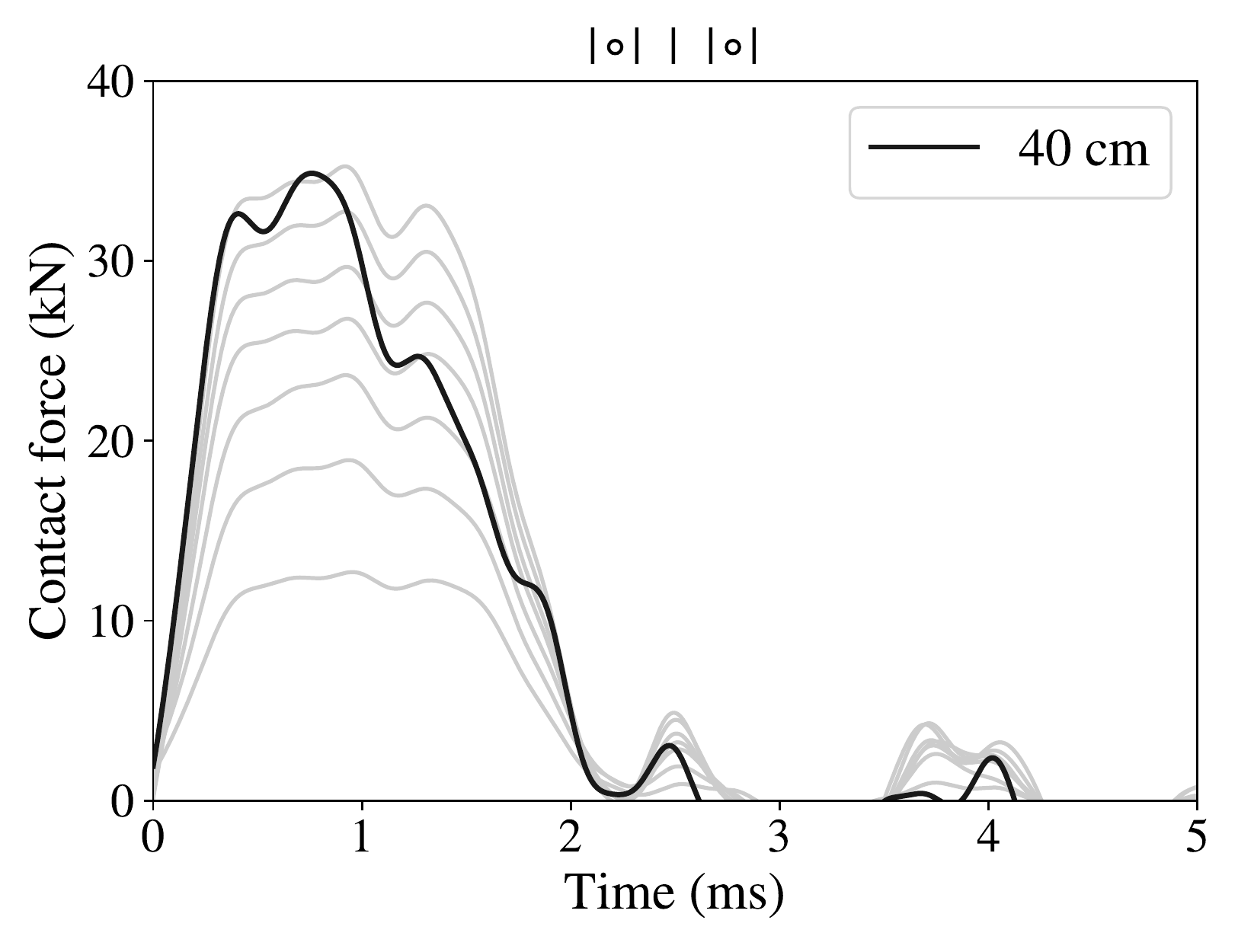}
         &  
         \includegraphics[width=0.35\textwidth]{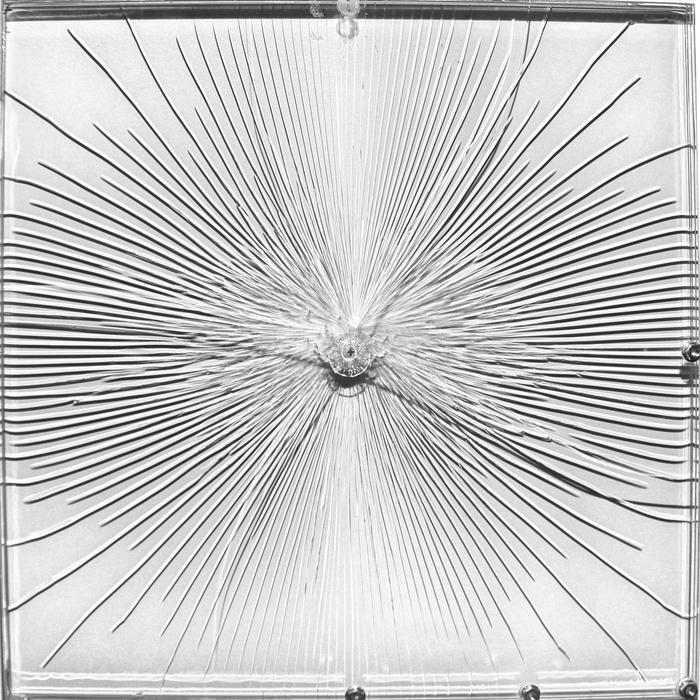}
         \\
         \\
         \includegraphics[width=0.475\textwidth]{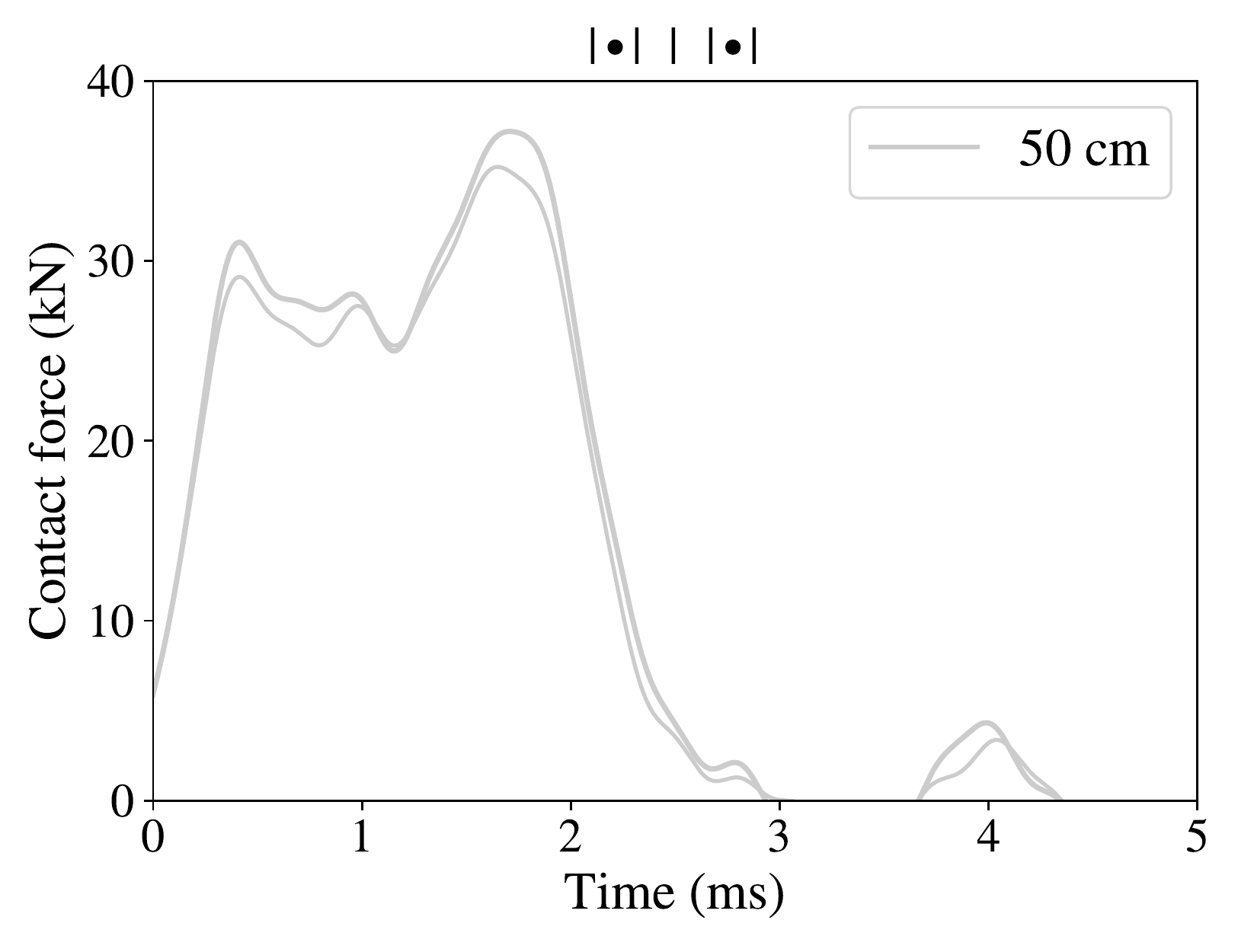}
&
         \includegraphics[width=0.35\textwidth]{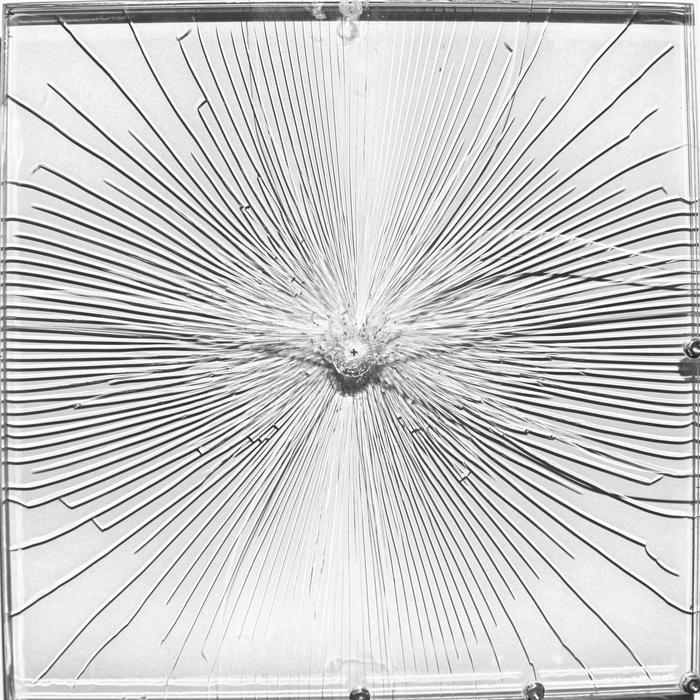}
    \end{tabular}
    \caption{Evolution of contact forces (2$\times$ filter CFC~1000) and corresponding fracture patterns for 7LG--3.  Black lines denote the contact forces during the destructive tests, whereas the grey ones correspond to the previous sequence of impact events at lower impact heights (increased by 5~cm for each following impact event).}
    \label{fig:cf_7LG-3}
\end{figure}

\begin{figure}[hp]
    \centering
\begin{tabular}{ccc}
    \includegraphics[width=0.3\textwidth]{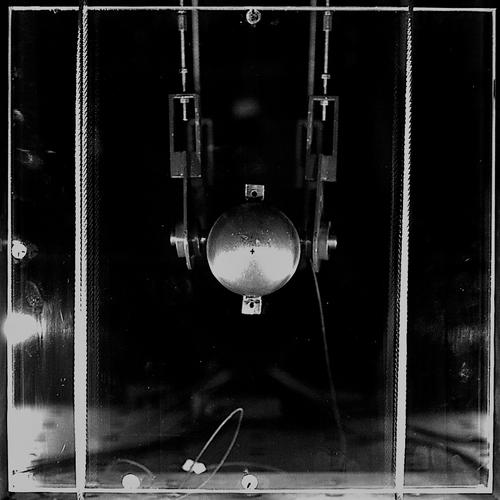}
     &  
    \includegraphics[width=0.3\textwidth]{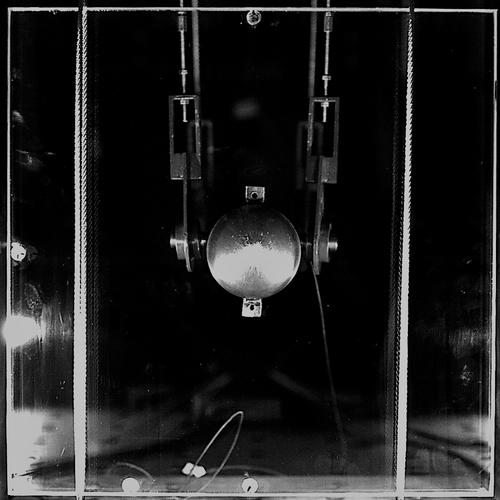}
    &
    \includegraphics[width=0.3\textwidth]{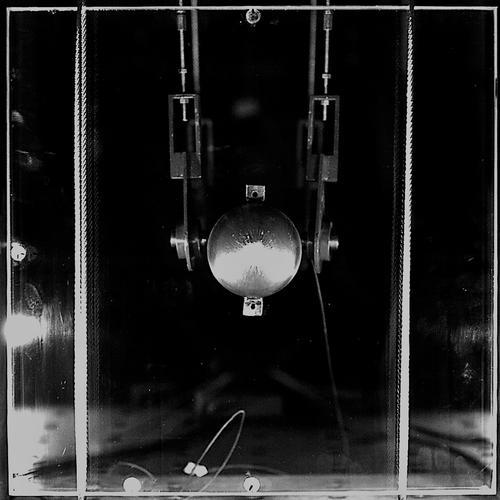}
     \\
\\
\includegraphics[width=0.3\textwidth]{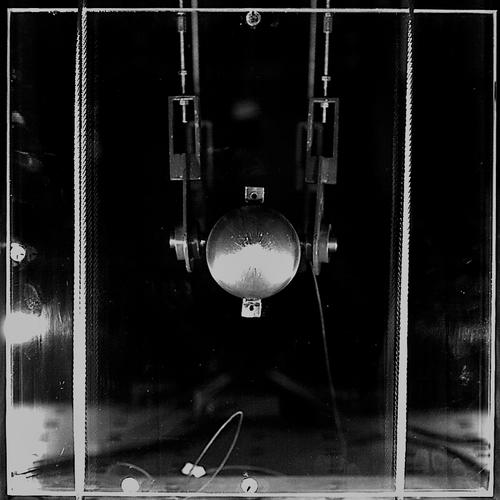}
    &
    \includegraphics[width=0.3\textwidth]{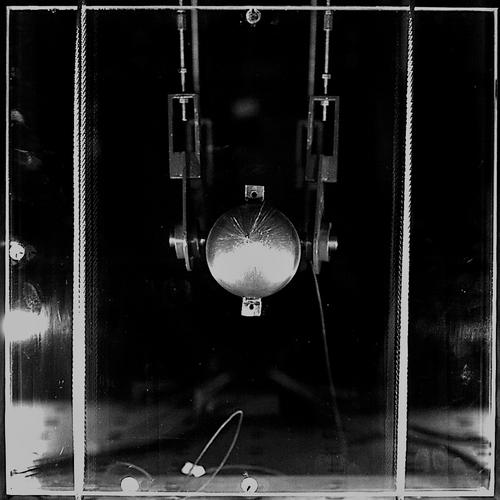}
    &
    \includegraphics[width=0.3\textwidth]{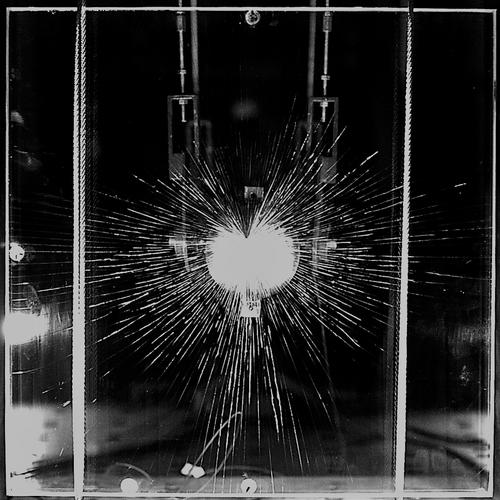}
    \\
     & 
    \footnotesize 40~cm 
\textbar$\ccirc$\textbar~\textbar~\textbar$\ccirc$\textbar 
    &
    \\
    \\
    \includegraphics[width=0.3\textwidth]{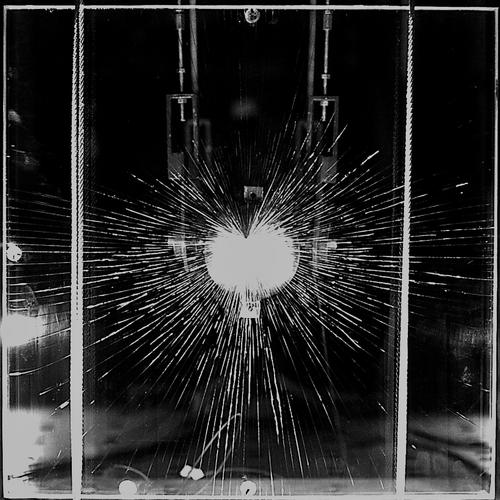}
    &
    \includegraphics[width=0.3\textwidth]{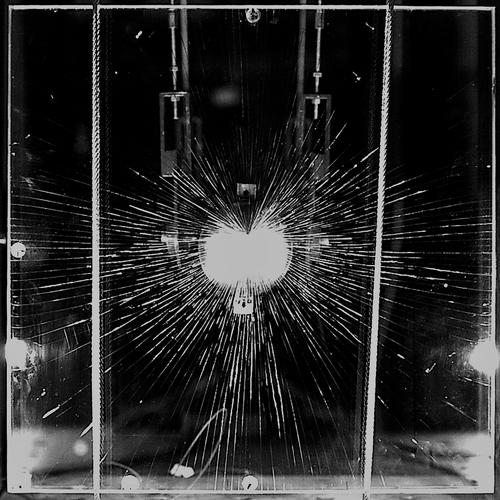}
    &
    \includegraphics[width=0.3\textwidth]{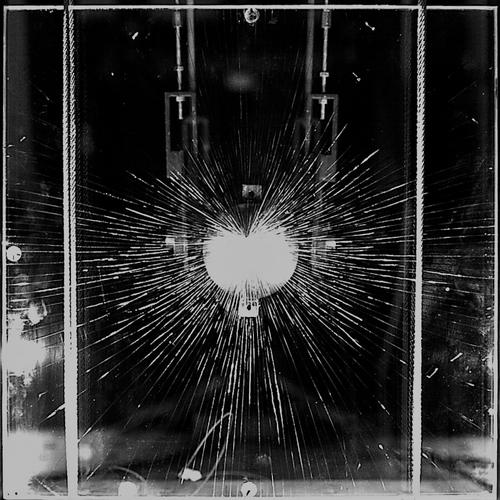}
    \\
    \footnotesize 40~cm 
\textbar$\sbullet[.72]$\textbar~\textbar~\textbar$\sbullet[.72]$\textbar
    &
    \footnotesize 45~cm 
\textbar$\sbullet[.72]$\textbar~\textbar~\textbar$\sbullet[.72]$\textbar
    &
    \footnotesize 50~cm 
\textbar$\sbullet[.72]$\textbar~\textbar~\textbar$\sbullet[.72]$\textbar
\end{tabular}
    \caption{Evolution of cracks in 7LG-3 sample (frame rate 5,000 frames per second, i.e., time step 0.2~ms between two adjacent photos in a row).}
    \label{fig:crack_ev_7LG3}
\end{figure}

\begin{figure}[hp]
    \centering
    \begin{tabular}[t]{cc}
         \includegraphics[width=0.475\textwidth]{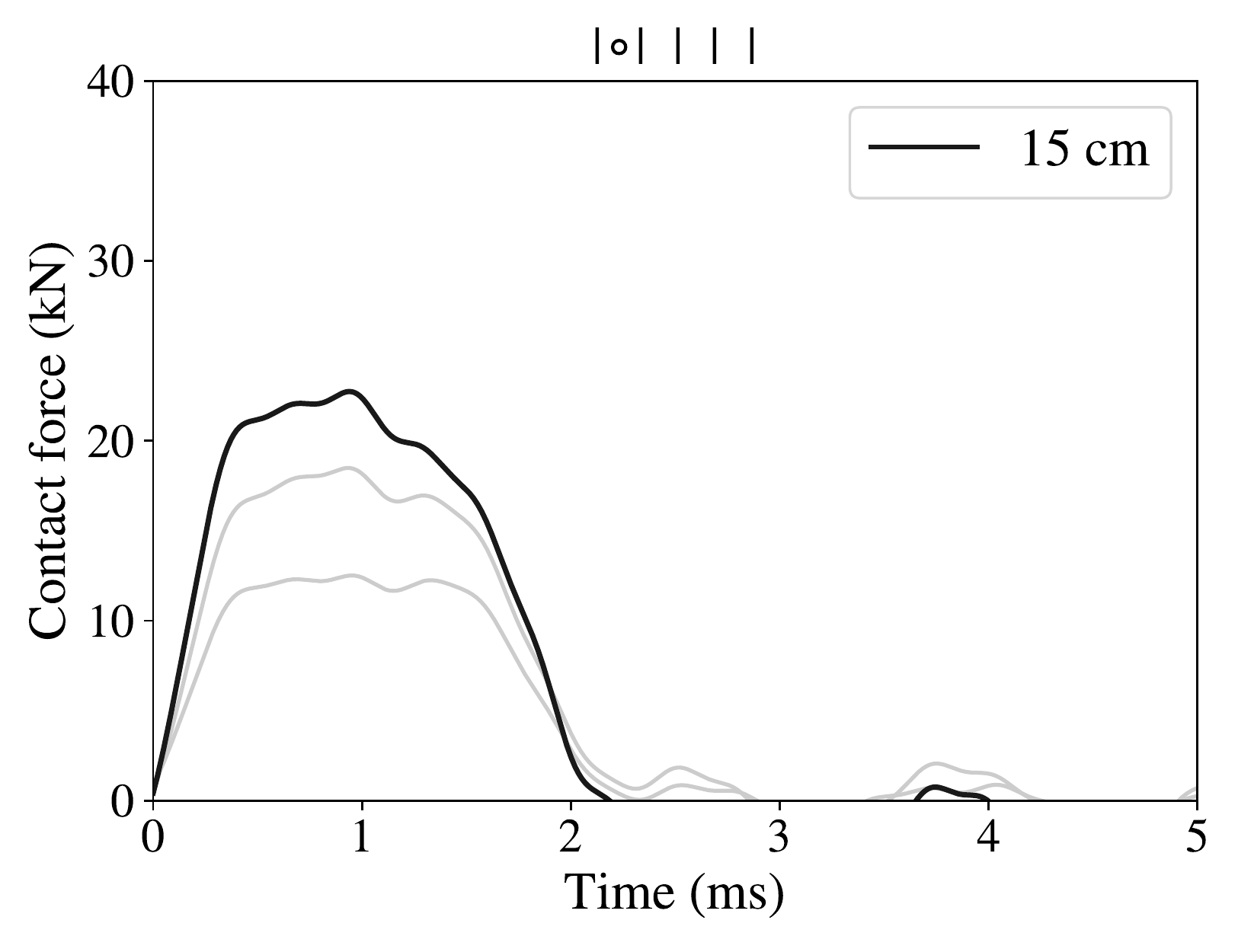}
         &  
         \includegraphics[width=0.35\textwidth]{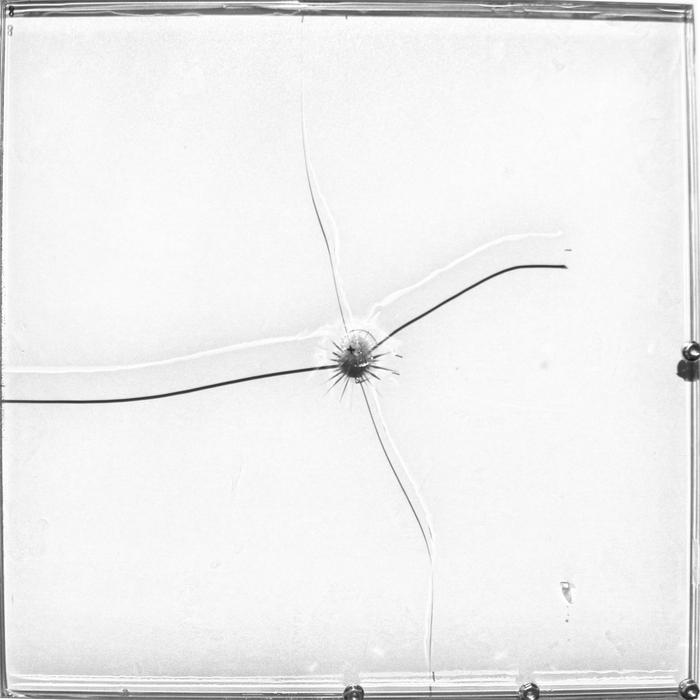}
         \\
         \\
         \includegraphics[width=0.475\textwidth]{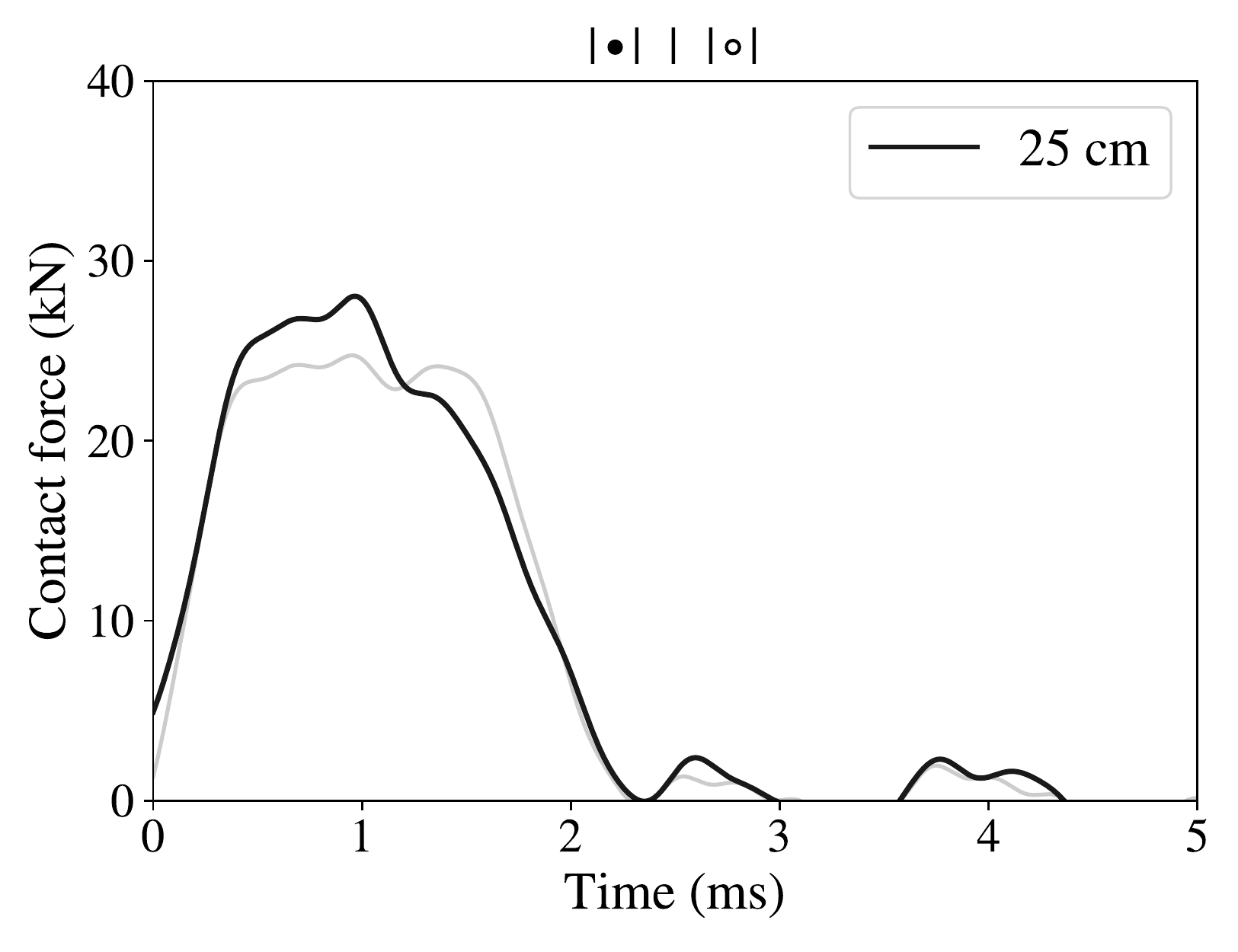}
         & 
         \includegraphics[width=0.35\textwidth]{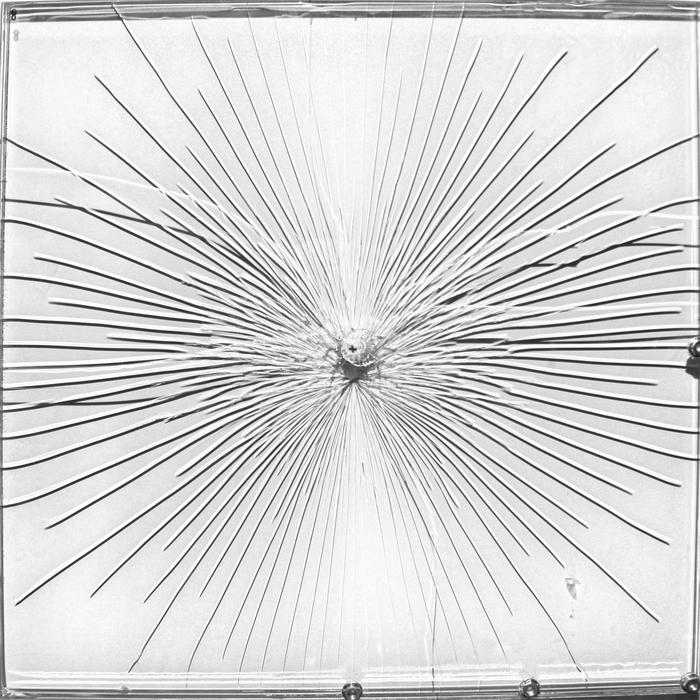}
         \\
         \\
         \includegraphics[width=0.475\textwidth]{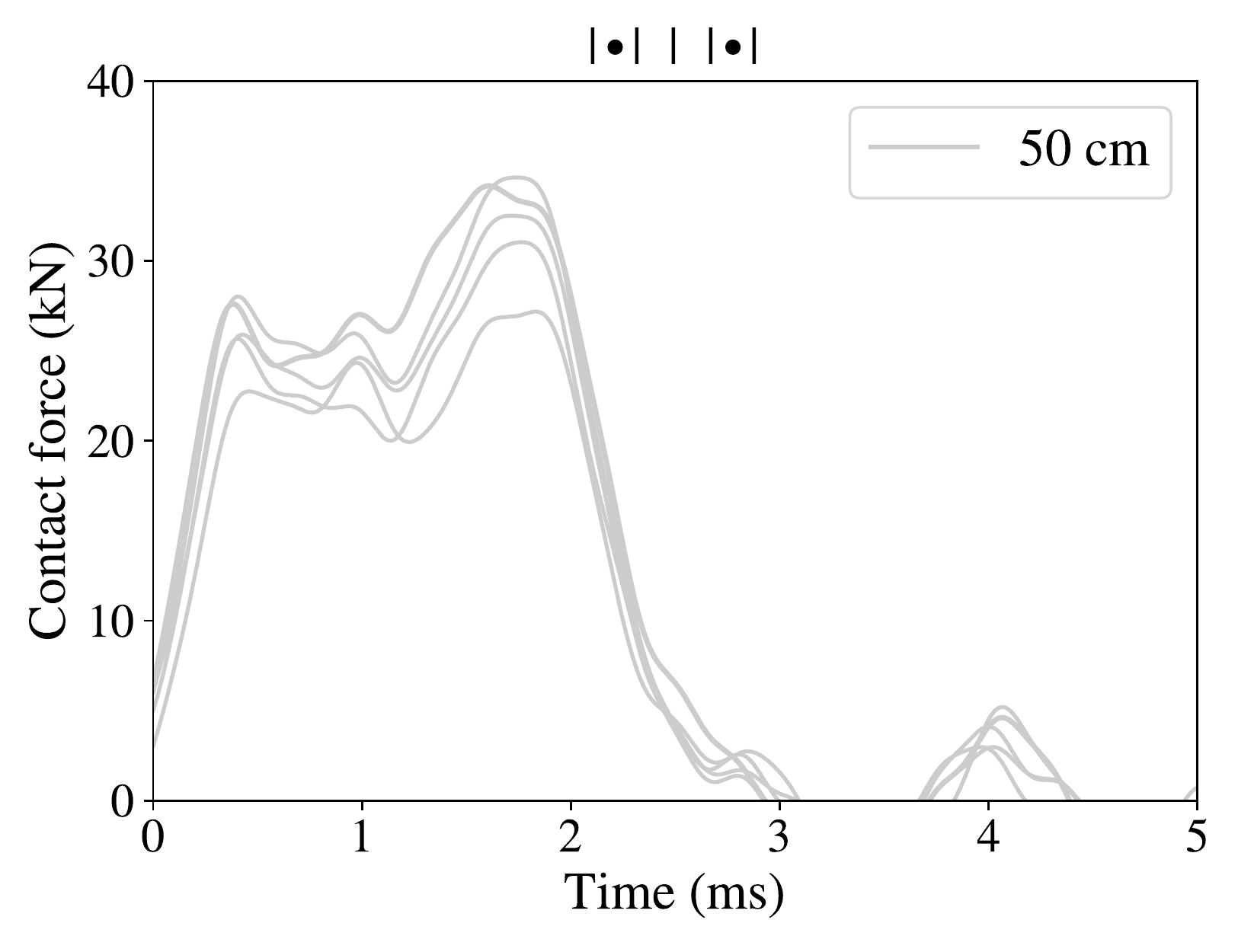}
         &
         \includegraphics[width=0.35\textwidth]{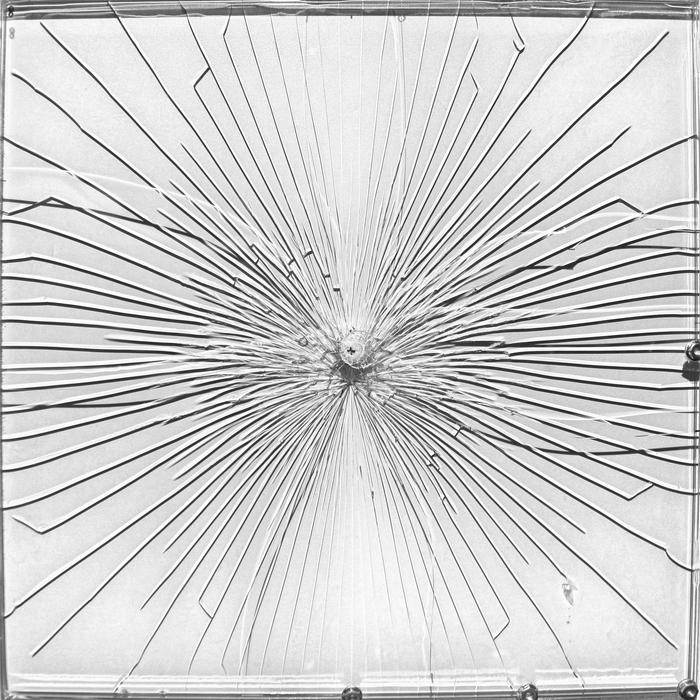}
    \end{tabular}
    \caption{Evolution of contact forces (2$\times$ filter CFC~1000) and corresponding fracture patterns for 7LG--4. Black lines denote the contact forces during the destructive tests, whereas the grey ones correspond to the previous sequence of impact events at lower impact heights (increased by 5~cm for each following impact event).}
    \label{fig:cf_7LG-4}
\end{figure}

The same strategy as in the previous section was 
employed 
to determine the contact forces during the destructive tests. We plot the results of this analysis in terms of contact forces for each sample independently (Figures~\ref{fig:cf_5LG-1}--\ref{fig:cf_7LG-4}) and complement them with distinctive fracture patterns. Recall that the evolution of contact forces exhibits the same trend for all non-destructive tests, their magnitudes only differ with an increasing impact height (\Fref{fig:cof_rep}). When a previously intact glass layer fractured, the contact force deviates from the formerly consistent path.
Then, a new shape of contact force can be observed, and its trend is again consistent for the increasing impact heights until the subsequent fracture of another glass layer(s). The contact duration was longer for samples with fractured layers as their modal response had changed.

\begin{figure}[hp]
    \centering
    \footnotesize
    \begin{tabular}{cc}
        5LG--2 & 5LG--4\\ 
         \includegraphics[width=0.45\textwidth]{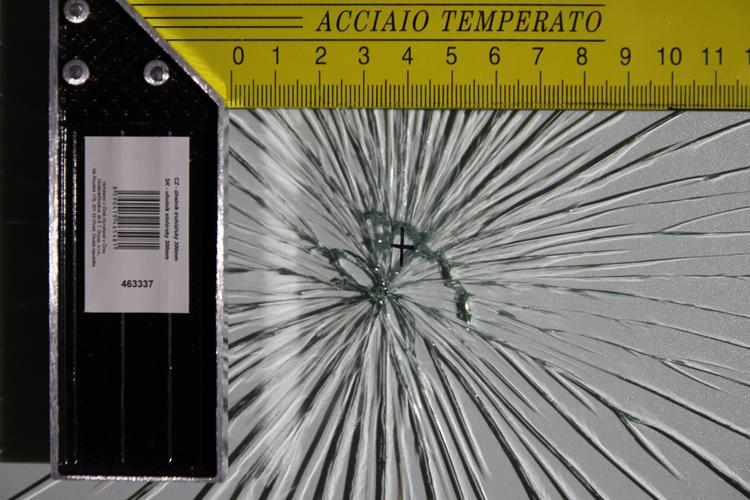}
         &  
         \includegraphics[width=0.45\textwidth]{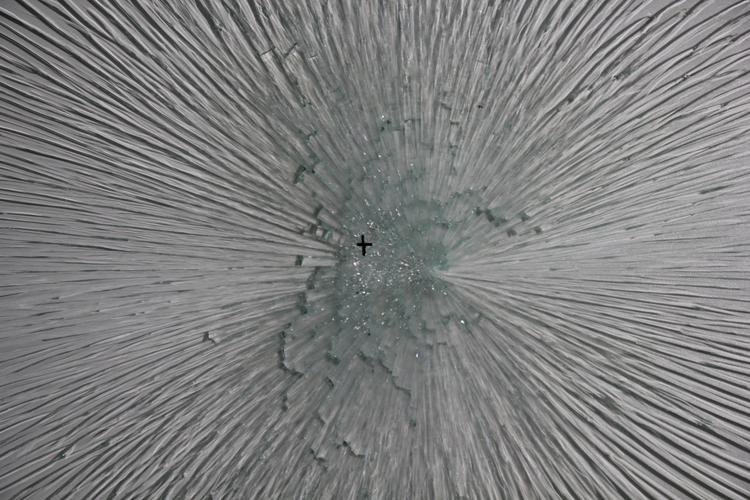}
         \\
         10~cm $\rightarrow$  \textbar~\textbar~\textbar$\sbullet[.72]$\textbar & 70~cm $\rightarrow$  \textbar~\textbar~\textbar$\sbullet[.72]$\textbar 
         \\
         \\
         \includegraphics[width=0.45\textwidth]{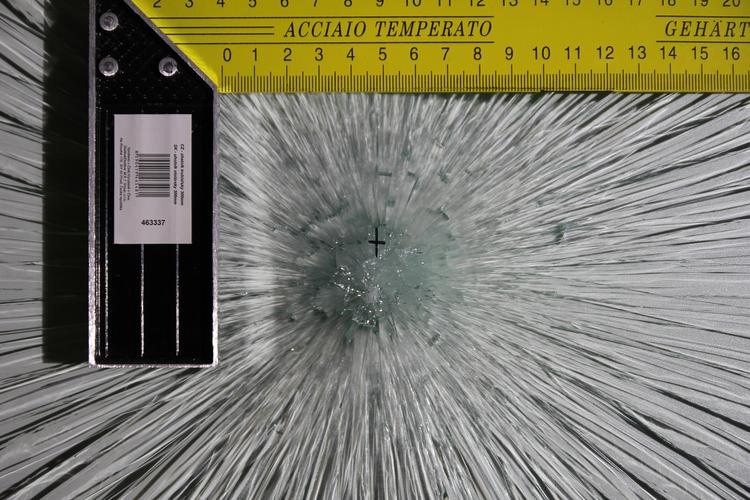}
         & 
         \includegraphics[width=0.45\textwidth]{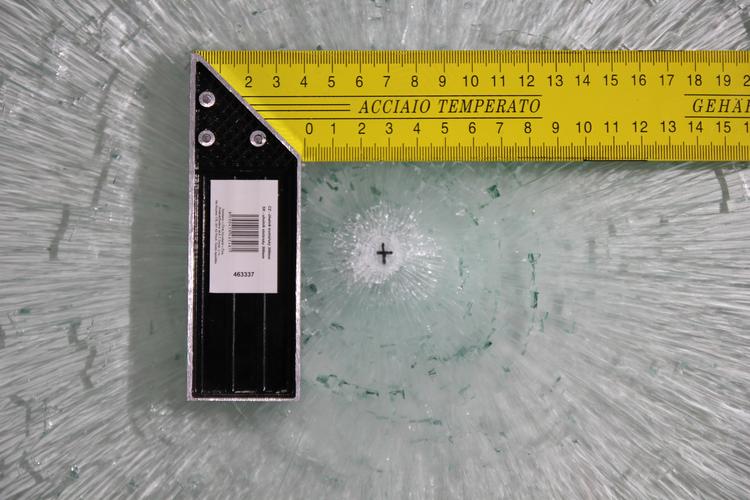}
         \\
          25~cm $\rightarrow$  \textbar~\textbar$\sbullet[.72]$\textbar$\sbullet[.72]$\textbar & 80~cm $\rightarrow$  \textbar$\sbullet[.72]$\textbar$\sbullet[.72]$\textbar$\sbullet[.72]$\textbar
         \\
         \\
         \includegraphics[width=0.45\textwidth]{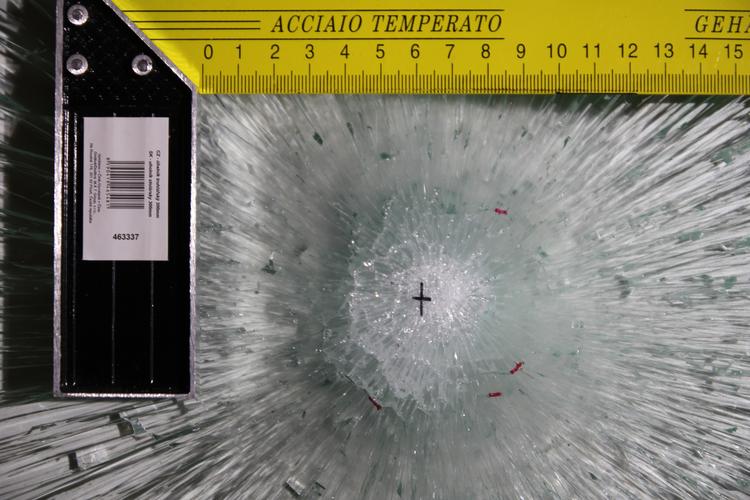}
         &
         \includegraphics[width=0.45\textwidth]{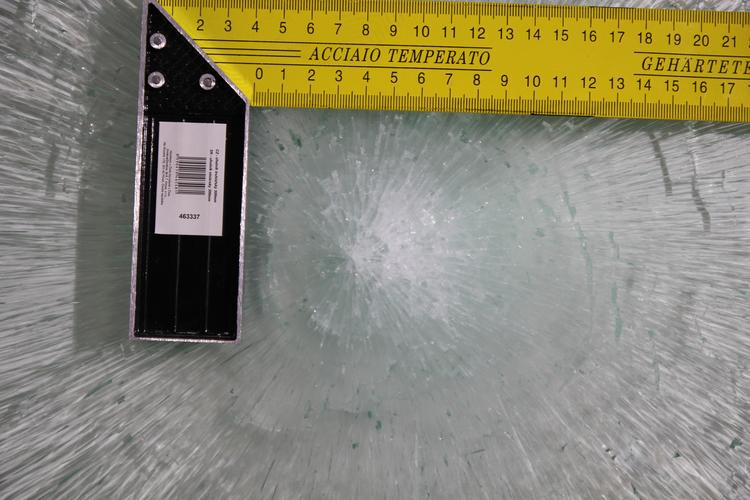}
         \\
         40~cm $\rightarrow$  \textbar$\sbullet[.72]$\textbar$\sbullet[.72]$\textbar$\sbullet[.72]$\textbar & 80~cm $\rightarrow$  \textbar$\sbullet[.72]$\textbar$\sbullet[.72]$\textbar$\sbullet[.72]$\textbar, back view \\
    \end{tabular}
    \caption{Details of crack patterns under the impact points for 5LG-samples seen from the impacted (no label) or non-impacted side (referred to as back view).}
    \label{fig:cracks_details_5LG}
\end{figure}

\begin{figure}[hp]
    \centering
    \footnotesize
    \begin{tabular}{cc}
        7LG--3 & 7LG--4\\ 
        
         \includegraphics[width=0.45\textwidth]{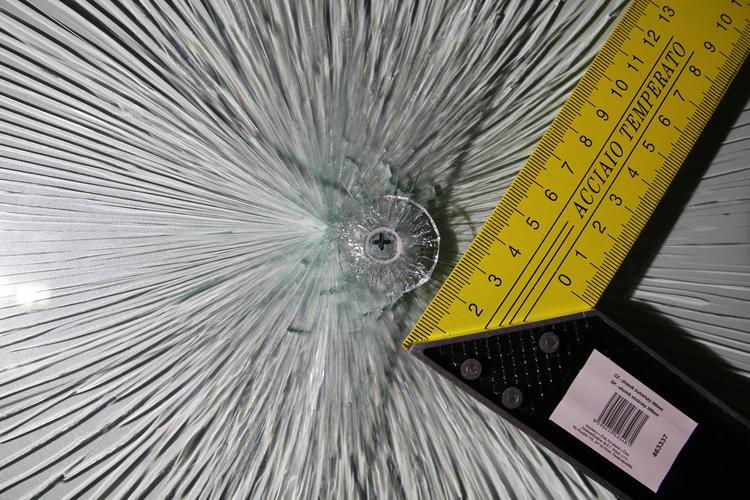}
         &  
         \includegraphics[width=0.45\textwidth]{figure23a}
         \\
         40~cm $\rightarrow$  \textbar$\sbullet[.72]$\textbar~\textbar~\textbar$\sbullet[.72]$\textbar & 
        20~cm $\rightarrow$  \textbar$\sbullet[.72]$\textbar~\textbar~\textbar~\textbar 
         \\
         \\
         \includegraphics[width=0.45\textwidth]{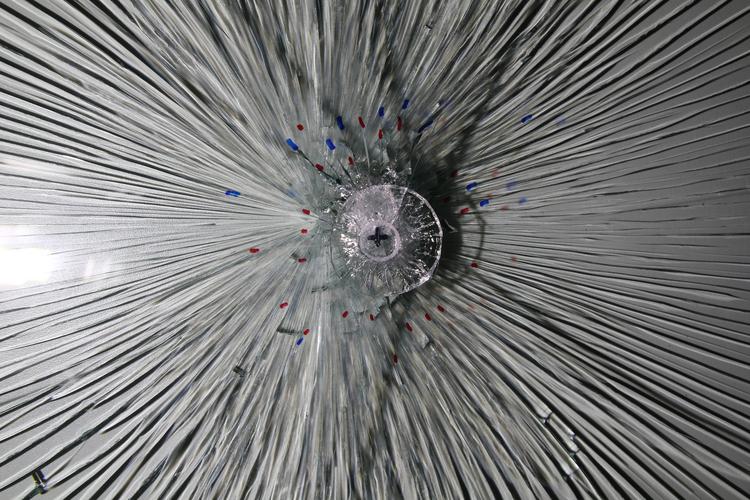}
         & 
         \includegraphics[width=0.45\textwidth]{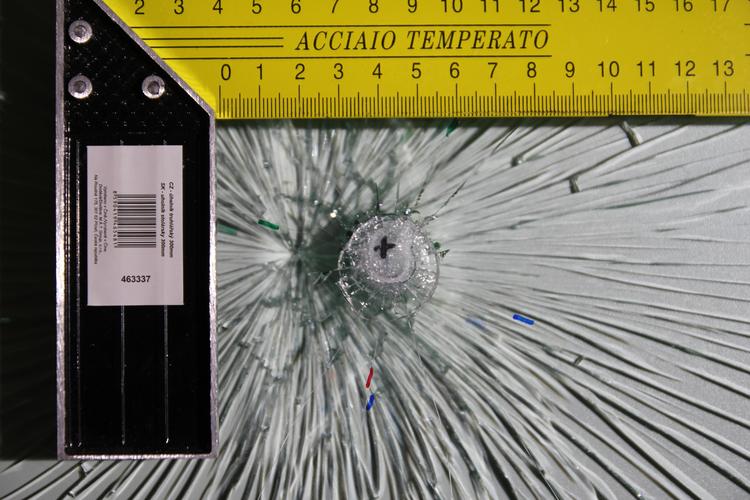}
         \\
         45~cm $\rightarrow$  \textbar$\sbullet[.72]$\textbar~\textbar~\textbar$\sbullet[.72]$\textbar & 
         30~cm $\rightarrow$  \textbar$\sbullet[.72]$\textbar~\textbar~\textbar$\sbullet[.72]$\textbar
         \\
         \\
         \includegraphics[width=0.45\textwidth]{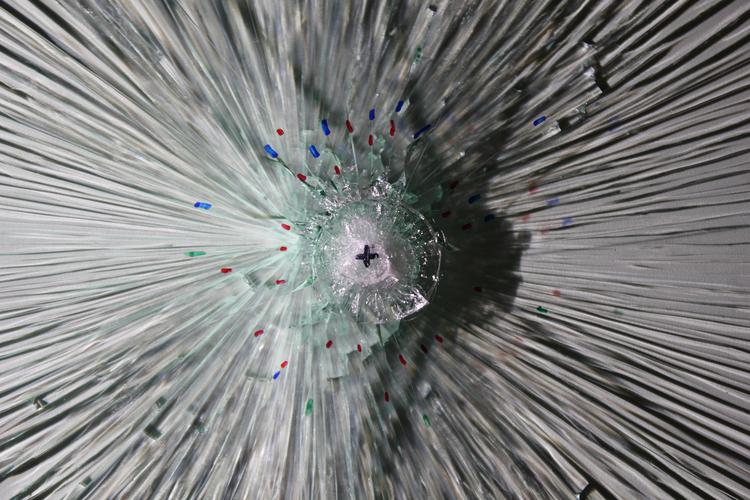}
         &
         \includegraphics[width=0.45\textwidth]{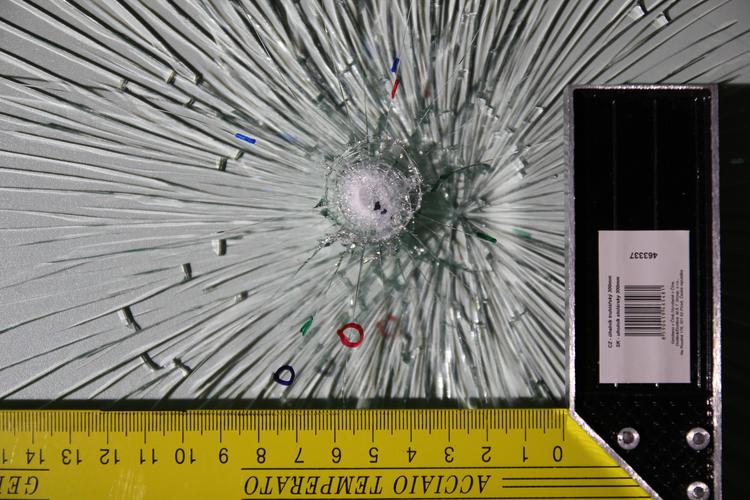}
         \\
         50~cm $\rightarrow$  \textbar$\sbullet[.72]$\textbar~\textbar~\textbar$\sbullet[.72]$\textbar & 50~cm $\rightarrow$  \textbar$\sbullet[.72]$\textbar~\textbar~\textbar$\sbullet[.72]$\textbar \\
    \end{tabular}
    \caption{Details of crack patterns under the impact points for 7LG-samples seen from the impacted side.}
    \label{fig:cracks_details_7LG}
\end{figure}

The typical fracture pattern consisted of several radial cracks of different lengths, in some samples seldomly
connected by a short parts of secondary cracks (e.g., Figures~\ref{fig:cf_5LG-1}, \ref{fig:cf_5LG-4}, \ref{fig:cf_7LG-3}, or \ref{fig:cf_7LG-4}). The area under the impact point was often covered by a finer net of cracks especially on the impacted glass layer (see ahead to Figures~\ref{fig:cracks_details_5LG} and \ref{fig:cracks_details_7LG}).
As expected, the density of fracture pattern increased with the impact energy.

\begin{figure}[hp]
    \centering
\begin{tabular}{ccc}
    \includegraphics[width=0.3\textwidth]{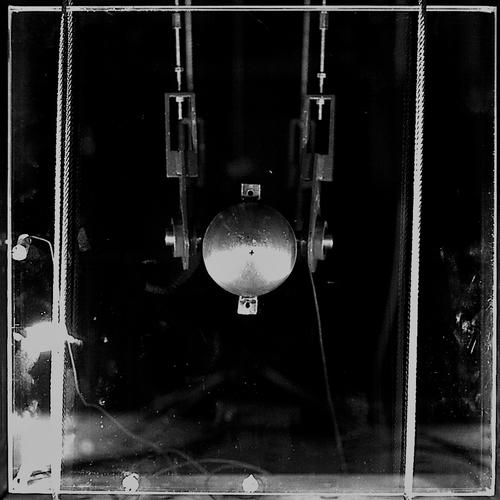}
     &  
    \includegraphics[width=0.3\textwidth]{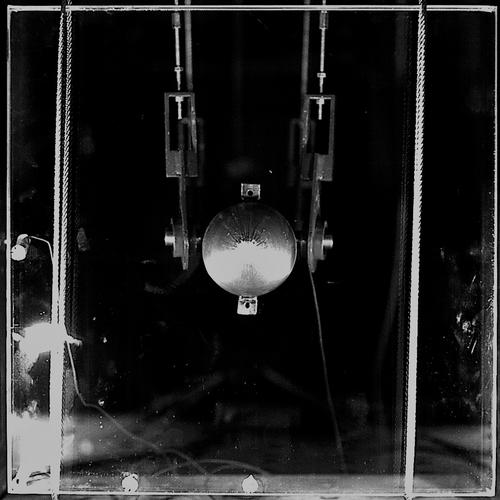}
    &
    \includegraphics[width=0.3\textwidth]{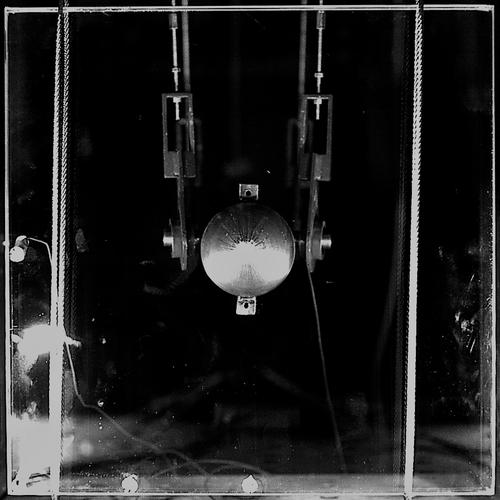}
     \\
     & 
    \footnotesize 15~cm 
\textbar$\ccirc$\textbar~\textbar~\textbar~\textbar 
    &
     \\
     \\
    \includegraphics[width=0.3\textwidth]{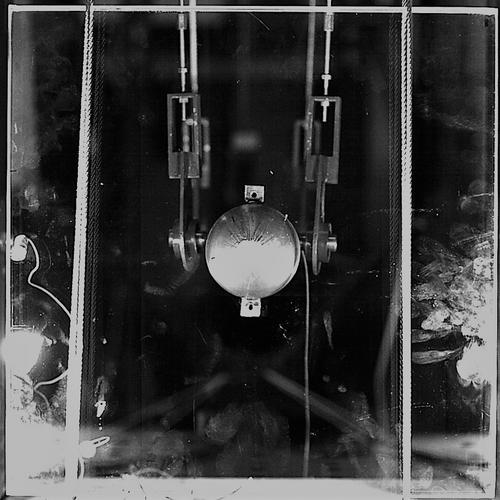}
    &
    \includegraphics[width=0.3\textwidth]{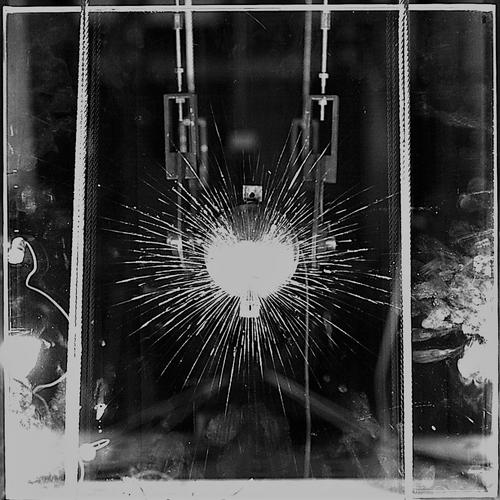}
    &
    \includegraphics[width=0.3\textwidth]{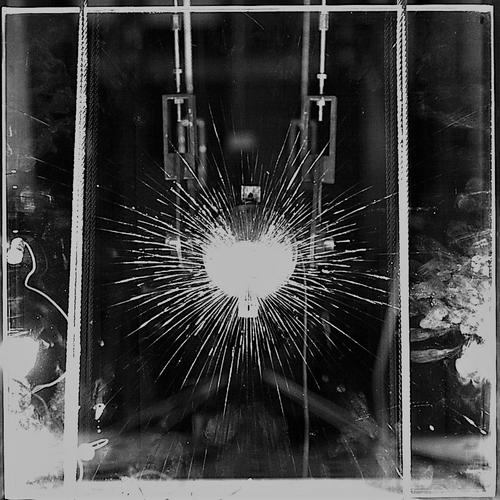}
    \\
     & 
    \footnotesize 25~cm 
\textbar$\sbullet[.72]$\textbar~\textbar~\textbar$\ccirc$\textbar 
    &
    \\
    \\
    \includegraphics[width=0.3\textwidth]{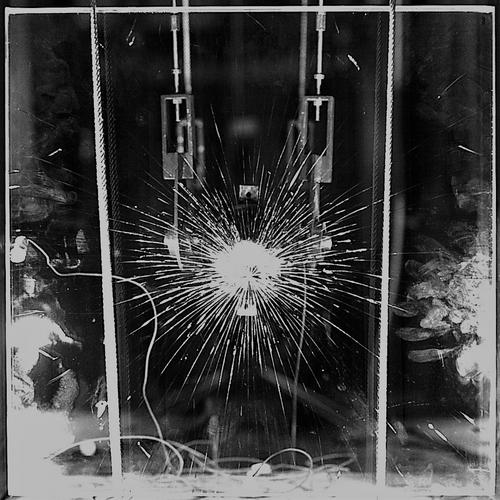}
    &
    \includegraphics[width=0.3\textwidth]{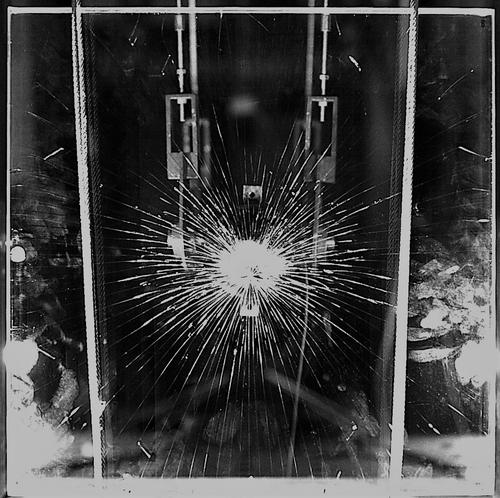}
    &
    \includegraphics[width=0.3\textwidth]{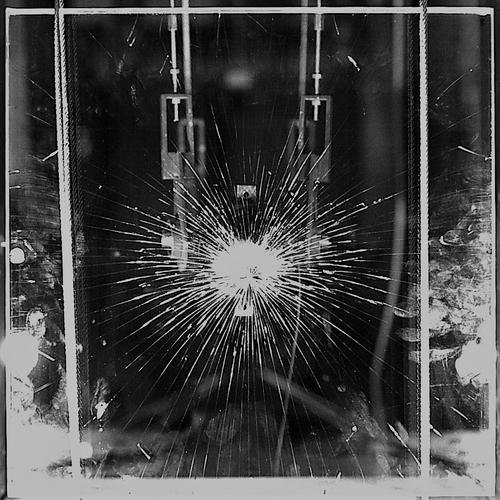}
    \\
    \footnotesize 30~cm 
\textbar$\sbullet[.72]$\textbar~\textbar~\textbar$\sbullet[.72]$\textbar
    &
    \footnotesize 40~cm 
\textbar$\sbullet[.72]$\textbar~\textbar~\textbar$\sbullet[.72]$\textbar
    &
    \footnotesize 50~cm 
\textbar$\sbullet[.72]$\textbar~\textbar~\textbar$\sbullet[.72]$\textbar
\end{tabular}
    \caption{Evolution of cracks in 7LG-4 sample (frame rate 5,000 frames per second, i.e., time step 0.2~ms between two adjacent photos in a row).}
    \label{fig:crack_ev_7LG4}
\end{figure}

For four  samples (two per the same geometry), the evolution of fracture pattern was captured by a high-speed camera (Figures~\ref{fig:crack_ev_5LG2}, \ref{fig:crack_ev_5LG3}, \ref{fig:crack_ev_7LG3}, and \ref{fig:crack_ev_7LG4}) with a 0.2~ms time intervals between two successive
frames.
The visibility of cracks depended on their opening during the vibration of a laminated glass plate. Therefore, we encourage the reader to compare the high-speed camera photos with the final fracture patterns captured on a white background after the impacts to detect all cracks (via their shadows) that developed during the contact. 

The details of starburst crack patterns under the impact point are summarised for four samples in~Figures~\ref{fig:cracks_details_5LG} and~\ref{fig:cracks_details_7LG}. We can observe that the centres of crack patterns in non-impacted glass layers do not coincide with the intended impact point (area) whereas the origins of the crack patterns in the impacted layers are significantly closer to this point (Figures~\ref{fig:cracks_details_5LG} and~\ref{fig:cracks_details_7LG}).  A sequence of images (\Fref{fig:crack_ev_7LG3}) also confirms that despite the impactor hit the centre of the plate with the tolerance less than 0.5~cm, the origin of the pattern in the back glass layer is located approximately 1.5~cm from the centre (\Fref{fig:cracks_details_7LG}). Therefore, the reason is not the eccentric impact but more likely some invisible microflaw on the surface under tension. The changes in crack patterns under the impactor were minor as highlighted for some cracks and consecutive impact tests by different colours or marks in \Fref{fig:cracks_details_7LG}.

\newpage

\section{Comparison of results and discussion}

\subsection{Comparison of crack patterns}

\begin{figure}[hp]
    \centering
    \begin{tabular}{p{0mm}ccc}
{\rotatebox[origin=c]{90}{\footnotesize  5LG--1}}
        &
        \raisebox{-0.5\height}{\includegraphics[width=0.28\textwidth]{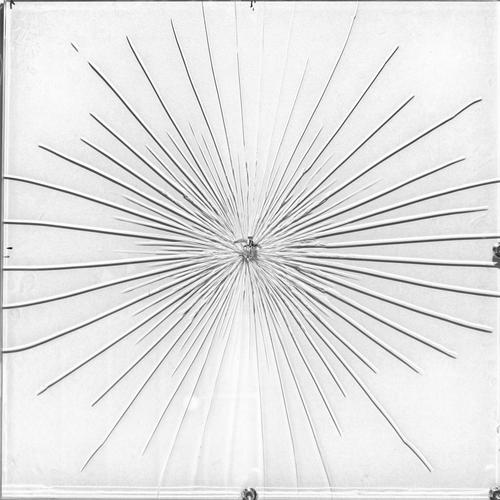}} & 
        \raisebox{-0.5\height}{\includegraphics[width=0.28\textwidth]{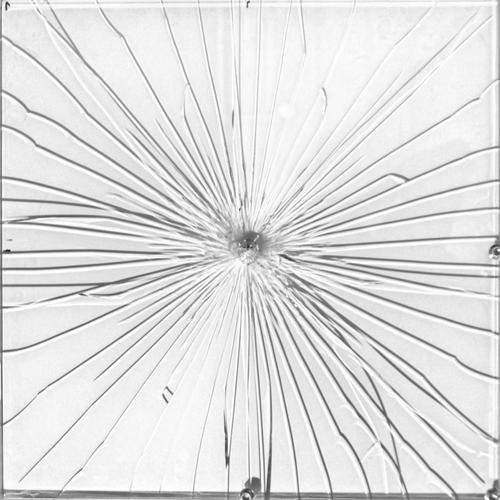}}  &
        \raisebox{-0.5\height}{\includegraphics[width=0.28\textwidth]{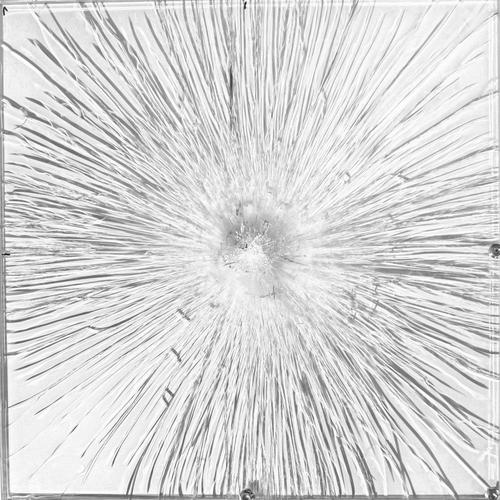}}\\
        &
        \footnotesize 15~cm $\rightarrow$ \textbar~\textbar~\textbar$\sbullet[.72]$\textbar &
        \footnotesize 25~cm $\rightarrow$ \textbar$\sbullet[.72]$\textbar~\textbar$\sbullet[.72]$\textbar &
        \footnotesize 45~cm $\rightarrow$ \textbar$\sbullet[.72]$\textbar$\sbullet[.72]$\textbar$\sbullet[.72]$\textbar 
        \\
{\rotatebox[origin=c]{90}{\footnotesize  5LG--2}} &
        \raisebox{-0.5\height}{\includegraphics[width=0.28\textwidth]{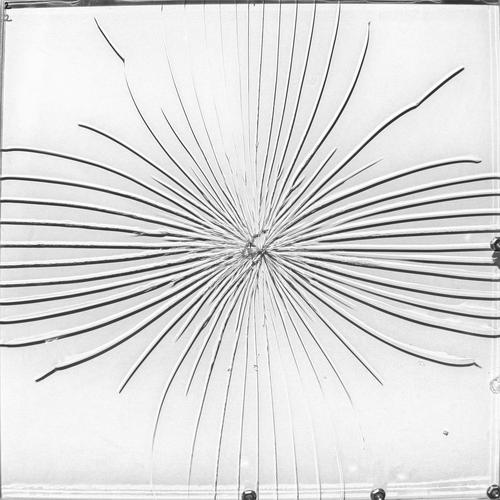}} & 
        \raisebox{-0.5\height}{\includegraphics[width=0.28\textwidth]{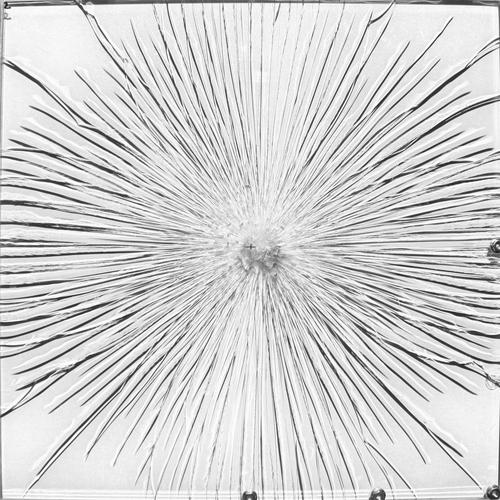}} &
        \raisebox{-0.5\height}{\includegraphics[width=0.28\textwidth]{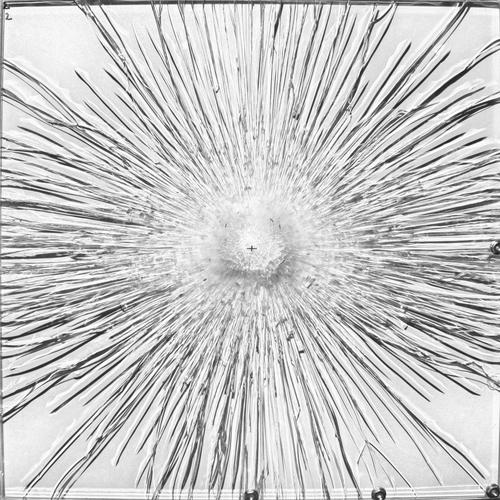}}
        \\
        &
        \footnotesize 10~cm $\rightarrow$ \textbar~\textbar~\textbar$\sbullet[.72]$\textbar & 
        \footnotesize 25~cm $\rightarrow$ \textbar~\textbar$\sbullet[.72]$\textbar$\sbullet[.72]$\textbar &
        \footnotesize 40~cm $\rightarrow$ \textbar$\sbullet[.72]$\textbar$\sbullet[.72]$\textbar$\sbullet[.72]$\textbar
        \\
{\rotatebox[origin=c]{90}{\footnotesize  5LG--3}} &
        \raisebox{-0.5\height}{\includegraphics[width=0.28\textwidth]{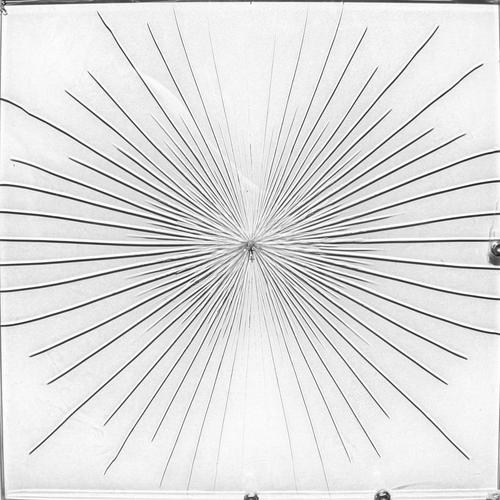}} &
        \raisebox{-0.5\height}{\includegraphics[width=0.28\textwidth]{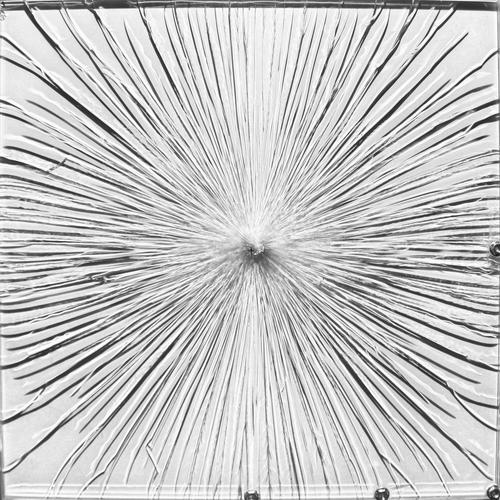}} &
        \raisebox{-0.5\height}{\includegraphics[width=0.28\textwidth]{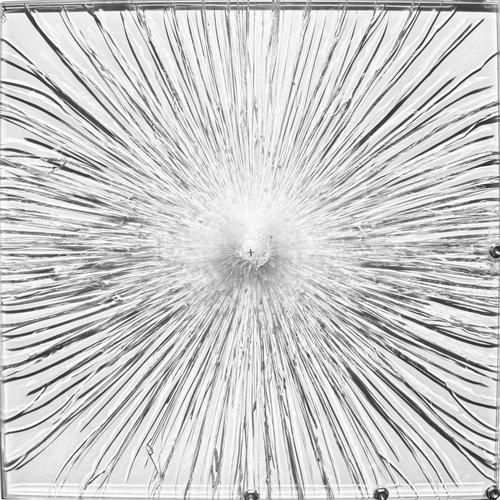}}
        \\
        & 
        \footnotesize 20~cm $\rightarrow$ \textbar~\textbar~\textbar$\sbullet[.72]$\textbar & 
        \footnotesize 25~cm $\rightarrow$ \textbar~\textbar$\sbullet[.72]$\textbar$\sbullet[.72]$\textbar&
        \footnotesize 30~cm $\rightarrow$ \textbar$\sbullet[.72]$\textbar$\sbullet[.72]$\textbar$\sbullet[.72]$\textbar 
        \\
{\rotatebox[origin=c]{90}{\footnotesize  5LG--4}} &
        \raisebox{-0.5\height}{\includegraphics[width=0.28\textwidth]{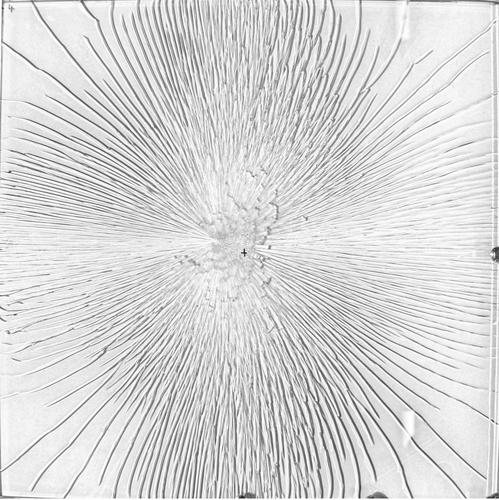}} &
        \raisebox{-0.5\height}{\includegraphics[width=0.28\textwidth]{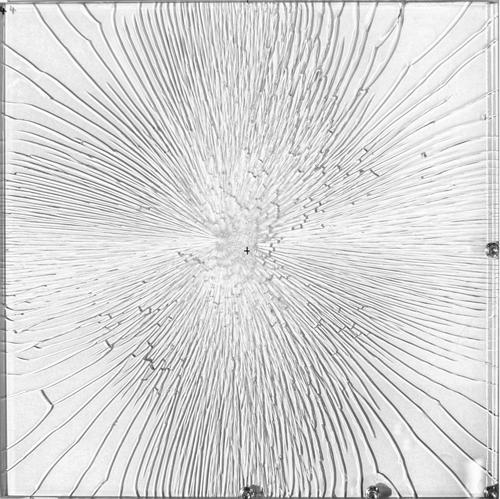}} &
        \raisebox{-0.5\height}{\includegraphics[width=0.28\textwidth]{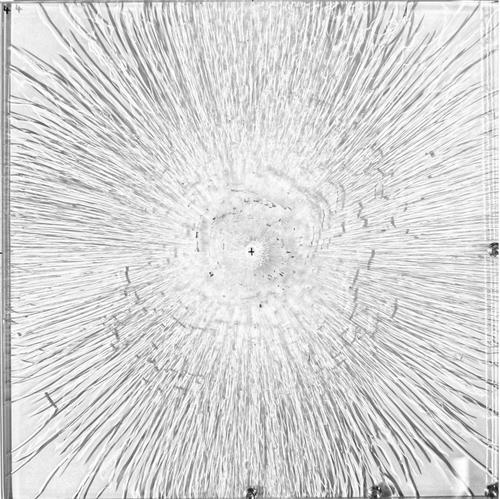}} 
        \\
        &
        \footnotesize 70~cm $\rightarrow$ \textbar~\textbar~\textbar$\sbullet[.72]$\textbar & 
        \footnotesize 75~cm $\rightarrow$ \textbar~\textbar~\textbar$\sbullet[.72]$\textbar &
         \footnotesize 80~cm $\rightarrow$ \textbar$\sbullet[.72]$\textbar$\sbullet[.72]$\textbar$\sbullet[.72]$\textbar
    \end{tabular}
    \caption{Fracture patterns for 5LG-samples corresponding to different impact heights complemented with the scheme of the cross-section with fractured layers marked by $\sbullet[.72]$. Photos are taken from the impacted side.}
    \label{fig:photos_fracture_5LG}
\end{figure}

\begin{figure}[hp]
    \centering
    \begin{tabular}{p{0mm}ccc}
        {\rotatebox[origin=c]{90}{\footnotesize  7LG--1}}
        &
        \raisebox{-0.5\height}{ \includegraphics[width=0.28\textwidth]{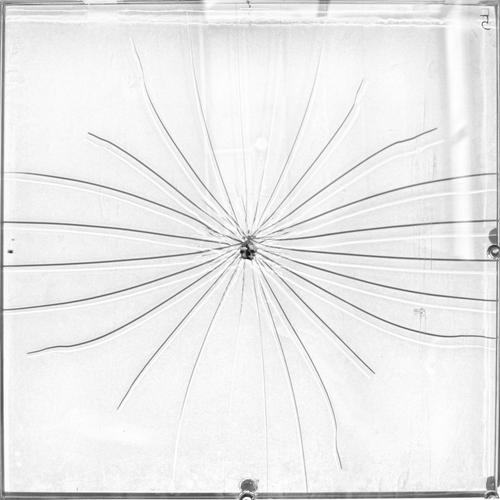}} & 
        \raisebox{-0.5\height}{\includegraphics[width=0.28\textwidth]{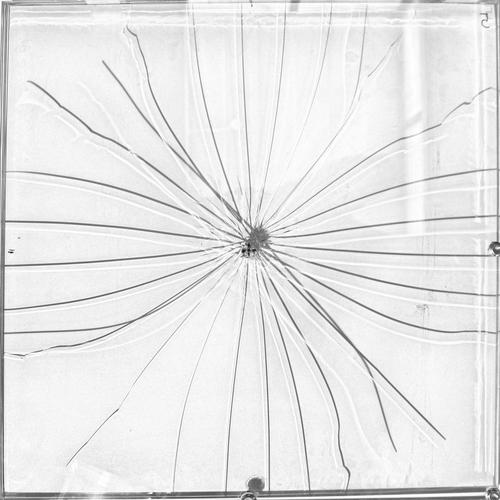} }  &
        \raisebox{-0.5\height}{\includegraphics[width=0.28\textwidth]{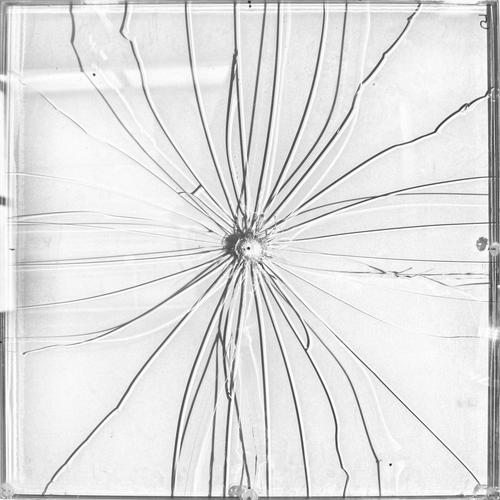}}\\
        &
        \footnotesize 5~cm $\rightarrow$ \textbar~\textbar~\textbar~\textbar$\sbullet[.72]$\textbar & 
        \footnotesize 10~cm $\rightarrow$ \textbar$\sbullet[.72]$\textbar~\textbar~\textbar$\sbullet[.72]$\textbar 
         & \footnotesize 50~cm $\rightarrow$ \textbar$\sbullet[.72]$\textbar~\textbar~\textbar$\sbullet[.72]$\textbar
        \\
{\rotatebox[origin=c]{90}{\footnotesize  7LG--2}} &
        \raisebox{-0.5\height}{\includegraphics[width=0.28\textwidth]{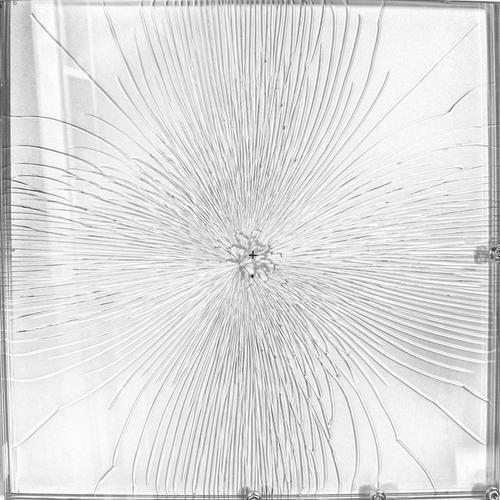}} & 
        \raisebox{-0.5\height}{\includegraphics[width=0.28\textwidth]{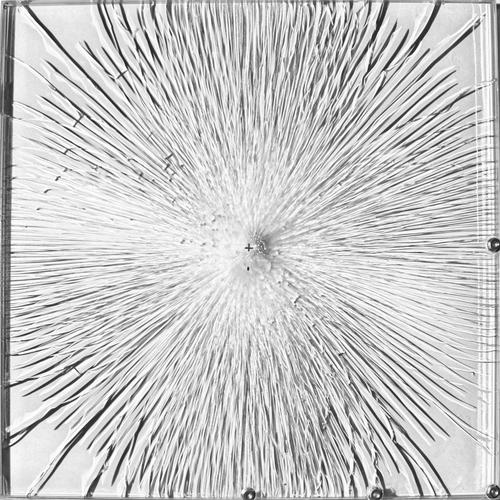}} &
        \raisebox{-0.5\height}{\includegraphics[width=0.28\textwidth]{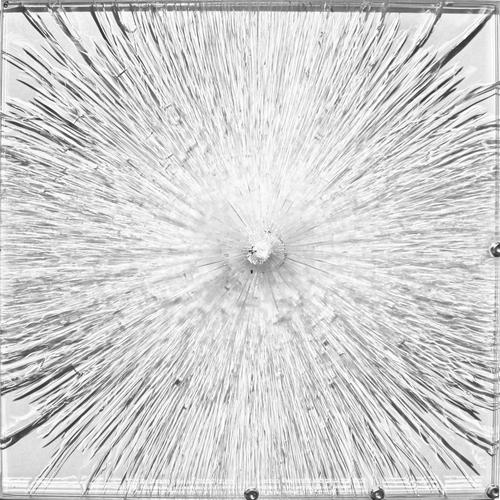}}
        \\
        &
        \footnotesize 40~cm $\rightarrow$ \textbar~\textbar~\textbar~\textbar$\sbullet[.72]$\textbar & 
        \footnotesize 45~cm $\rightarrow$ \textbar~\textbar~\textbar$\sbullet[.72]$\textbar$\sbullet[.72]$\textbar &
        \footnotesize 50~cm $\rightarrow$ \textbar$\sbullet[.72]$\textbar$\sbullet[.72]$\textbar$\sbullet[.72]$\textbar$\sbullet[.72]$\textbar
        \\
{\rotatebox[origin=c]{90}{\footnotesize  7LG--3}} &
        \raisebox{-0.5\height}{\includegraphics[width=0.28\textwidth]{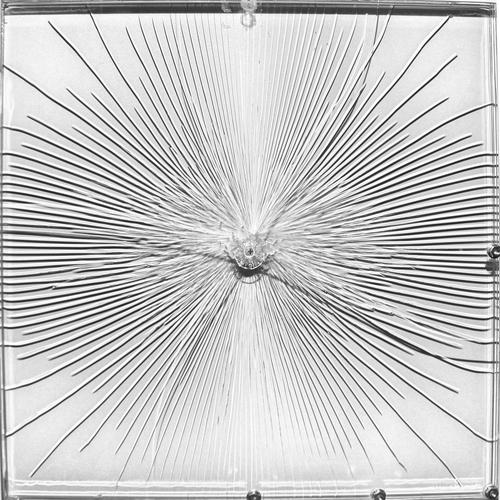}} &
        \raisebox{-0.5\height}{\includegraphics[width=0.28\textwidth]{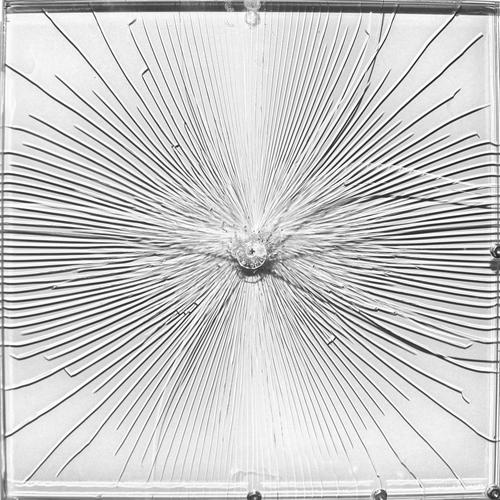}} &
        \raisebox{-0.5\height}{\includegraphics[width=0.28\textwidth]{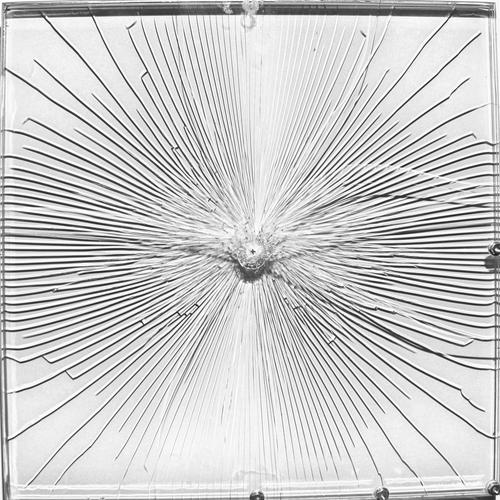}}
        \\
        & 
        \footnotesize 40~cm $\rightarrow$ \textbar$\sbullet[.72]$\textbar~\textbar~\textbar$\sbullet[.72]$\textbar & 
        \footnotesize 45~cm $\rightarrow$ \textbar$\sbullet[.72]$\textbar~\textbar~\textbar$\sbullet[.72]$\textbar &
        \footnotesize 50~cm $\rightarrow$ \textbar$\sbullet[.72]$\textbar~\textbar~\textbar$\sbullet[.72]$\textbar
        \\
{\rotatebox[origin=c]{90}{\footnotesize  7LG--4}} &
        \raisebox{-0.5\height}{\includegraphics[width=0.28\textwidth]{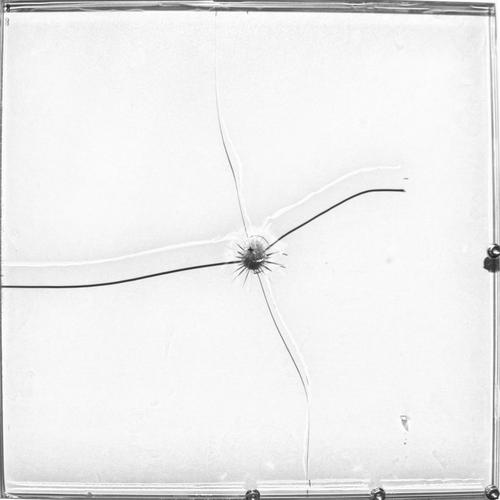}} &
        \raisebox{-0.5\height}{\includegraphics[width=0.28\textwidth]{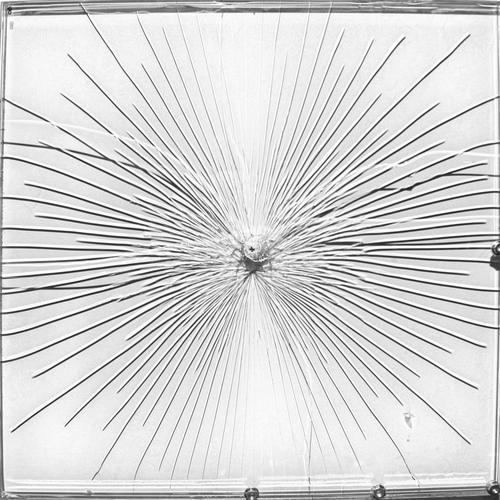}} &
        \raisebox{-0.5\height}{\includegraphics[width=0.28\textwidth]{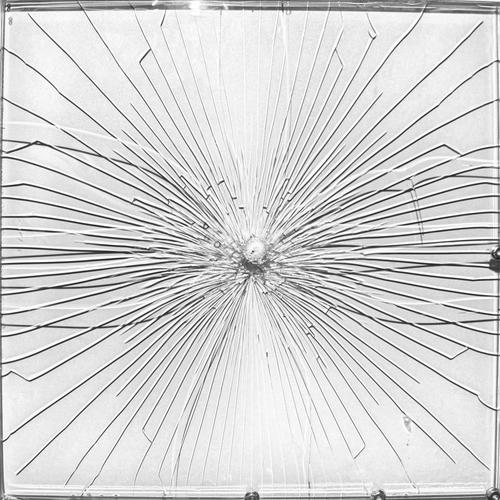}} 
        \\
        &
        \footnotesize 15~cm $\rightarrow$ \textbar$\sbullet[.72]$\textbar~\textbar~\textbar~\textbar
        & 
        \footnotesize 25~cm $\rightarrow$ \textbar$\sbullet[.72]$\textbar~\textbar~\textbar$\sbullet[.72]$\textbar
        & \footnotesize 50~cm $\rightarrow$ \textbar$\sbullet[.72]$\textbar~\textbar~\textbar$\sbullet[.72]$\textbar
    \end{tabular}
    \caption{Fracture patterns for 7LG-samples corresponding to different impact heights complemented with the scheme of the cross-section with fractured layers marked by $\sbullet[.72]$. Photos are taken from the impacted side.}
    \label{fig:photos_fracture_7LG}
\end{figure}

For both types of laminated glass, Figures~\ref{fig:photos_fracture_5LG} and~\ref{fig:photos_fracture_7LG} summarise all fracture patterns corresponding to the subsequent fracture of glass layers. The back layer fractured first for all 5LG-samples and the density of the radial cracks was comparable for three tested samples that fractured under the impact heights of 10, 15, and 20 cm. A significantly finer crack pattern on 5LG--4 corresponded to a more powerful impact. For 7LG-samples, the back glass layer fractured first in two cases and the front one in other two cases. Corresponding fracture patterns differ (first column in~\Fref{fig:photos_fracture_7LG}) which indicates that the way of cracking was different in both outer glass layers. 

The high-speed camera investigation also revealed that cracks spread quickly on the outer non-impacted surface and reached the edges within less than 0.4~ms (\Fref{fig:crack_ev_5LG2} or \ref{fig:crack_ev_5LG3}). If the impacted layer fractured first, the cracks were mostly localised around the impact point and spread only occasionally towards edges (\Fref{fig:cf_7LG-4}).
In many structural applications, the outer glass layers in multi-layer laminated glass are made of tempered glass due to its surface compressive stresses developed during the tempering process. This treatment leads to higher tensile strength, but it also modifies the fracture pattern of a glass sample that breaks into many small pieces. This behaviour is undesirable for the post-critical resistance because the fine net of crack would not remain localised under the impact point. 

The major changes in crack patterns occurred when the first glass layer(s) fractured. Further, the growth of cracks under consecutive impacts was small until the fracture of next glass layer; compare, e.g., fracture patterns for different impact heights but the same scheme of cracked layers (\Fref{fig:crack_ev_7LG3}).
A visual comparison of patterns for the impact height of 50~cm shows that the fewest cracks appeared in the 7LG--1 sample where both outer glass layers fractured at relatively low impact energies. This suggests that the density of cracks is given by the initial power of the impact leading to fracture, and the additional impacts did not change the density of cracks significantly. 

\begin{figure}[ht]
    \centering
    \begin{tabular}{cc}
    \footnotesize 5LG--2
        & 
        \footnotesize 5LG--4
        \\
    \includegraphics[height=22mm]{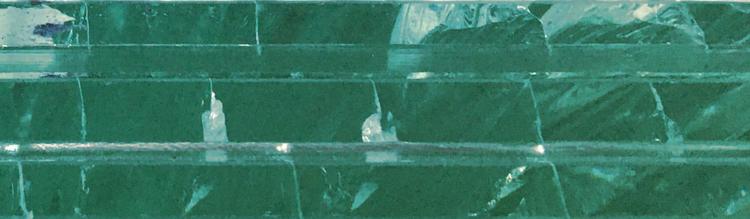}
         &  
        \includegraphics[height=22mm]{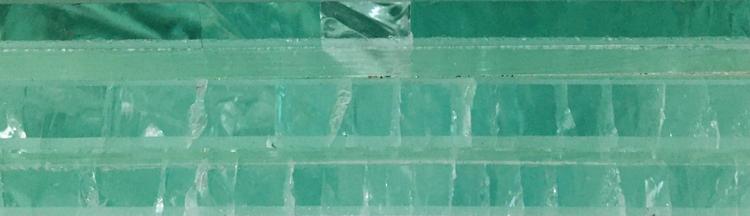}
        \\
        \\
        \footnotesize 7LG--2
        & 
        \footnotesize 7LG--4
        \\
        \includegraphics[height=22mm]{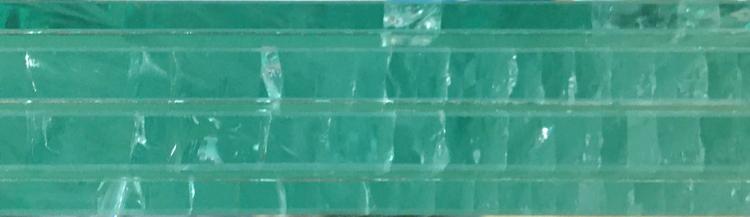}
         &  
        \includegraphics[height=22mm]{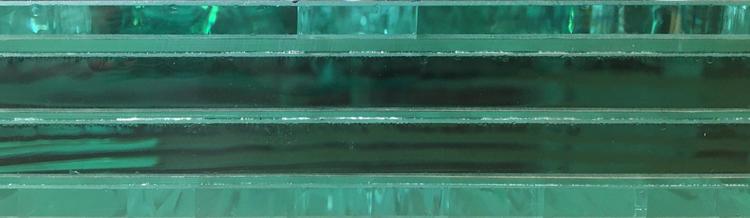}
        \\
    \end{tabular}
    \caption{Side views of the final fracture patterns visible on samples' edges after all impact tests (the top glass layer is the impacted one). }
    \label{fig:cracks_side}
\end{figure}

The driving mechanisms of cracking for three-layer laminated glass was studied in~\citep{chen2014different}. The authors observed that the radial cracks initiated in the supported glass layer, and after a short time interval, the cracks propagated at the same in-plane locations in the impacted glass layer. In our impact tests, all glass layers did not generally fracture at one impact event, so the cracks did not completely propagate across the layers as can be seen in the side views on samples' edges (\Fref{fig:cracks_side}). The more significant differences can be observed for the crack patterns on the impacted glass layer, which was underlied by a thicker PVB foil and where the cracks stayed mostly localised around the impact point (clearly visible, e.g., for 7LG--4 in Figures~\ref{fig:cf_7LG-4} and~\ref{fig:cracks_side}). The crack patterns partly differed also for the samples where glass layers fractured consecutively in the direction from the non-impacted surface to the impacted one (e.g., 5LG--2, 5LG--4, 7LG--2 in~\Fref{fig:cracks_side}).  

\subsection{Comparison of breakage forces}

A higher number of tested specimens would be needed for a precise statistical evaluation of the tensile strengths and the associated contact forces leading to the fracture of individual glass layers in laminated glass samples (called further also breakage forces). Even though the available tests are limited, we believe that except for their use for validating numerical simulations, they can also illustrate some features of impact response of laminated glass. 

\begin{figure}[hp]
    \centering
    \begin{tabular}{cc}
        \multicolumn{2}{c}{\footnotesize Contact forces corresponding to the first glass layer(s) fracture}\\
        \includegraphics[width=0.475\textwidth]{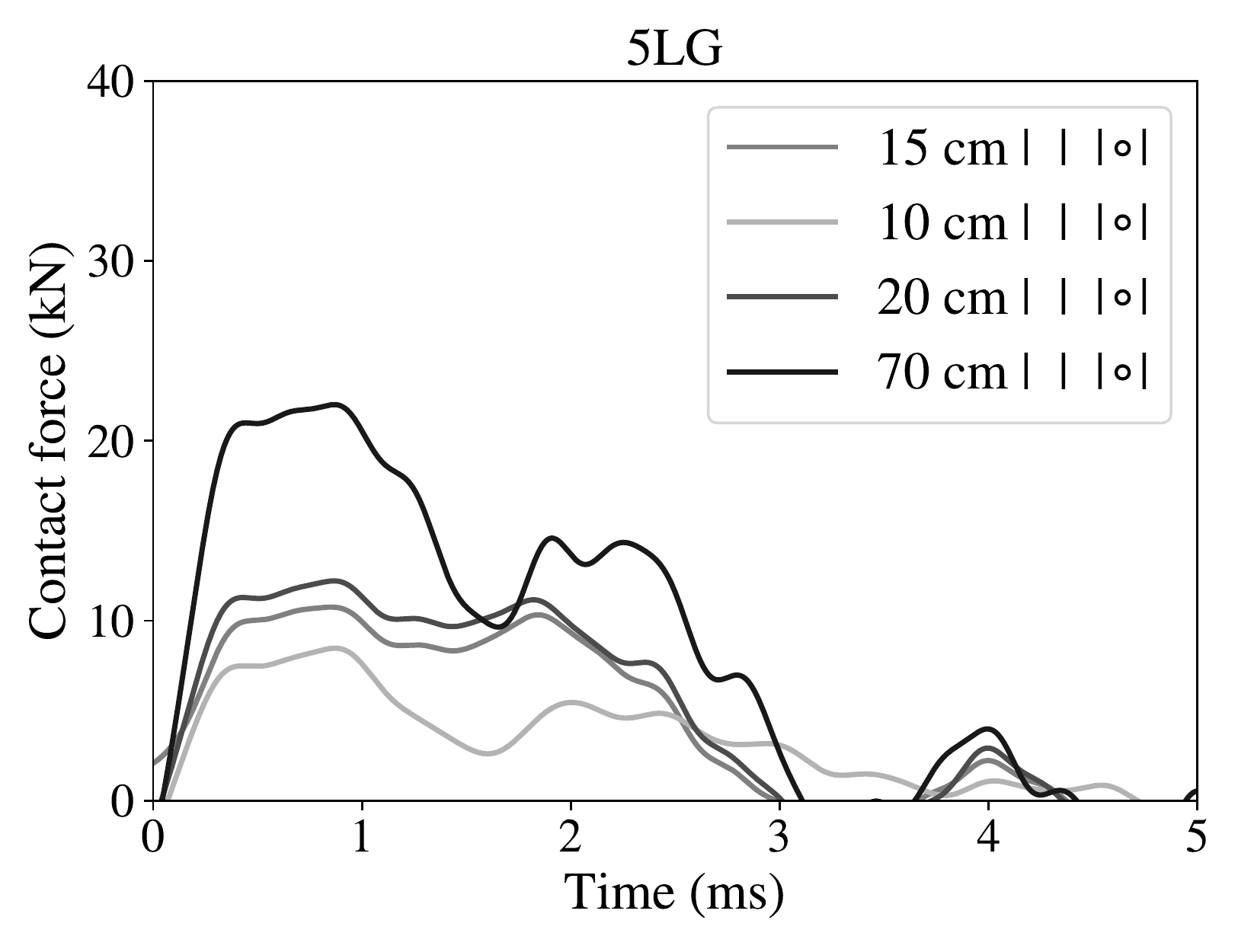}
         &
         \includegraphics[width=0.475\textwidth]{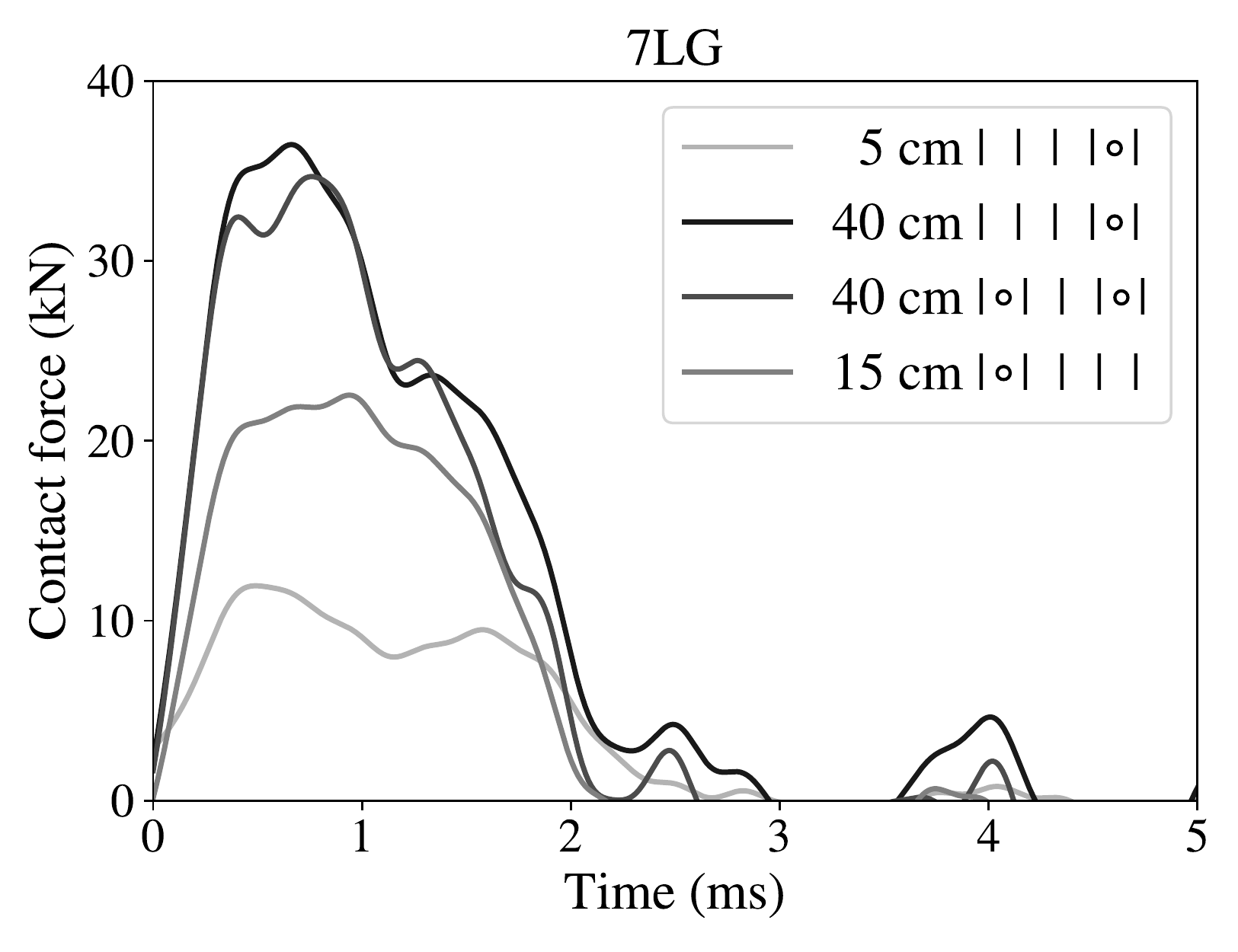}
         \\
         \multicolumn{2}{c}{\footnotesize Contact forces corresponding to a subsequent fracture of different glass layer(s)}
         \\
        \includegraphics[width=0.475\textwidth]{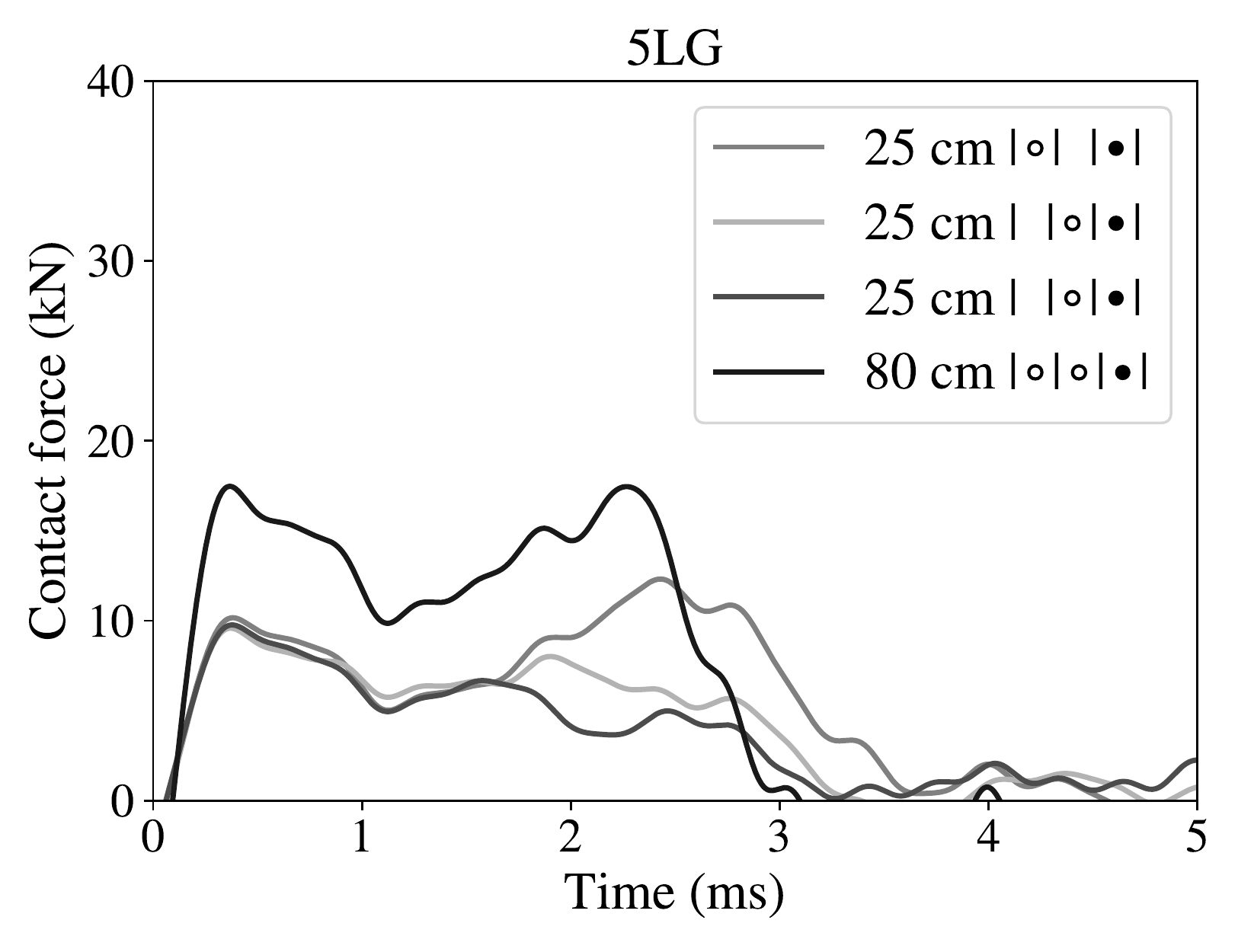}
         &
         \includegraphics[width=0.475\textwidth]{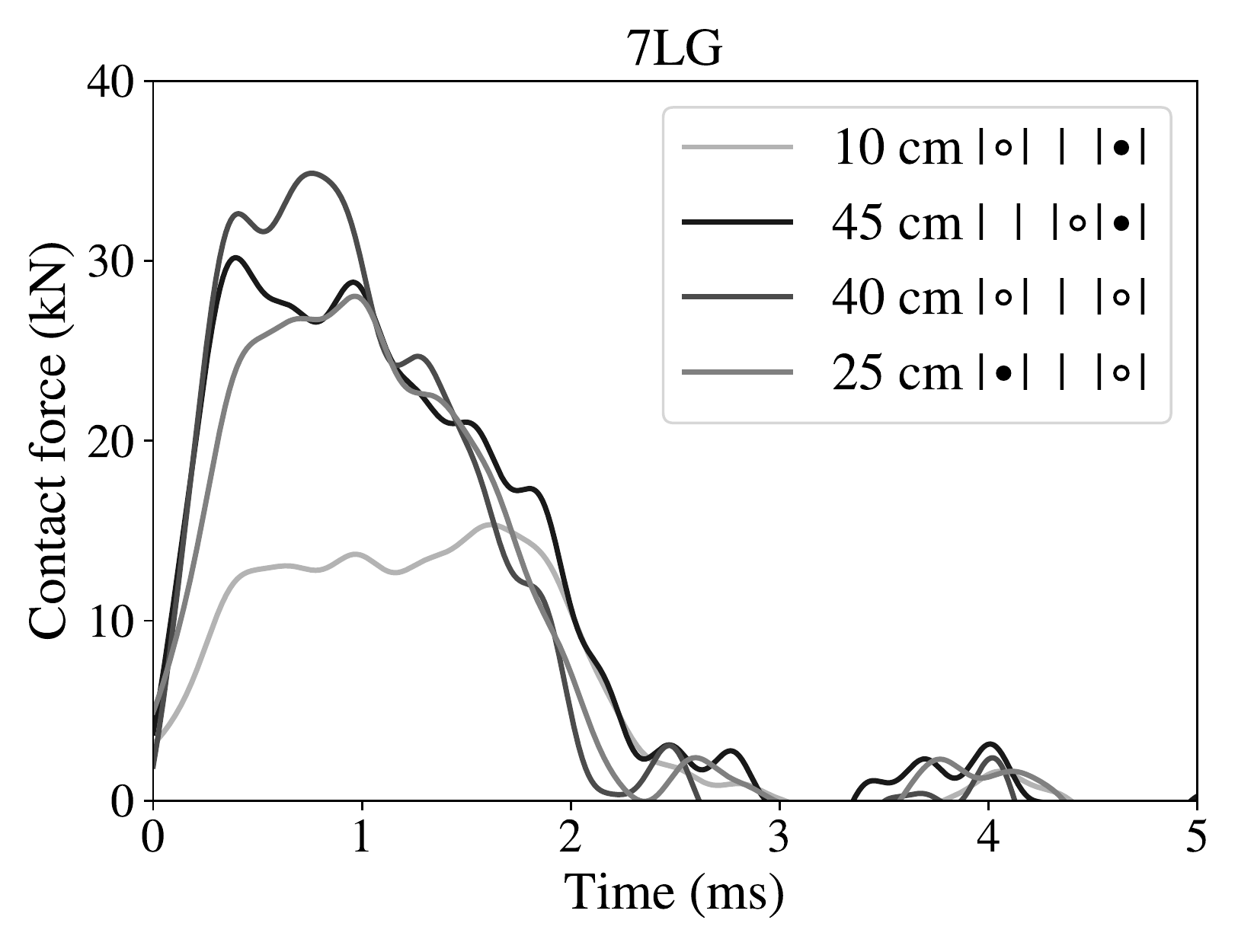}
         \\
         \multicolumn{2}{c}{\footnotesize Contact forces corresponding to the fracture of all glass layers}
         \\
         \includegraphics[width=0.475\textwidth]{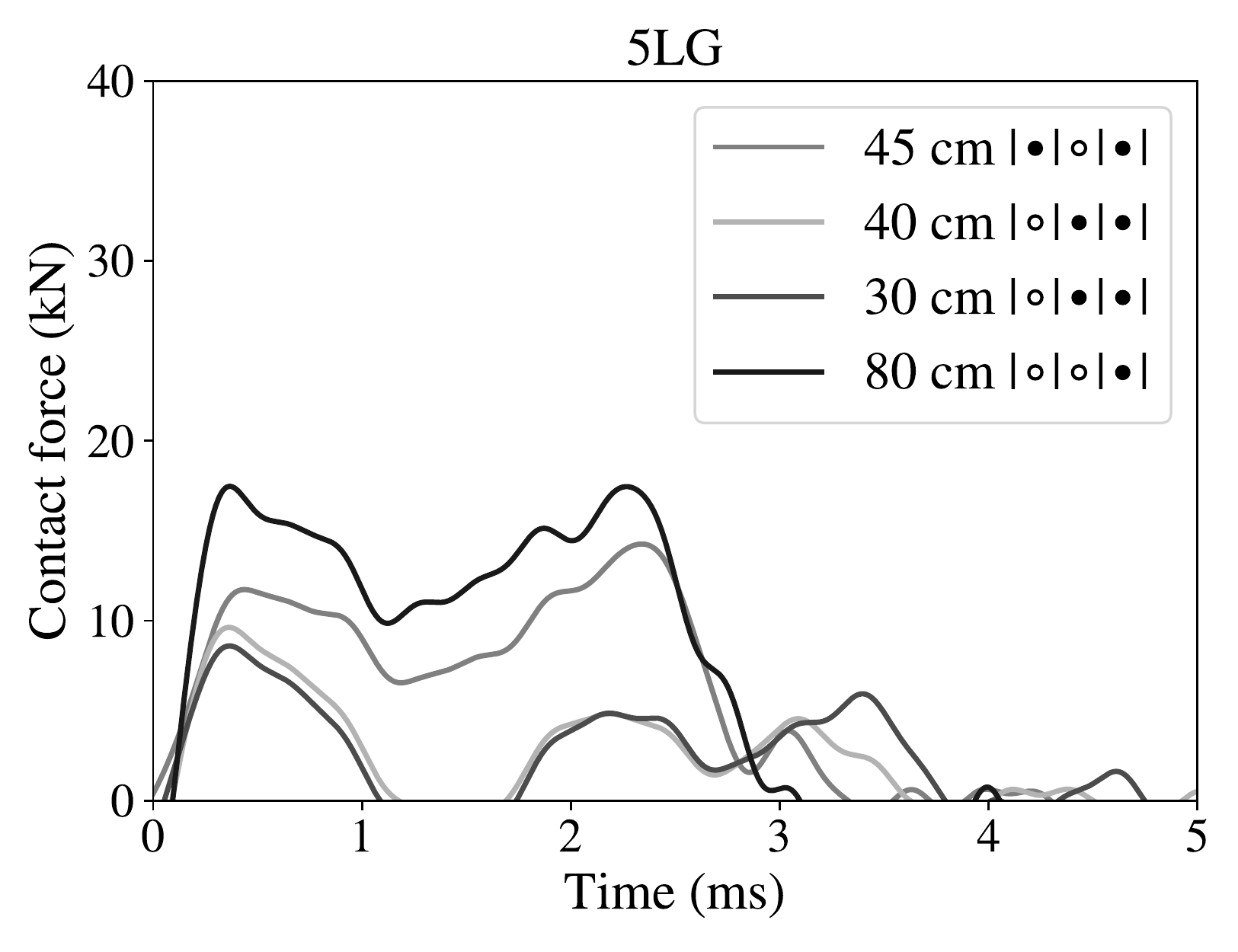}
         &
         \includegraphics[width=0.475\textwidth]{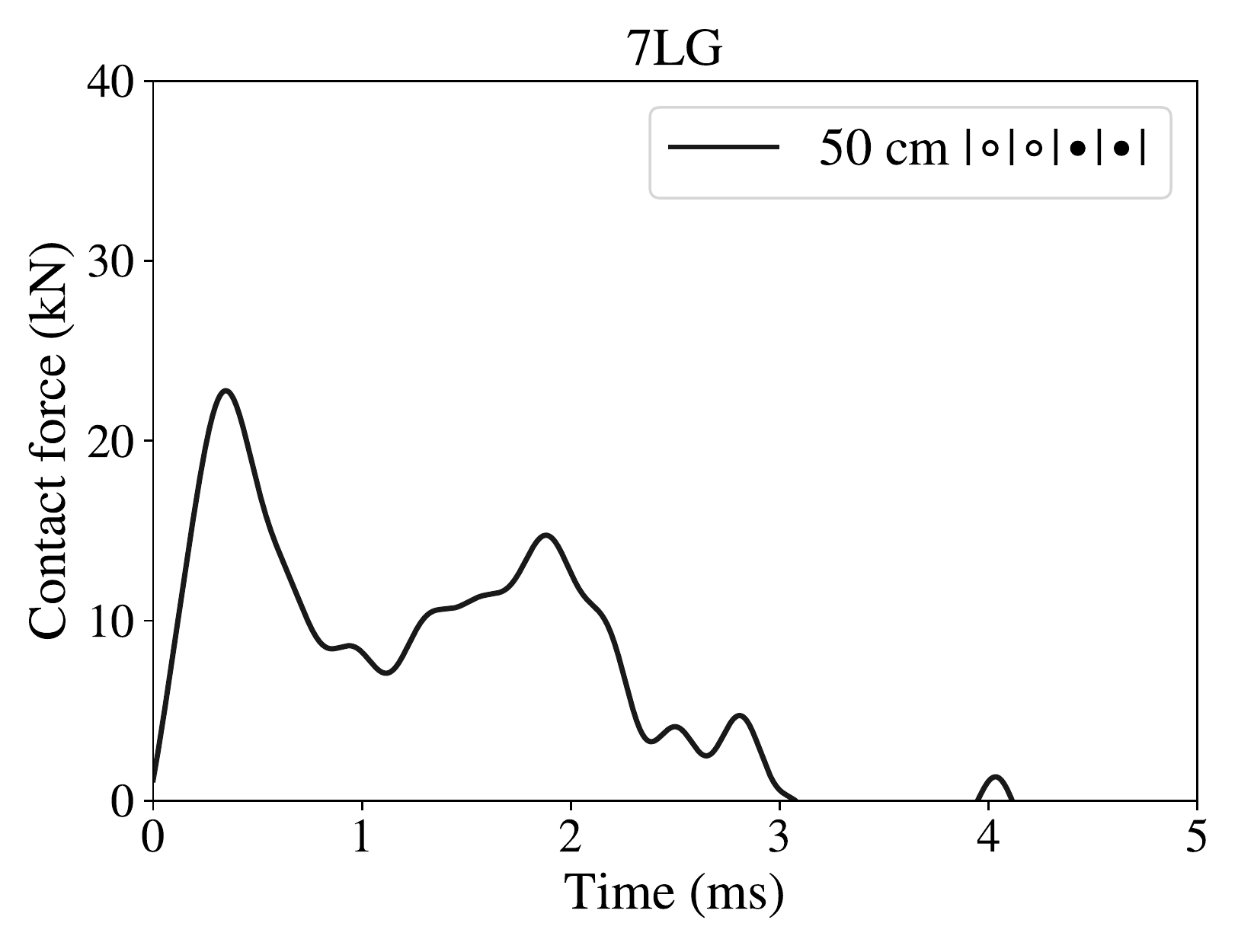}
    \end{tabular}
\caption{Comparison of contact forces (2$\times$ filter CFC~1000) corresponding to different stages of glass fracture. 
}
    \label{fig:com_force_frac}
\end{figure}

The overview of breakage forces (presented in~\Fref{fig:com_force_frac}) indicates that the 7LG-samples resisted mostly higher contact forces than the 5LG-sample. 
For both types of laminated glass, the scatter of forces is substantial due to the stochastic nature of tensile strength of glass. For 5LG-samples, the lowest peak value leading to the first glass layer fracture was less than 10~kN, whereas the largest value was over 20~kN. For 7LG-samples, the breakage force reached the value over 35~kN for two samples but was less than 15~kN in another case. Similar differences can be also seen for the peak contact forces leading to the fracture of subsequent glass layers. 

The experiments confirmed that the samples with one or more fractured layers were in many cases able to resist similar or even higher contact forces than those leading to the fracture of the first or the previous glass layer. This effect is most obvious for the sample 7LG--1, where both outer glass layers fractured but the inner two intact glass layers resisted more than two times higher force without any failure (\Fref{fig:cf_7LG-1}).
Notice also that the peaks of contact forces leading to the breakage of a glass layer were not always higher than the largest forces observed during previous less powerful impact events (samples 5LG--1, 5LG--2, and 5LG--4). This observation may indicate that some naked-eye invisible damage occurred during the previous contacts and/or the initial microflaws grew during previous impacts and caused subsequent failure~\citep{foraboschi2013hybrid}.    

\begin{figure}[hp]
    \centering
    \begin{tabular}{cc}
         \includegraphics[width=0.475\textwidth]{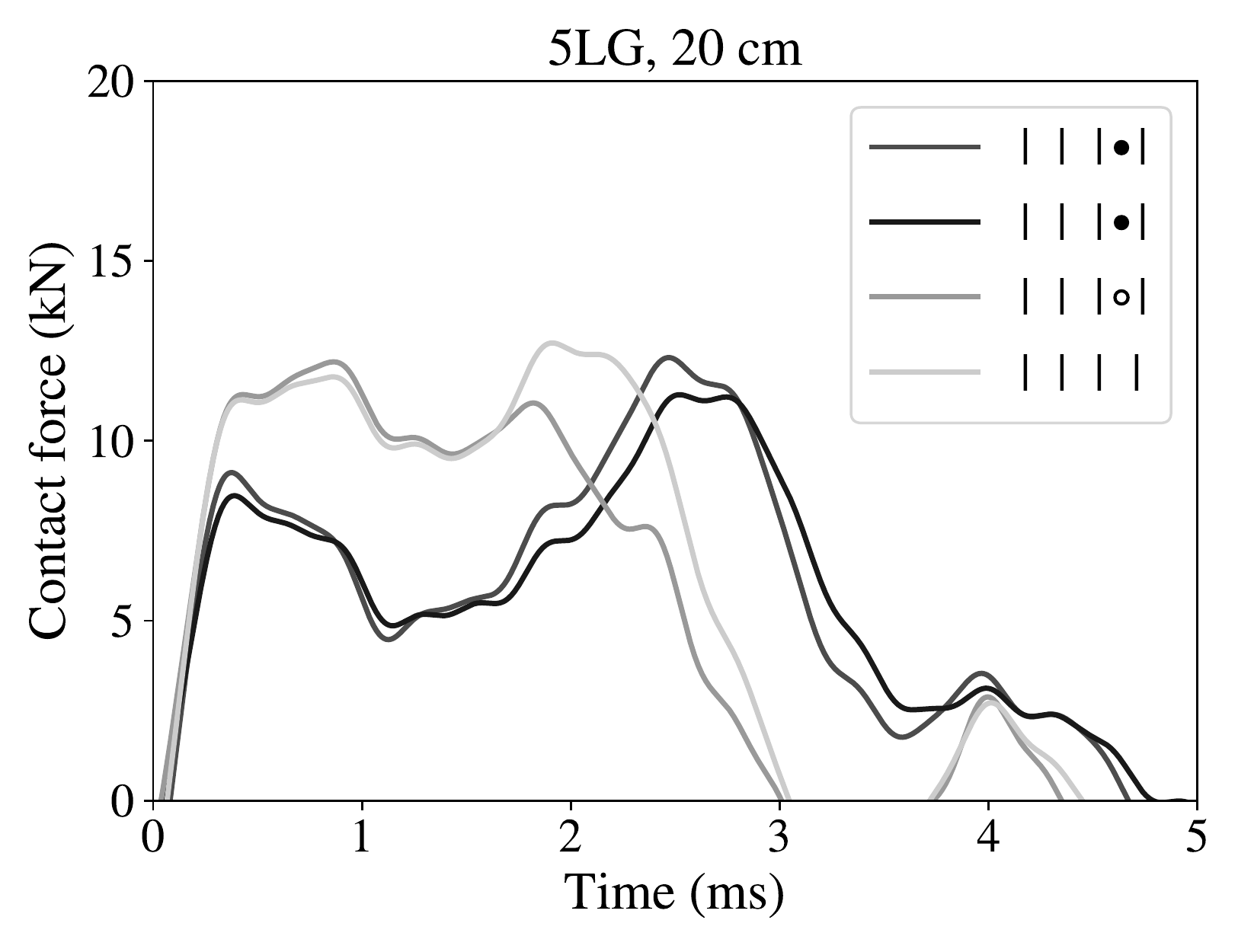}
         &
         \includegraphics[width=0.475\textwidth]{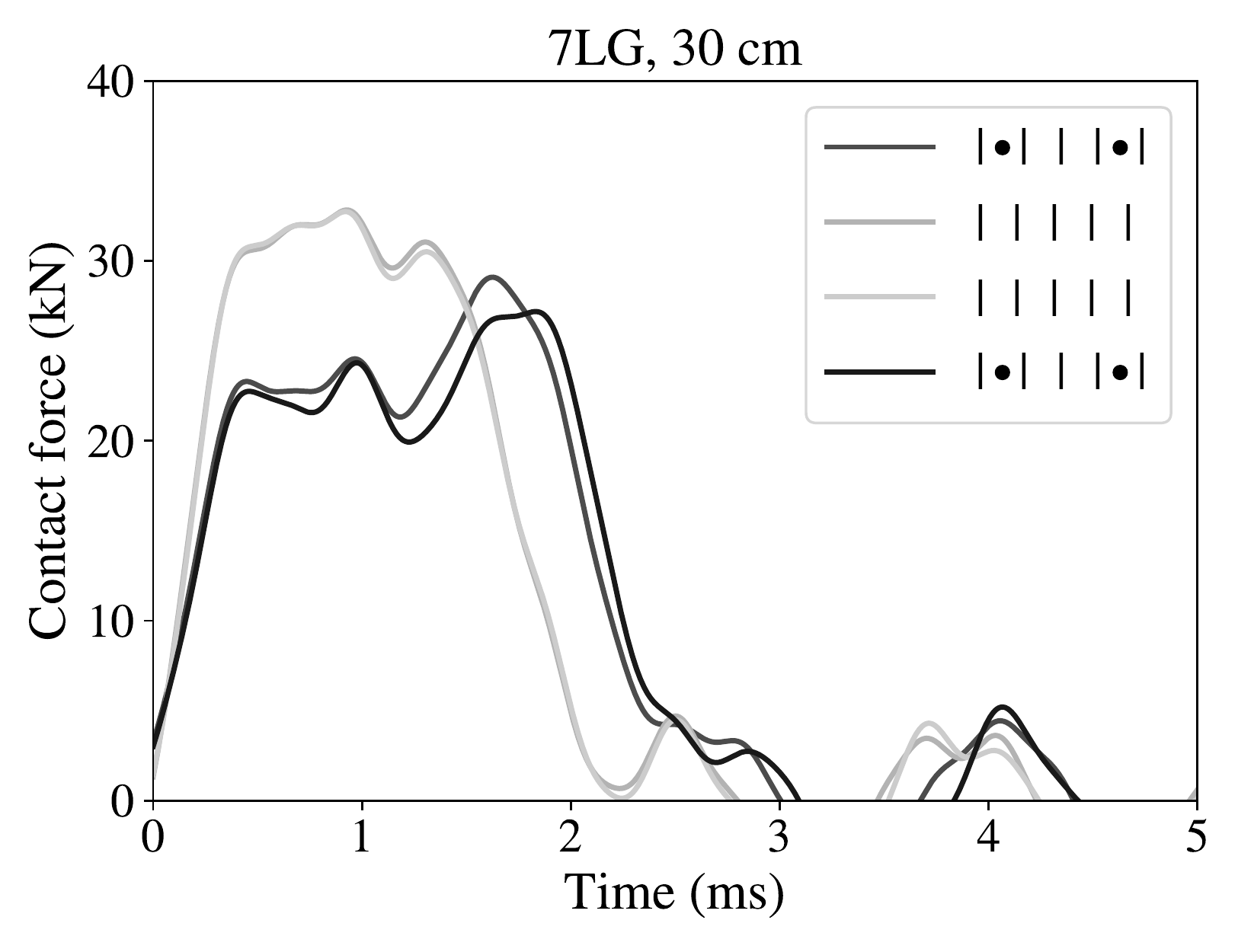}
    \end{tabular}
    \footnotesize
\begin{tabular}{cc}
         \includegraphics[width=0.475\textwidth]{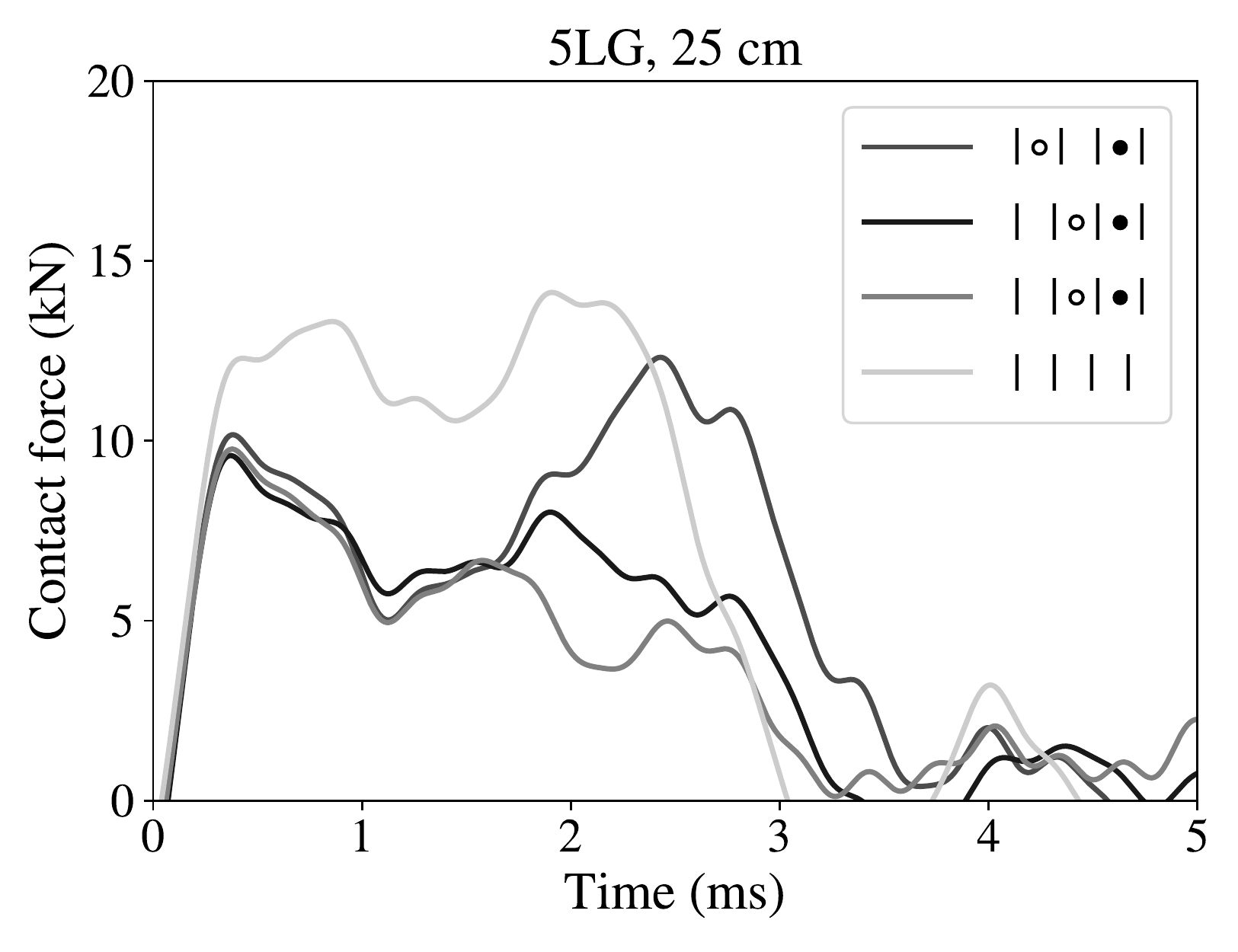}
         &
         \includegraphics[width=0.475\textwidth]{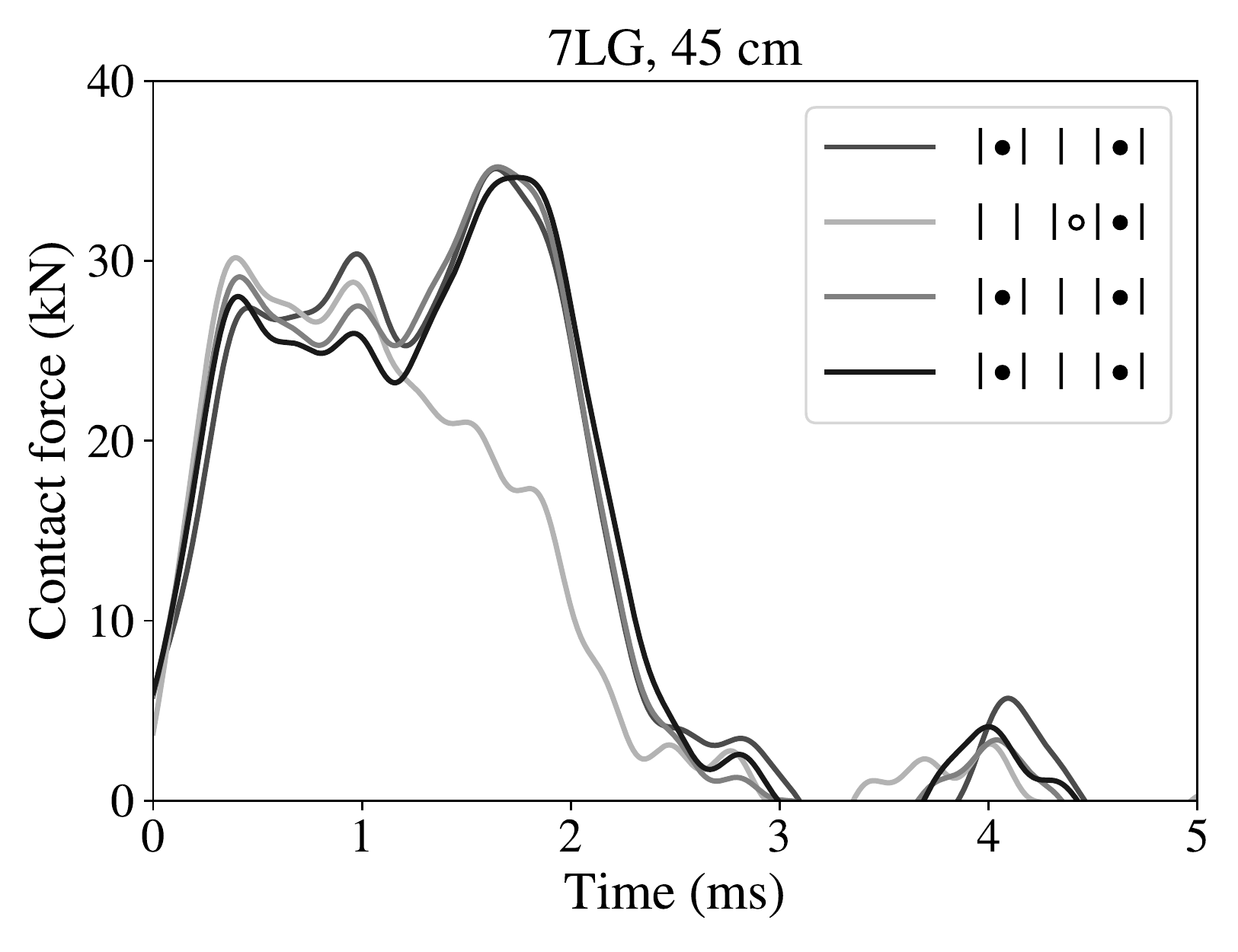}
    \end{tabular}
\caption{Comparison of contact forces (2$\times$ filter CFC~1000) for two different schemes of cracked layers in laminated glass samples. 
}
    \label{fig:com_force_frac_sim}
\end{figure}

Further, we tested if the same scheme of cracked layers provided the same evolution of contact forces. 
The very good level of conformity in the evolution of contact forces could already be seen for the post-fracture response of individual laminated glass samples (Figures~\ref{fig:cf_5LG-1}--\ref{fig:cf_7LG-4}).
The comparison of contact forces acting on different samples (\Fref{fig:com_force_frac_sim}) proves that the contact forces were consistent not only for non-destructive tests, but very similar evolutions developed also  for the same scheme of fractured layers. For 5LG-samples and the impact height of 20~cm, the contact forces differed only slightly, and the impact event resulted in approximately the same length of the contact and a similar peak value of the contact force. When the impact height was increased to 25~cm, the evolution of contact force followed initially the same path for three samples with the same scheme of cracked layers until the fracture of another glass layer. Almost identical contact forces can be also found for the 7LG-samples with both outer glass layers fractured (\Fref{fig:com_force_frac_sim}).

\subsection{Modal response of unfractured and partially fractured laminated glass samples}

Using the approach described in~\Sref{S:NOND}, we derived the natural frequencies for all samples from sensors $C$ placed at a glass corner (data available up to the impact height of 30~cm). The first natural frequencies were grouped by the fracture scheme  in~\Tref{tab:frequencies}. The vibration of unfractured samples was characterised with very similar first natural frequencies for 5LG-samples (differences in values less than 4\%) and with an identical first natural frequency for 7LG-samples. Not enough tests were often available for the fractured samples corresponding to the same scheme. However, the limited data indicate that the partially fractured samples with identical cracked layers vibrate at very similar frequencies. The differences in first natural frequencies were up to 10\%, with the largest error for the scheme with both outer glass layers cracked, where the density of cracks more than doubled (\Fref{fig:photos_fracture_7LG}). This shows that the post-critical response of partially fractured laminated glass samples is relatively consistent, and therefore, it could be estimated by a numerical model with reduced stiffness of fractured glass layers, similarly to~\citep{mohagheghian2018effect}.

\begin{table}[ht]
\caption{Natural frequencies for unfractured and partially fractured glass samples obtained from the FFT-based signal analysis of recorded accelerations (sensor $C$).}
    \centering
     \footnotesize
    \begin{tabular}{llll}
     \hline
         type & cracked layers & number of tests &
frequency (range) (Hz)
        \\
         \hline
5LG &
         \textbar~\textbar~\textbar~\textbar 
         &
12 values (4 samples)
        &
        383 -- 395 
         \\
         & \textbar~\textbar~\textbar$\sbullet[.72]$\textbar 
         &
         3 values (2 samples)
         &
         281 -- 281
\\
         & \textbar$\sbullet[.72]$\textbar~\textbar$\sbullet[.72]$\textbar
         &
         1 value  (1 sample)
         &
         191 \\
         & \textbar~\textbar$\sbullet[.72]$\textbar$\sbullet[.72]$\textbar
         &
         1 value  (1 sample)
         &
         153 \\
7LG &
         \textbar~\textbar~\textbar~\textbar~\textbar  
         & 14 values  (3 samples)
         &
         584 -- 584
\\
         & \textbar$\sbullet[.72]$\textbar~\textbar~\textbar~\textbar  
         &
         1 value  (1 sample)
         &
         520 
\\
         & \textbar$\sbullet[.72]$\textbar~\textbar~\textbar$\sbullet[.72]$\textbar  & 4 values (2 samples)
         &
         394 -- 432  
\\
          \hline
    \end{tabular}
    \label{tab:frequencies}
\end{table}

\begin{figure}[ht]
    \centering
    \begin{tabular}[t]{cc}
         \includegraphics[width=0.475\textwidth]{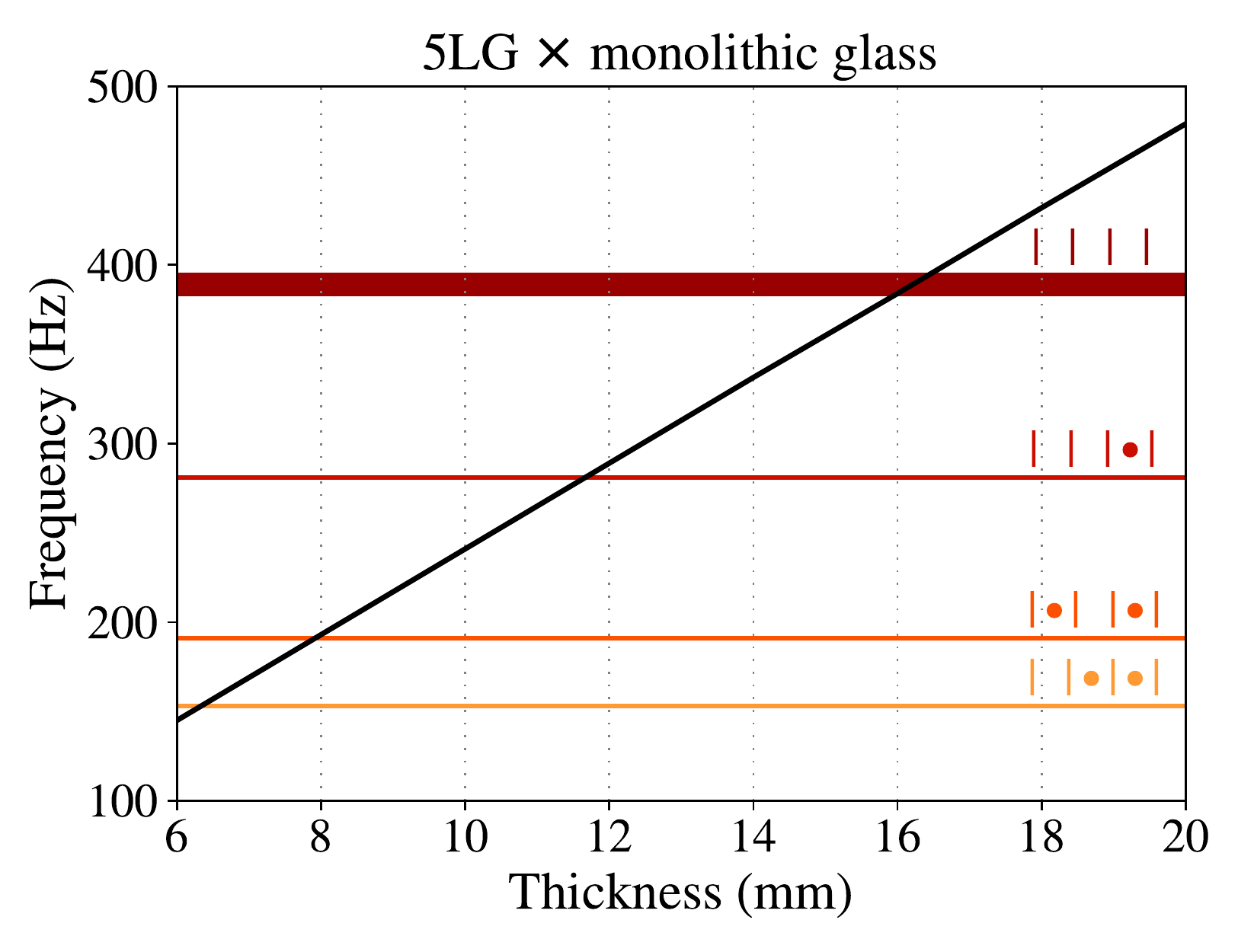}
         &
         \includegraphics[width=0.475\textwidth]{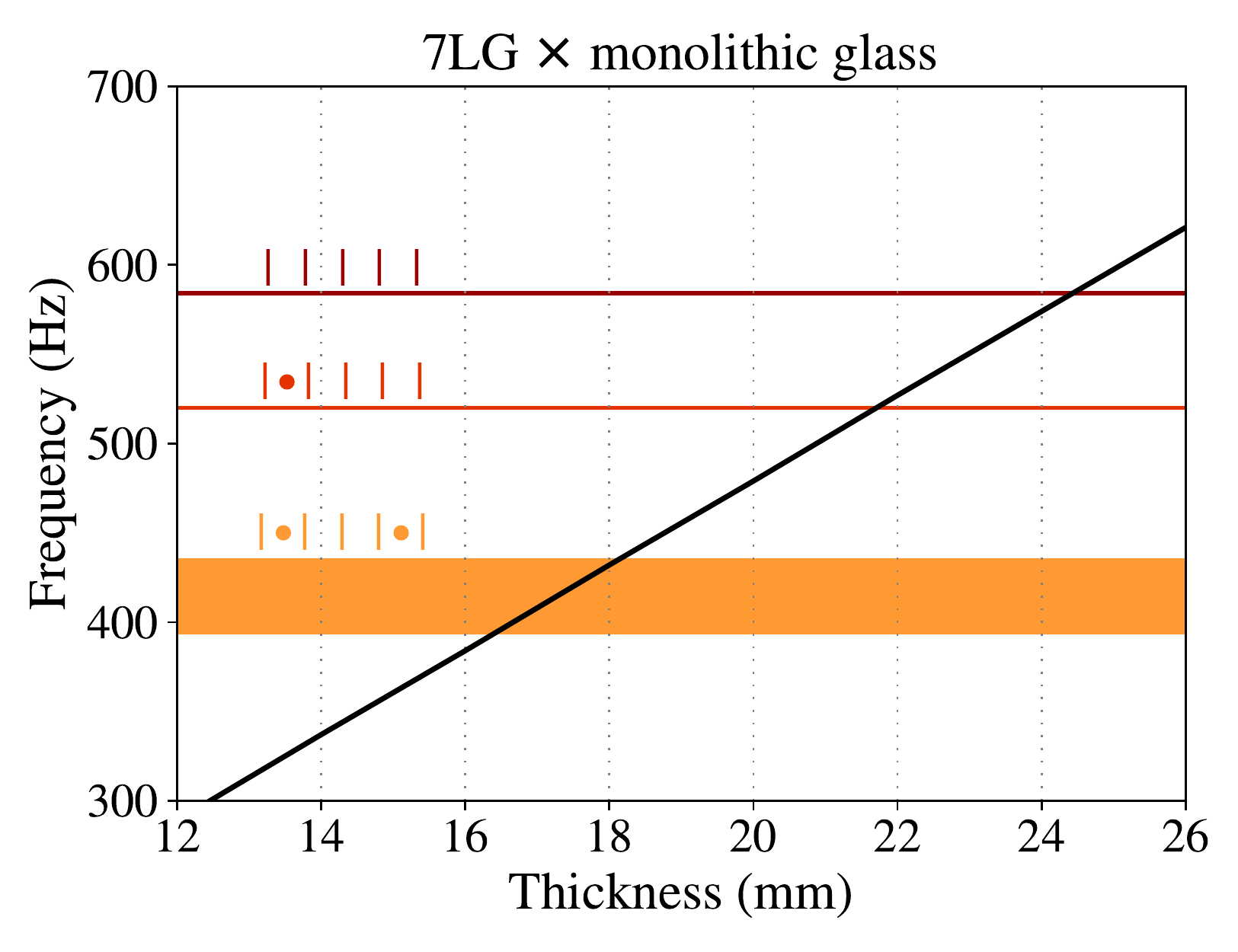}
    \end{tabular}
    \caption{Comparison of the natural frequencies corresponding to the first bi-symmetric mode shape derived from FFT-based analysis for laminated glass samples against numerical predictions for monolithic glass plate (black line; Young's modulus of glass 72~GPa, Poisson's ratio 0.22, density 2,500~kg$\cdot$m$^{-3}$).}
    \label{fig:freq_mon}
\end{figure}

Next, the thicknesses of equivalent monolithic glass plates (with the same first natural frequencies) can be obtained from~\Fref{fig:freq_mon}. The natural frequencies of monolithic glass plates were derived numerically (employing the Young's modulus of glass 72~GPa, Poisson's ratio 0.22, and density 2,500~kg$\cdot$m$^{-3}$).
The 5LG-samples vibrated similarly to a 16~mm-thick glass ply (the sum of the nominal thicknesses of glass layers together).
If the back glass layer 
fractured in 5LG-samples, the natural frequencies corresponded to a monolithic glass plate thicker than 11~mm (the sum of nominal thicknesses of both unfractured layers). A difference in the first natural frequencies can be seen for 5LG-samples with only undamaged middle or front (impacted) glass layer. The first scheme provided the response comparable to an 8~mm-thick glass plate (the middle layer was 6 mm thick), whereas the second one to an approximately  6~mm-thick glass sample (the impacted layer was 5 mm thick).

For the undamaged 7LG-samples, the dynamic response corresponded to a {24.5~mm-thick} glass plate (less than 26~mm---the nominal thicknesses of glass layers in a 7LG-sample together). 
If the impacted glass layer fractured, the response was comparable to a 21.5~mm-thick glass plate (the total thickness of unfractured glass layers was 21~mm). A wider range of obtained natural frequencies can be seen in~\Fref{fig:freq_mon} for 7LG-samples with both outer glass layer damaged, when their modal response corresponded to a monolithic glass with the thickness of 16--18 mm.

This comparison of laminated samples with monolithic glass plates illustrated that the PVB interlayers provided a stiff shear connection between glass layers during the low-velocity impact tests with the contact duration less than 10~ms. The frequencies of laminated glass samples were mainly higher than those of monolithic glass plates with the thickness corresponding to the overall nominal thickness of unfractured glass layers. 

\section{Conclusions}
\label{S:Conclusions}

This study summarised an experimental data set for validating numerical models of impacted plates made of multi-layer laminated glass and included a description of their post-critical response. Namely:

\begin{itemize}
    \item Experimental data for multi-layer laminated glass plates with two different geometries (four samples for each type) were provided and compared. 
    \item Good consistency for the non-destructive responses of impacted laminated glass samples, demonstrated in terms of velocities and contact forces, proved the repeatability of the experiments and the consistency of the acquired data.
    \item Significant differences in impact energies leading to the fracture of glass layers were observed for individual specimens. Thus, a higher number of tests would be needed to evaluate the breakage forces statistically. However, the results indicated that specimens with one or more fractured layers were in many cases able to resist similar or even higher contact forces than those leading to the fracture of the previous glass layer.
    \item The damage of laminated glass samples initiated in the back glass layer for 5LG-samples or in one of the outer glass layers for 7LG-samples. Then, the sequence of fractured glass layers differed. In 7LG-samples, both the outer glass layers seem to be more prone to fracture earlier. The ways of their cracking were different (the cracks were mainly localised around the impact point and spread only rarely towards edges in the impacted layer). \item The response of 7LG--1 indicated that the fracture of outer glass layers under lower impact energies did not predetermine lower breakage forces leading to the fracture of the rest of undamaged glass layers. 
\item Despite the large scatter of breakage forces, the post-critical vibrations of partially fractured samples exhibited similar first natural frequencies. The PVB interlayers provided a stiff shear connection between glass layers during the impact; the frequencies of laminated glass samples were mostly higher than those of monolithic glass plates with its thickness corresponding to the overall nominal thickness of unfractured glass layers.
\end{itemize}

\section*{Acknowledgements}
This work was supported by the Czech Science Foundation, grant No.~{19-15326S}.


\begin{thebibliography}{50}
\expandafter\ifx\csname natexlab\endcsname\relax\def\natexlab#1{#1}\fi
\providecommand{\url}[1]{\texttt{#1}}
\providecommand{\href}[2]{#2}
\providecommand{\path}[1]{#1}
\providecommand{\DOIprefix}{doi:}
\providecommand{\ArXivprefix}{arXiv:}
\providecommand{\URLprefix}{URL: }
\providecommand{\Pubmedprefix}{pmid:}
\providecommand{\doi}[1]{\href{http://dx.doi.org/#1}{\path{#1}}}
\providecommand{\Pubmed}[1]{\href{pmid:#1}{\path{#1}}}
\providecommand{\bibinfo}[2]{#2}
\ifx\xfnm\relax \def\xfnm[#1]{\unskip,\space#1}\fi
\bibitem[{Mori et~al.(2016)Mori, Masiello, and Mamone}]{mori2016design}
\bibinfo{author}{M.~Mori}, \bibinfo{author}{G.~Masiello},
  \bibinfo{author}{V.~Mamone},
\newblock \bibinfo{title}{Design of a glass walkway in the historical city of
  {Pisa} using jumbo glass slabs},
\newblock in: \bibinfo{booktitle}{Challenging Glass Conference Proceedings},
  volume~\bibinfo{volume}{5}, \bibinfo{year}{2016}, pp.
  \bibinfo{pages}{527--536}.
\bibitem[{Wever et~al.(2016)Wever, Haarhuis, ten Brincke, and
  Nijsse}]{wever2016swimming}
\bibinfo{author}{T.~Wever}, \bibinfo{author}{K.~Haarhuis},
  \bibinfo{author}{E.~ten Brincke}, \bibinfo{author}{R.~Nijsse},
\newblock \bibinfo{title}{Swimming pools of glass},
\newblock in: \bibinfo{booktitle}{Challenging Glass Conference Proceedings},
  volume~\bibinfo{volume}{5}, \bibinfo{year}{2016}, pp.
  \bibinfo{pages}{557--564}.
\bibitem[{Dotan(2016)}]{dotan2016zhangjiajie}
\bibinfo{author}{H.~Dotan},
\newblock \bibinfo{title}{Zhangjiajie {Grand} {Canyon} glass bridge},
\newblock in: \bibinfo{booktitle}{Challenging Glass Conference Proceedings},
  volume~\bibinfo{volume}{5}, \bibinfo{year}{2016}, pp. \bibinfo{pages}{3--12}.
\bibitem[{Overend et~al.(2007)Overend, De~Gaetano, and
  Haldimann}]{overend2007diagnostic}
\bibinfo{author}{M.~Overend}, \bibinfo{author}{S.~De~Gaetano},
  \bibinfo{author}{M.~Haldimann},
\newblock \bibinfo{title}{Diagnostic interpretation of glass failure},
\newblock \bibinfo{journal}{Structural Engineering International}
  \bibinfo{volume}{17} (\bibinfo{year}{2007}) \bibinfo{pages}{151--158}.
\bibitem[{Backman and Goldsmith(1978)}]{backman1978mechanics}
\bibinfo{author}{M.~E. Backman}, \bibinfo{author}{W.~Goldsmith},
\newblock \bibinfo{title}{The mechanics of penetration of projectiles into
  targets},
\newblock \bibinfo{journal}{International Journal of Engineering Science}
  \bibinfo{volume}{16} (\bibinfo{year}{1978}) \bibinfo{pages}{1--99}.
\bibitem[{Mohagheghian et~al.(2017)Mohagheghian, Wang, Jiang, Zhang, Guo, Yan,
  Kinloch, and Dear}]{mohagheghian2017quasi}
\bibinfo{author}{I.~Mohagheghian}, \bibinfo{author}{Y.~Wang},
  \bibinfo{author}{L.~Jiang}, \bibinfo{author}{X.~Zhang},
  \bibinfo{author}{X.~Guo}, \bibinfo{author}{Y.~Yan},
  \bibinfo{author}{A.~Kinloch}, \bibinfo{author}{J.~Dear},
\newblock \bibinfo{title}{Quasi-static bending and low velocity impact
  performance of monolithic and laminated glass windows employing chemically
  strengthened glass},
\newblock \bibinfo{journal}{European Journal of Mechanics-A/Solids}
  \bibinfo{volume}{63} (\bibinfo{year}{2017}) \bibinfo{pages}{165--186}.
\bibitem[{Larcher et~al.(2012)Larcher, Solomos, Casadei, and
  Gebbeken}]{larcher2012experimental}
\bibinfo{author}{M.~Larcher}, \bibinfo{author}{G.~Solomos},
  \bibinfo{author}{F.~Casadei}, \bibinfo{author}{N.~Gebbeken},
\newblock \bibinfo{title}{Experimental and numerical investigations of
  laminated glass subjected to blast loading},
\newblock \bibinfo{journal}{International Journal of Impact Engineering}
  \bibinfo{volume}{39} (\bibinfo{year}{2012}) \bibinfo{pages}{42--50}.
\bibitem[{Zhang et~al.(2013)Zhang, Hao, and Ma}]{zhang2013parametric}
\bibinfo{author}{X.~Zhang}, \bibinfo{author}{H.~Hao}, \bibinfo{author}{G.~Ma},
\newblock \bibinfo{title}{Parametric study of laminated glass window response
  to blast loads},
\newblock \bibinfo{journal}{Engineering Structures} \bibinfo{volume}{56}
  (\bibinfo{year}{2013}) \bibinfo{pages}{1707--1717}.
\bibitem[{Pelfrene et~al.(2016)Pelfrene, Kuntsche, Van~Dam, Van~Paepegem, and
  Schneider}]{pelfrene2016critical}
\bibinfo{author}{J.~Pelfrene}, \bibinfo{author}{J.~Kuntsche},
  \bibinfo{author}{S.~Van~Dam}, \bibinfo{author}{W.~Van~Paepegem},
  \bibinfo{author}{J.~Schneider},
\newblock \bibinfo{title}{Critical assessment of the post-breakage performance
  of blast loaded laminated glazing: experiments and simulations},
\newblock \bibinfo{journal}{International Journal of Impact Engineering}
  \bibinfo{volume}{88} (\bibinfo{year}{2016}) \bibinfo{pages}{61--71}.
\bibitem[{Liu et~al.(2016)Liu, Xu, Xu, Wang, Sun, and Li}]{liu2016energy}
\bibinfo{author}{B.~Liu}, \bibinfo{author}{T.~Xu}, \bibinfo{author}{X.~Xu},
  \bibinfo{author}{Y.~Wang}, \bibinfo{author}{Y.~Sun}, \bibinfo{author}{Y.~Li},
\newblock \bibinfo{title}{Energy absorption mechanism of polyvinyl butyral
  laminated windshield subjected to head impact: Experiment and numerical
  simulations},
\newblock \bibinfo{journal}{International Journal of Impact Engineering}
  \bibinfo{volume}{90} (\bibinfo{year}{2016}) \bibinfo{pages}{26--36}.
\bibitem[{Chen et~al.(2021)Chen, Chen, and Li}]{chen2021experimental}
\bibinfo{author}{X.~Chen}, \bibinfo{author}{S.~Chen}, \bibinfo{author}{G.-Q.
  Li},
\newblock \bibinfo{title}{Experimental investigation on the blast resistance of
  framed {PVB}-laminated glass},
\newblock \bibinfo{journal}{International Journal of Impact Engineering}
  \bibinfo{volume}{149} (\bibinfo{year}{2021}) \bibinfo{pages}{103788}.
\bibitem[{Osnes et~al.(2019{\natexlab{a}})Osnes, Dey, Hopperstad, and
  B{\o}rvik}]{osnes2019dynamic}
\bibinfo{author}{K.~Osnes}, \bibinfo{author}{S.~Dey}, \bibinfo{author}{O.~S.
  Hopperstad}, \bibinfo{author}{T.~B{\o}rvik},
\newblock \bibinfo{title}{On the dynamic response of laminated glass exposed to
  impact before blast loading},
\newblock \bibinfo{journal}{Experimental Mechanics} \bibinfo{volume}{59}
  (\bibinfo{year}{2019}{\natexlab{a}}) \bibinfo{pages}{1033--1046}.
\bibitem[{Osnes et~al.(2019{\natexlab{b}})Osnes, Holmen, Hopperstad, and
  B{\o}rvik}]{osnes2019fracture}
\bibinfo{author}{K.~Osnes}, \bibinfo{author}{J.~K. Holmen},
  \bibinfo{author}{O.~S. Hopperstad}, \bibinfo{author}{T.~B{\o}rvik},
\newblock \bibinfo{title}{Fracture and fragmentation of blast-loaded laminated
  glass: An experimental and numerical study},
\newblock \bibinfo{journal}{International Journal of Impact Engineering}
  \bibinfo{volume}{132} (\bibinfo{year}{2019}{\natexlab{b}})
  \bibinfo{pages}{103334}.
\bibitem[{Osnes et~al.(2021)Osnes, Holmen, Grue, and
  B{\o}rvik}]{osnes2021perforation}
\bibinfo{author}{K.~Osnes}, \bibinfo{author}{J.~K. Holmen},
  \bibinfo{author}{T.~Grue}, \bibinfo{author}{T.~B{\o}rvik},
\newblock \bibinfo{title}{Perforation of laminated glass: An experimental and
  numerical study},
\newblock \bibinfo{journal}{International Journal of Impact Engineering}
  (\bibinfo{year}{2021}) \bibinfo{pages}{103922}.
\bibitem[{Zhao et~al.(2006)Zhao, Dharani, Chai, and Barbat}]{zhao2006analysis}
\bibinfo{author}{S.~Zhao}, \bibinfo{author}{L.~R. Dharani},
  \bibinfo{author}{L.~Chai}, \bibinfo{author}{S.~D. Barbat},
\newblock \bibinfo{title}{Analysis of damage in laminated automotive glazing
  subjected to simulated head impact},
\newblock \bibinfo{journal}{Engineering Failure Analysis} \bibinfo{volume}{13}
  (\bibinfo{year}{2006}) \bibinfo{pages}{582--597}.
\bibitem[{Untaroiu et~al.(2007)Untaroiu, Shin, and
  Crandall}]{untaroiu2007design}
\bibinfo{author}{C.~D. Untaroiu}, \bibinfo{author}{J.~Shin},
  \bibinfo{author}{J.~R. Crandall},
\newblock \bibinfo{title}{A design optimization approach of vehicle hood for
  pedestrian protection},
\newblock \bibinfo{journal}{International Journal of Crashworthiness}
  \bibinfo{volume}{12} (\bibinfo{year}{2007}) \bibinfo{pages}{581--589}.
\bibitem[{Timmel et~al.(2007)Timmel, Kolling, Osterrieder, and
  Du~Bois}]{timmel2007finite}
\bibinfo{author}{M.~Timmel}, \bibinfo{author}{S.~Kolling},
  \bibinfo{author}{P.~Osterrieder}, \bibinfo{author}{P.~Du~Bois},
\newblock \bibinfo{title}{A finite element model for impact simulation with
  laminated glass},
\newblock \bibinfo{journal}{International Journal of Impact Engineering}
  \bibinfo{volume}{34} (\bibinfo{year}{2007}) \bibinfo{pages}{1465--1478}.
\bibitem[{Xu et~al.(2010)Xu, Li, Chen, Yan, Ge, Zhu, and
  Liu}]{xu2010characteristics}
\bibinfo{author}{J.~Xu}, \bibinfo{author}{Y.~Li}, \bibinfo{author}{X.~Chen},
  \bibinfo{author}{Y.~Yan}, \bibinfo{author}{D.~Ge}, \bibinfo{author}{M.~Zhu},
  \bibinfo{author}{B.~Liu},
\newblock \bibinfo{title}{Characteristics of windshield cracking upon low-speed
  impact: numerical simulation based on the extended finite element method},
\newblock \bibinfo{journal}{Computational Materials Science}
  \bibinfo{volume}{48} (\bibinfo{year}{2010}) \bibinfo{pages}{582--588}.
\bibitem[{Pyttel et~al.(2011)Pyttel, Liebertz, and Cai}]{pyttel2011failure}
\bibinfo{author}{T.~Pyttel}, \bibinfo{author}{H.~Liebertz},
  \bibinfo{author}{J.~Cai},
\newblock \bibinfo{title}{Failure criterion for laminated glass under impact
  loading and its application in finite element simulation},
\newblock \bibinfo{journal}{International Journal of Impact Engineering}
  \bibinfo{volume}{38} (\bibinfo{year}{2011}) \bibinfo{pages}{252--263}.
\bibitem[{Chen et~al.(2017)Chen, Zang, Wang, Yoshimura, and
  Yamada}]{chen2017numerical}
\bibinfo{author}{S.~Chen}, \bibinfo{author}{M.~Zang},
  \bibinfo{author}{D.~Wang}, \bibinfo{author}{S.~Yoshimura},
  \bibinfo{author}{T.~Yamada},
\newblock \bibinfo{title}{Numerical analysis of impact failure of automotive
  laminated glass: a review},
\newblock \bibinfo{journal}{Composites Part B: Engineering}
  \bibinfo{volume}{122} (\bibinfo{year}{2017}) \bibinfo{pages}{47--60}.
\bibitem[{Alter et~al.(2017)Alter, Kolling, and Schneider}]{alter2017enhanced}
\bibinfo{author}{C.~Alter}, \bibinfo{author}{S.~Kolling},
  \bibinfo{author}{J.~Schneider},
\newblock \bibinfo{title}{An enhanced non--local failure criterion for
  laminated glass under low velocity impact},
\newblock \bibinfo{journal}{International Journal of Impact Engineering}
  \bibinfo{volume}{109} (\bibinfo{year}{2017}) \bibinfo{pages}{342--353}.
\bibitem[{Zhang et~al.(2013)Zhang, Hao, and Ma}]{zhang2013laboratory}
\bibinfo{author}{X.~Zhang}, \bibinfo{author}{H.~Hao}, \bibinfo{author}{G.~Ma},
\newblock \bibinfo{title}{Laboratory test and numerical simulation of laminated
  glass window vulnerability to debris impact},
\newblock \bibinfo{journal}{International Journal of Impact Engineering}
  \bibinfo{volume}{55} (\bibinfo{year}{2013}) \bibinfo{pages}{49--62}.
\bibitem[{Behr et~al.(1999)Behr, Kremer, Dharani, Ji, and
  Kaiser}]{behr1999dynamic}
\bibinfo{author}{R.~A. Behr}, \bibinfo{author}{P.~A. Kremer},
  \bibinfo{author}{L.~R. Dharani}, \bibinfo{author}{F.~Ji},
  \bibinfo{author}{N.~Kaiser},
\newblock \bibinfo{title}{Dynamic strains in architectural laminated glass
  subjected to low velocity impacts from small projectiles},
\newblock \bibinfo{journal}{Journal of Materials Science} \bibinfo{volume}{34}
  (\bibinfo{year}{1999}) \bibinfo{pages}{5749--5756}.
\bibitem[{Flocker and Dharani(1997)}]{flocker1997stresses}
\bibinfo{author}{F.~W. Flocker}, \bibinfo{author}{L.~R. Dharani},
\newblock \bibinfo{title}{Stresses in laminated glass subject to low velocity
  impact},
\newblock \bibinfo{journal}{Engineering Structures} \bibinfo{volume}{19}
  (\bibinfo{year}{1997}) \bibinfo{pages}{851--856}.
\bibitem[{Zhang et~al.(2019)Zhang, Mohammed, Zheng, Wu, Mohagheghian, Zhang,
  Yan, and Dear}]{zhang2019temperature}
\bibinfo{author}{X.~Zhang}, \bibinfo{author}{I.~K. Mohammed},
  \bibinfo{author}{M.~Zheng}, \bibinfo{author}{N.~Wu},
  \bibinfo{author}{I.~Mohagheghian}, \bibinfo{author}{G.~Zhang},
  \bibinfo{author}{Y.~Yan}, \bibinfo{author}{J.~P. Dear},
\newblock \bibinfo{title}{Temperature effects on the low velocity impact
  response of laminated glass with different types of interlayer materials},
\newblock \bibinfo{journal}{International Journal of Impact Engineering}
  \bibinfo{volume}{124} (\bibinfo{year}{2019}) \bibinfo{pages}{9--22}.
\bibitem[{Zhang et~al.(2020)Zhang, Liu, Maharaj, Zheng, Mohagheghian, Zhang,
  Yan, and Dear}]{zhang2020impact}
\bibinfo{author}{X.~Zhang}, \bibinfo{author}{H.~Liu},
  \bibinfo{author}{C.~Maharaj}, \bibinfo{author}{M.~Zheng},
  \bibinfo{author}{I.~Mohagheghian}, \bibinfo{author}{G.~Zhang},
  \bibinfo{author}{Y.~Yan}, \bibinfo{author}{J.~P. Dear},
\newblock \bibinfo{title}{Impact response of laminated glass with varying
  interlayer materials},
\newblock \bibinfo{journal}{International Journal of Impact Engineering}
  \bibinfo{volume}{139} (\bibinfo{year}{2020}) \bibinfo{pages}{103505}.
\bibitem[{Grant et~al.(1998)Grant, Cantwell, McKenzie, and
  Corkhill}]{grant1998damage}
\bibinfo{author}{P.~Grant}, \bibinfo{author}{W.~Cantwell},
  \bibinfo{author}{H.~McKenzie}, \bibinfo{author}{P.~Corkhill},
\newblock \bibinfo{title}{The damage threshold of laminated glass structures},
\newblock \bibinfo{journal}{International Journal of Impact Engineering}
  \bibinfo{volume}{21} (\bibinfo{year}{1998}) \bibinfo{pages}{737--746}.
\bibitem[{Fr{\"o}ling et~al.(2014)Fr{\"o}ling, Persson, and
  Austrell}]{froling2014reduced}
\bibinfo{author}{M.~Fr{\"o}ling}, \bibinfo{author}{K.~Persson},
  \bibinfo{author}{P.-E. Austrell},
\newblock \bibinfo{title}{A reduced model for the design of glass structures
  subjected to dynamic impulse load},
\newblock \bibinfo{journal}{Engineering Structures} \bibinfo{volume}{80}
  (\bibinfo{year}{2014}) \bibinfo{pages}{53--60}.
\bibitem[{Pelfrene et~al.(2016)Pelfrene, Van~Dam, Kuntsche, and
  Van~Paepegem}]{pelfrene2016numerical}
\bibinfo{author}{J.~Pelfrene}, \bibinfo{author}{S.~Van~Dam},
  \bibinfo{author}{J.~Kuntsche}, \bibinfo{author}{W.~Van~Paepegem},
\newblock \bibinfo{title}{Numerical simulation of the {EN} 12600 pendulum test
  for structural glass},
\newblock in: \bibinfo{booktitle}{Challenging Glass Conference Proceedings},
  volume~\bibinfo{volume}{5}, \bibinfo{year}{2016}, pp.
  \bibinfo{pages}{429--438}.
\bibitem[{Koz{\l}owski(2019)}]{kozlowski2019experimental}
\bibinfo{author}{M.~Koz{\l}owski},
\newblock \bibinfo{title}{Experimental and numerical assessment of structural
  behaviour of glass balustrade subjected to soft body impact},
\newblock \bibinfo{journal}{Composite Structures} \bibinfo{volume}{229}
  (\bibinfo{year}{2019}) \bibinfo{pages}{111380}.
\bibitem[{Bez et~al.(2021)Bez, Bedon, Manara, Amadio, and
  Lori}]{bez2021calibrated}
\bibinfo{author}{A.~Bez}, \bibinfo{author}{C.~Bedon},
  \bibinfo{author}{G.~Manara}, \bibinfo{author}{C.~Amadio},
  \bibinfo{author}{G.~Lori},
\newblock \bibinfo{title}{Calibrated numerical approach for the dynamic
  analysis of glass curtain walls under spheroconical bag impact},
\newblock \bibinfo{journal}{Buildings} \bibinfo{volume}{11}
  (\bibinfo{year}{2021}) \bibinfo{pages}{154}.
\bibitem[{Viviani et~al.(2021)Viviani, Consolaro, Maffeis, and
  Royer-Carfagni}]{viviani2021engineered}
\bibinfo{author}{L.~Viviani}, \bibinfo{author}{A.~Consolaro},
  \bibinfo{author}{M.~Maffeis}, \bibinfo{author}{G.~Royer-Carfagni},
\newblock \bibinfo{title}{Engineered modelling of the soft-body impact test on
  glazed surfaces},
\newblock \bibinfo{journal}{Engineering Structures} \bibinfo{volume}{226}
  (\bibinfo{year}{2021}) \bibinfo{pages}{111315}.
\bibitem[{Overend and Zammit(2012)}]{overend2012computer}
\bibinfo{author}{M.~Overend}, \bibinfo{author}{K.~Zammit},
\newblock \bibinfo{title}{A computer algorithm for determining the tensile
  strength of float glass},
\newblock \bibinfo{journal}{Engineering Structures} \bibinfo{volume}{45}
  (\bibinfo{year}{2012}) \bibinfo{pages}{68--77}.
\bibitem[{Kaiser et~al.(2000)Kaiser, Behr, Minor, Dharani, Ji, and
  Kremer}]{kaiser2000impact}
\bibinfo{author}{N.~D. Kaiser}, \bibinfo{author}{R.~A. Behr},
  \bibinfo{author}{J.~E. Minor}, \bibinfo{author}{L.~R. Dharani},
  \bibinfo{author}{F.~Ji}, \bibinfo{author}{P.~A. Kremer},
\newblock \bibinfo{title}{Impact resistance of laminated glass using
  “sacrificial ply” design concept},
\newblock \bibinfo{journal}{Journal of Architectural Engineering}
  \bibinfo{volume}{6} (\bibinfo{year}{2000}) \bibinfo{pages}{24--34}.
\bibitem[{Saxe et~al.(2002)Saxe, Behr, Minor, Kremer, and
  Dharani}]{saxe2002effects}
\bibinfo{author}{T.~J. Saxe}, \bibinfo{author}{R.~A. Behr},
  \bibinfo{author}{J.~E. Minor}, \bibinfo{author}{P.~A. Kremer},
  \bibinfo{author}{L.~R. Dharani},
\newblock \bibinfo{title}{Effects of missile size and glass type on impact
  resistance of “sacrificial ply” laminated glass},
\newblock \bibinfo{journal}{Journal of Architectural Engineering}
  \bibinfo{volume}{8} (\bibinfo{year}{2002}) \bibinfo{pages}{24--39}.
\bibitem[{Foraboschi(2013)}]{foraboschi2013hybrid}
\bibinfo{author}{P.~Foraboschi},
\newblock \bibinfo{title}{Hybrid laminated-glass plate: design and assessment},
\newblock \bibinfo{journal}{Composite Structures} \bibinfo{volume}{106}
  (\bibinfo{year}{2013}) \bibinfo{pages}{250--263}.
\bibitem[{Wang et~al.(2020)Wang, Yang, Chong, Qiao, Peng, and
  Huang}]{wang2020post}
\bibinfo{author}{X.~Wang}, \bibinfo{author}{J.~Yang}, \bibinfo{author}{W.~T.~A.
  Chong}, \bibinfo{author}{P.~Qiao}, \bibinfo{author}{S.~Peng},
  \bibinfo{author}{X.~Huang},
\newblock \bibinfo{title}{Post-fracture performance of laminated glass panels
  under consecutive hard body impacts},
\newblock \bibinfo{journal}{Composite Structures} \bibinfo{volume}{254}
  (\bibinfo{year}{2020}) \bibinfo{pages}{112777}.
\bibitem[{Wang et~al.(2021)Wang, Meng, Yang, Huang, Wang, and
  Xu}]{wang2021optimal}
\bibinfo{author}{X.-e. Wang}, \bibinfo{author}{Y.~Meng},
  \bibinfo{author}{J.~Yang}, \bibinfo{author}{X.~Huang},
  \bibinfo{author}{F.~Wang}, \bibinfo{author}{H.~Xu},
\newblock \bibinfo{title}{Optimal kernel extreme learning machine model for
  predicting the fracture state and impact response of laminated glass panels},
\newblock \bibinfo{journal}{Thin-Walled Structures} \bibinfo{volume}{162}
  (\bibinfo{year}{2021}) \bibinfo{pages}{107541}.
\bibitem[{Wang et~al.(2018)Wang, Yang, Liu, and Zhao}]{wang2018experimental}
\bibinfo{author}{X.~Wang}, \bibinfo{author}{J.~Yang}, \bibinfo{author}{Q.~Liu},
  \bibinfo{author}{C.~Zhao},
\newblock \bibinfo{title}{Experimental investigations into sgp laminated glass
  under low velocity impact},
\newblock \bibinfo{journal}{International Journal of Impact Engineering}
  \bibinfo{volume}{122} (\bibinfo{year}{2018}) \bibinfo{pages}{91--108}.
\bibitem[{Zhao et~al.(2019)Zhao, Yang, Wang, and Azim}]{zhao2019experimental}
\bibinfo{author}{C.~Zhao}, \bibinfo{author}{J.~Yang},
  \bibinfo{author}{X.~Wang}, \bibinfo{author}{I.~Azim},
\newblock \bibinfo{title}{Experimental investigation into the post-breakage
  performance of pre-cracked laminated glass plates},
\newblock \bibinfo{journal}{Construction and Building Materials}
  \bibinfo{volume}{224} (\bibinfo{year}{2019}) \bibinfo{pages}{996--1006}.
\bibitem[{Flocker and Dharani(1998)}]{flocker1998low}
\bibinfo{author}{F.~W. Flocker}, \bibinfo{author}{L.~R. Dharani},
\newblock \bibinfo{title}{Low velocity impact resistance of laminated
  architectural glass},
\newblock \bibinfo{journal}{Journal of Architectural Engineering}
  \bibinfo{volume}{4} (\bibinfo{year}{1998}) \bibinfo{pages}{12--17}.
\bibitem[{Zemanová(2021)}]{git_data}
\bibinfo{author}{A.~Zemanová}, \bibinfo{title}{Supplementary data files for
  {Experimental} study on the gradual fracture of layers in multi-layer
  laminated glass plates under low-velocity impact},
  \bibinfo{howpublished}{\url{https://gitlab.com/zemanova.alena/e_mlg_lvi}},
  \bibinfo{year}{2021}.
\bibitem[{Nie et~al.(2009)Nie, Chen, Wereszczak, and Templeton}]{nie2009effect}
\bibinfo{author}{X.~Nie}, \bibinfo{author}{W.~W. Chen}, \bibinfo{author}{A.~A.
  Wereszczak}, \bibinfo{author}{D.~W. Templeton},
\newblock \bibinfo{title}{Effect of loading rate and surface conditions on the
  flexural strength of borosilicate glass},
\newblock \bibinfo{journal}{Journal of the American Ceramic Society}
  \bibinfo{volume}{92} (\bibinfo{year}{2009}) \bibinfo{pages}{1287--1295}.
\bibitem[{Datsiou and Overend(2018)}]{datsiou2018weibull}
\bibinfo{author}{K.~C. Datsiou}, \bibinfo{author}{M.~Overend},
\newblock \bibinfo{title}{Weibull parameter estimation and goodness-of-fit for
  glass strength data},
\newblock \bibinfo{journal}{Structural Safety} \bibinfo{volume}{73}
  (\bibinfo{year}{2018}) \bibinfo{pages}{29--41}.
\bibitem[{Osnes et~al.(2020)Osnes, Hopperstad, and B{\o}rvik}]{osnes2020rate}
\bibinfo{author}{K.~Osnes}, \bibinfo{author}{O.~S. Hopperstad},
  \bibinfo{author}{T.~B{\o}rvik},
\newblock \bibinfo{title}{Rate dependent fracture of monolithic and laminated
  glass: Experiments and simulations},
\newblock \bibinfo{journal}{Engineering Structures} \bibinfo{volume}{212}
  (\bibinfo{year}{2020}) \bibinfo{pages}{110516}.
\bibitem[{Zaccaria and Overend(2016)}]{zaccaria2016thermal}
\bibinfo{author}{M.~Zaccaria}, \bibinfo{author}{M.~Overend},
\newblock \bibinfo{title}{Thermal healing of realistic flaws in glass},
\newblock \bibinfo{journal}{Journal of Materials in Civil Engineering}
  \bibinfo{volume}{28} (\bibinfo{year}{2016}) \bibinfo{pages}{04015127}.
\bibitem[{Newmark(1959)}]{newmark1959method}
\bibinfo{author}{N.~M. Newmark},
\newblock \bibinfo{title}{A method of computation for structural dynamics},
\newblock \bibinfo{journal}{Journal of the Engineering Mechanics Division}
  \bibinfo{volume}{85} (\bibinfo{year}{1959}) \bibinfo{pages}{67--94}.
\bibitem[{Lenci et~al.(2015)Lenci, Consolini, and
  Clementi}]{lenci2015experimental}
\bibinfo{author}{S.~Lenci}, \bibinfo{author}{L.~Consolini},
  \bibinfo{author}{F.~Clementi},
\newblock \bibinfo{title}{On the experimental determination of dynamical
  properties of laminated glass},
\newblock \bibinfo{journal}{Annals of Solid and Structural Mechanics}
  \bibinfo{volume}{7} (\bibinfo{year}{2015}) \bibinfo{pages}{27--43}.
\bibitem[{Chen et~al.(2014)Chen, Xu, Yao, Xu, Liu, and Li}]{chen2014different}
\bibinfo{author}{J.~Chen}, \bibinfo{author}{J.~Xu}, \bibinfo{author}{X.~Yao},
  \bibinfo{author}{X.~Xu}, \bibinfo{author}{B.~Liu}, \bibinfo{author}{Y.~Li},
\newblock \bibinfo{title}{Different driving mechanisms of in-plane cracking on
  two brittle layers of laminated glass},
\newblock \bibinfo{journal}{International Journal of Impact Engineering}
  \bibinfo{volume}{69} (\bibinfo{year}{2014}) \bibinfo{pages}{80--85}.
\bibitem[{Mohagheghian et~al.(2018)Mohagheghian, Charalambides, Wang, Jiang,
  Zhang, Yan, Kinloch, and Dear}]{mohagheghian2018effect}
\bibinfo{author}{I.~Mohagheghian}, \bibinfo{author}{M.~Charalambides},
  \bibinfo{author}{Y.~Wang}, \bibinfo{author}{L.~Jiang},
  \bibinfo{author}{X.~Zhang}, \bibinfo{author}{Y.~Yan},
  \bibinfo{author}{A.~Kinloch}, \bibinfo{author}{J.~Dear},
\newblock \bibinfo{title}{Effect of the polymer interlayer on the high-velocity
  soft impact response of laminated glass plates},
\newblock \bibinfo{journal}{International Journal of Impact Engineering}
  \bibinfo{volume}{120} (\bibinfo{year}{2018}) \bibinfo{pages}{150--170}.

\end{thebibliography}
\end{document}